\documentclass{aa}
\usepackage[varg]{txfonts}
\usepackage[utf8]{inputenc}
\usepackage{natbib}
\bibpunct{(}{)}{;}{a}{}{,}
\usepackage{siunitx}
\usepackage{graphicx}
\usepackage{amsmath}
\usepackage{booktabs}
\usepackage{float}
\usepackage{makecell}
\usepackage{afterpage}
\usepackage{color}
\usepackage[usenames, dvipsnames]{xcolor}
\usepackage{pdflscape}
\usepackage{longtable}

\usepackage{hyperref}
\hypersetup{
    colorlinks=true,
	breaklinks=true,
	linkcolor=blue,
    citecolor=blue,
	filecolor=blue,
	urlcolor=blue,
	unicode=false,
	pdftoolbar=true,
	pdfmenubar=true,
	pdffitwindow=false,
	pdfstartview={Fit},
	pdftitle={Orion 3D dynamics},
	pdfauthor={Josefa Grossschedl},
	pdfsubject={Orion 3D dynamics},   					
	pdfcreator={Josefa Grossschedl},
	pdfkeywords={}, 
	pdfnewwindow=true,
	pdfdisplaydoctitle=true
}

\begin{document}

\title{3D dynamics of the Orion cloud complex}
\subtitle{Discovery of coherent radial gas motions at the 100-pc scale}

\titlerunning{3D dynamics of the Orion cloud complex}

\author{Josefa~E.~Gro\ss schedl\inst{1}
          \and
          Jo\~ao~Alves\inst{1,2}
          \and
          Stefan~Meingast\inst{1}
          \and
          Gabor~Herbst-Kiss\inst{1}
        }%
  
   \authorrunning{J.~Gro\ss schedl}

   \institute{University of Vienna, Department of Astrophysics, T\"urkenschanzstra{\ss}e 17, 1180 Wien, Austria, \\
            \email{josefa.elisabeth.grossschedl@univie.ac.at}
        \and
        University of Vienna, Data Science at Uni Vienna Research Platform, Austria
             }%
             
\date{Submitted to A\&A on July 13, 2020; Accepted on November 17, 2020}

\abstract{
We present the first study of the three-dimensional (3D) dynamics of the gas in the entire southern Orion cloud complex. We used the  parallaxes and proper motions of young stellar objects (YSOs) from \textit{Gaia} DR2 as a proxy for gas distance and proper motion, and the gas radial velocities from archival CO data, to compute the space motions of the different star-forming clouds in the complex, including subregions in Orion\,A, Orion\,B, and two outlying cometary clouds. From the analysis of the clouds' orbits in space and time, we find that they were closest about 6 Myr ago and are moving radially away from roughly the same region in space. This coherent 100-pc scale radial motion supports a scenario where the entire complex is reacting to a major feedback event, which we name the Orion-BB (big blast) event. This event, which we tentatively associate with the recently discovered Orion\,X stellar population, shaped the distribution and kinematics of the gas we observe today, although it is unlikely to have been the sole major feedback event in the region. We argue that the dynamics of most of the YSOs carry the memory of the feedback-driven star formation history in Orion and that the majority of the young stars in this complex are a product of large-scale triggering, which can raise the star formation rate by at least an order of magnitude, as for the head of Orion\,A (the Integral Shape Filament). 
Our results imply that a feedback, compression, and triggering process lies at the genesis of the Orion Nebula Cluster and NGC~2023/2024 in Orion\,B, thus confirming broadly the classical feedback-driven scenario proposed in \citet{Elmegreen1977}. The space motions of the well-known young compact clusters, $\sigma$\,Orionis and NGC\,1977, are consistent with this scenario. 
A momentum estimate suggests that the energy of a few to several supernovae is needed to power the coherent 3D gas motion we measure in this paper.}


\maketitle

\defcitealias{Grossschedl2018}{Paper\,I}
\defcitealias{Nishimura2015}{N15}
\defcitealias{Guieu2010}{G10}
\defcitealias{Kerr1986}{KL86}

\newcommand*\vhel{$v_\mathrm{HEL}$\xspace}
\newcommand*\vlsr{$v_\mathrm{LSR}$\xspace}
\newcommand*\galpy{\textit{galpy}\xspace}
\newcommand*\oriona{\texttt{Orion\,A}\xspace}
\newcommand*\orionb{\texttt{Orion\,B}\xspace}

\section{Introduction} \label{Intro}

Nearby molecular clouds are the only places where observations with the necessary detail can test star formation theories and infer the physics behind this fundamental process. The \object{Orion} star-formation complex \citep{Bally2008} is one of these regions since it is the closest region with ongoing massive star formation and it has a wide variety of different star formation environments. Much is known about this well-studied region such as mass estimates (atomic and molecular gas content); the magnetic field environment; the stellar populations (young stellar objects, YSOs, OB associations, and clusterings); several fundamental statistics such as the initial mass function or star formation rates; and line of sight dynamics \citep[e.g.,][]{Bally1987, Brown1994, Muench2002, Briceno2007b, Briceno2008, Reipurth2008, Muench2008, ODell2008, Meyer2008, Alves2012, Megeath2012, Furlan2016, Nishimura2015, Ochsendorf2015, Soler2018, Hacar2018, Grossschedl2019A, Kong2019, Feddersen2019}. However, all of these studies used projected two-dimensional (2D) observations, and many assumptions are necessary to derive physical properties that depend on the depth along the line of sight, the third dimension. With the deployment of \textit{Gaia} \citep{Prusti2016}, we can begin to extend the analysis of this benchmark region into three-dimensional (3D) space. \textit{Gaia}, especially its second data release \citep[DR2,][]{Brown2018}, provides high-quality parallaxes and proper motions for billions of stars for the first time.

Recently, we used \textit{Gaia} DR2 parallaxes to determine distances to the giant molecular cloud (GMC) \object{Orion A} \citep[][hereafter, Paper\,I]{Grossschedl2018}, by using YSO parallaxes as proxy for cloud distances. This analysis revealed a striking distance gradient from ``head'' to ``tail''\footnote{We refer to the high-mass star-forming parts of the cloud as head, including the Integral Shaped Filament \citep[ISF,][]{Bally1987}, the Orion Nebula \citep[\object{M42},][]{Muench2008, ODell2008}, and the Orion Nebula Cluster \citep[\object{ONC},][]{Hillenbrand1997,Hillenbrand2000,Muench2002}, and the low-mass star-forming parts as tail, including \object{L1641} and \object{L1647} \citep[][]{Allen2008}.}, resulting in an almost 100\,pc long structure, meaning that the cloud is at least twice as long as previously assumed. 
This distance gradient was already suggested by previous studies using other methods, for example, by \citet{Brown1994}, \citet{Schlafly2014}, or \citet{Kounkel2017a}, and then confirmed with \textit{Gaia} data by \citet{Grossschedl2018}, \citet{Kounkel2018}, \citet{Zucker2020}, \citet{Leike2020}, or \citet{Rezaei2020}.
The cloud's 3D structure analysis in \citetalias{Grossschedl2018} also revealed that the head of the cloud seems to be ``bent'' with respect to its tail, suggesting that external forces have shaped the region in the past. Knowing a cloud's 3D spatial structure allows one to break fundamental degeneracies, such as the interpretation of molecular line data \citep[position-position-velocity, PPV, e.g.,][]{Zucker2018b}. 
For example, it has been known for a long time from molecular line observations of the Orion\,A cloud that there is a ``jump'' in radial velocities at the location of the Orion Nebula Cluster (ONC) \citep[$\Delta v \sim \SI{2.5}{km/s}$, see][]{Tobin2009} and a mystifying velocity gradient from head to tail \citep[e.g.,][]{Kutner1977, Maddalena1986, Bally1987, Dame2001, Nishimura2015}. The gradient's origin has been attributed to either rotation \citep{Kutner1977, Maddalena1986} or large-scale expansion due to stellar winds \citep{Bally1987}. The third spatial dimension promises to test current Orion\,A models by disentangling radial velocity from 3D shape (PPPV).

This paper investigates if an external feedback event could be responsible for the inferred 3D shape of the cloud and its bulk motion in the Orion complex. Such feedback mechanisms, from previous generations of nearby massive stars, were already proposed in the past, for example, to explain the Orion-Eridanus superbubble \citep[e.g.,][]{Heiles1976, ReynoldsOgden1979, Bally1987, Brown1994, Brown1995, Ogura1998, LeeChen2009, Bally2010, Ochsendorf2015, Pon2016}. Alternatively, \citet{Fukui2018} propose that a cloud-cloud collision shaped the Orion\,A GMC near the ONC, which could also explain the observed bent head \citep[see also][]{Nakamura2012}. 
The head of the cloud produced about a factor of ten more stars than the tail, within the last 3 to 5\,Myr, as inferred from the distribution of the YSOs along the cloud \citep[][]{Grossschedl2018B}. Such increased star formation activity would fit a picture of triggered star formation by an external event at one end of the cloud while explaining the cloud's 3D shape.

A crucial piece of information needed to disentangle the various structure formation scenarios in Orion is its 3D space motion, requiring measurements of the unknown cloud's proper motions. An analysis of the 3D motions of individual subregions in Orion\,A would ideally reveal the physical status of the cloud (collapse, contraction, rotation, collision, passing) and be a useful discriminant between the various scenarios or even provide new insights into a new interpretation of the observables.

\begin{figure*}[!ht]
    \centering
        \includegraphics[width=1\linewidth]{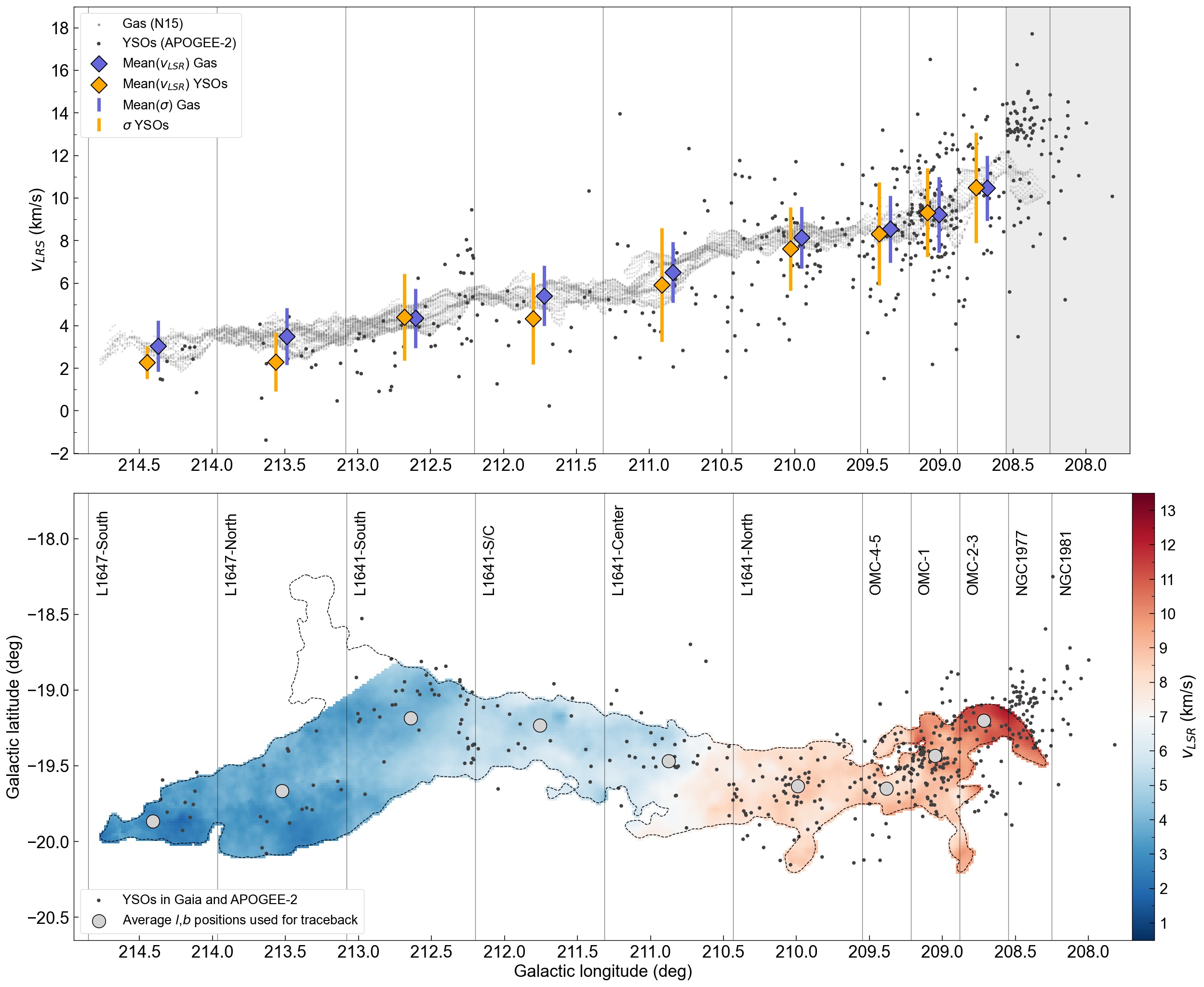}
    \caption{%
    Determination of average line of sight motions for Orion\,A.
    \textit{Top:} PV-diagram (\vlsr vs $l$) for gas and YSOs in Orion\,A. The gas velocities (gray small dots) were extracted as the 1st\,Moment of the $^{12}$CO(2-1) map from \citet{Nishimura2015}. Only pixels within a smoothed column-density contour of $A_\mathrm{K} > \SI{0.5}{mag}$ were used to avoid contamination by background emission. The black dots show the selected YSO sample that satisfies the \textit{Gaia} and \mbox{APOGEE-2} quality criteria.  The averages per region for YSOs and gas are shown with orange and blue diamonds, respectively, with the error bars representing the velocity dispersion per region (see also Fig.~\ref{fig:pv-sigma-oriona}). 
    The $l$-positions of these symbols were shifted from mid-bin-positions for better visibility.
    \textit{Bottom:} Velocity map as derived from the CO measurements per pixel.
    The pixel values are given in \vlsr, as shown by the color-scale, and were used for the PV-diagram (top panel).
    The nine subregions are labeled within the bins at the top of the frame, and their average $l,b$ positions are shown as gray circles, which are used for the traceback. The bins enclosing NGC\,1977 and NGC\,1981 were excluded (gray shaded area, top panel), since the gas and YSOs appear decoupled.
    }
    \label{fig:pv-gas-ysos}
\end{figure*}

\begin{figure}[!ht]
    \centering
        \includegraphics[width=1.02\linewidth]{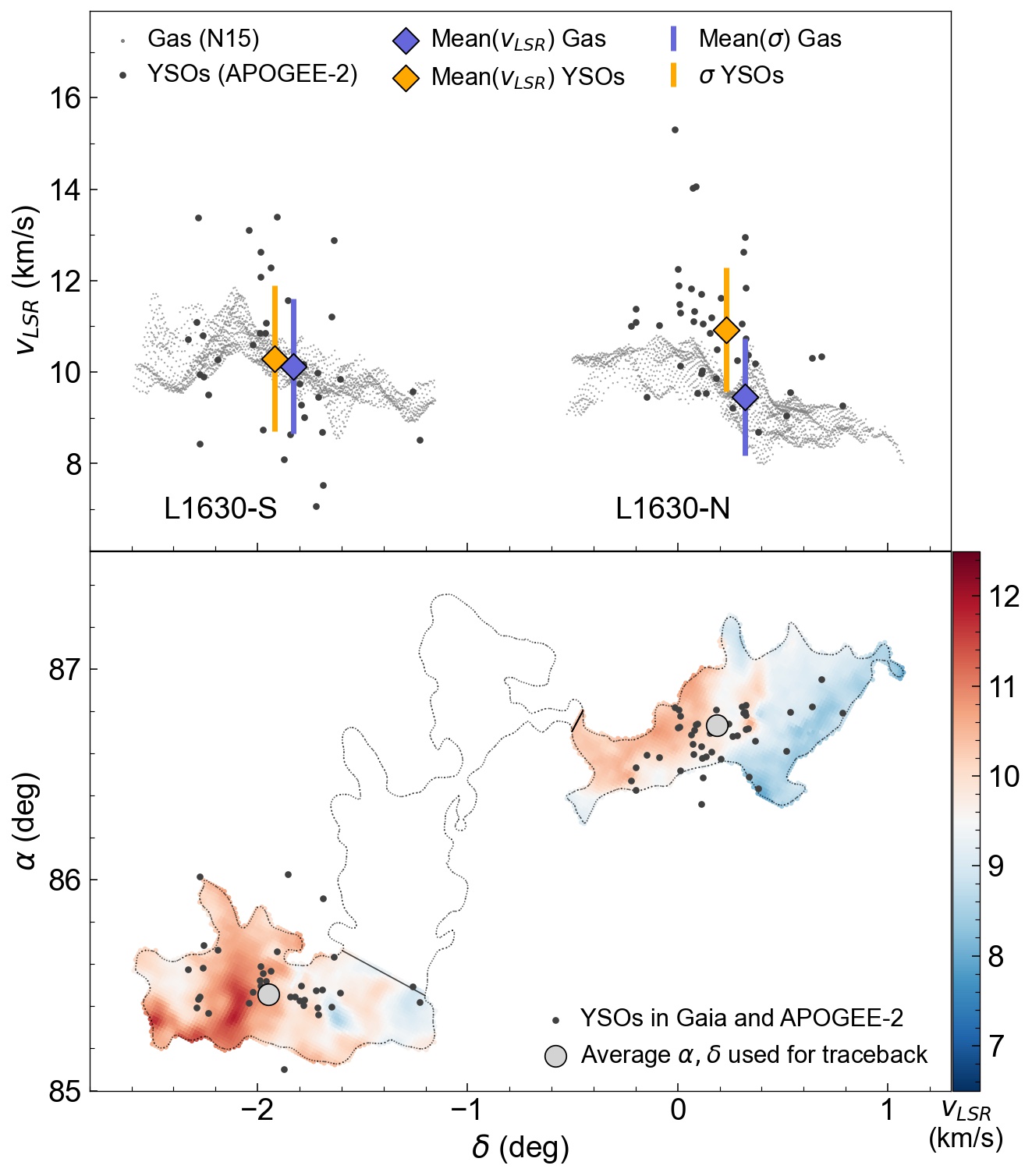}
    \caption{%
    Determination of average line of sight motions for Orion\,B. Similar as Fig.~\ref{fig:pv-gas-ysos}. 
    \textit{Top:} PV-diagram ($\delta$ vs \vlsr) for gas and YSOs in Orion\,B. 
    \textit{Bottom:} Map of the selected L1630-S/N molecular clouds. The dotted line is a smoothed column-density contour ($A_K > \SI{0.4}{mag}$), to avoid contamination by background emission. The middle part of L1630 (empty contour) was excluded due to lower intensity CO measurements, little star formation activity, and uncertain distance estimates. The gray circles mark the average positions of the two subregions. See Fig.~\ref{fig:pv-gas-ysos} and text for more explanations.
    }
    \label{fig:pv-gas-ysos_orionb}
\end{figure}

Directly measuring proper motions of diffuse objects such as clouds is virtually impossible. However, one can equate cloud proper motion with the average proper motion of the youngest embedded sources inside a cloud. 
Using YSOs as a proxy for cloud proper motions is, to first order, justified because (1) these objects are still very young and close to their birth sites \citep[e.g.,][]{Dunham2015, Heiderman2015, Grossschedl2019A} and (2) there is solid evidence that the YSOs share, on average, the same velocity properties as their parental cloud. For example, the YSOs in Orion\,A share the same radial velocity as the molecular gas  \citep[e.g.,][and Fig.~\ref{fig:pv-gas-ysos}]{Furesz2008, Tobin2009, Hacar2016b}, also seen in Orion\,B \citep[e.g.,][and Fig.~\ref{fig:pv-gas-ysos_orionb}]{Kounkel2017b}. It is then very likely that, on average, YSOs have the same proper motion as the gas from which they formed.
Until recently, there were no proper motions available for a statistically significant sample of young sources in the Orion molecular clouds. There have been estimates of proper motions of a handful of young embedded sources from VLBI radio observations \citep[e.g.,][]{Menten2007, Kounkel2017a, Reid2014, Reid2016, Reid2019a}, but they often do not agree with each other. 
A possible reason for this situation is the sample size and as a result the possibility that peculiar motions could dominate a small sample, including observations of multiple stellar systems. 
Proper motions of less embedded YSOs observed by \textit{Gaia} have more than an order of magnitude of better statistics, hence are the best probe currently available to infer average gas motion.

The goal of this paper is to derive, for the first time, the 3D space motions of subregions in the Orion cloud complex to analyze the clouds' large-scale dynamics and possibly illuminate the star formation history and existing formation mechanism scenarios for this benchmark region. We first describe the necessary steps to combine \textit{Gaia} DR2 parallaxes and proper motions of YSOs with radial velocity measurements from spectroscopic surveys and molecular line observations (Sect.~\ref{Data}) to achieve an estimate of the space motion of the gas. The methods are presented in Sect.~\ref{Methods} and the results in Sect.~\ref{Results}. We discuss our results and their implications in Sect.~\ref{Discussion} and we summarize our work in Sect.~\ref{Summary}.

\begin{figure*}[!t]
\small
    \centering
        \begin{minipage}[t]{0.47\linewidth}
        \centering
        \includegraphics[width=\linewidth]{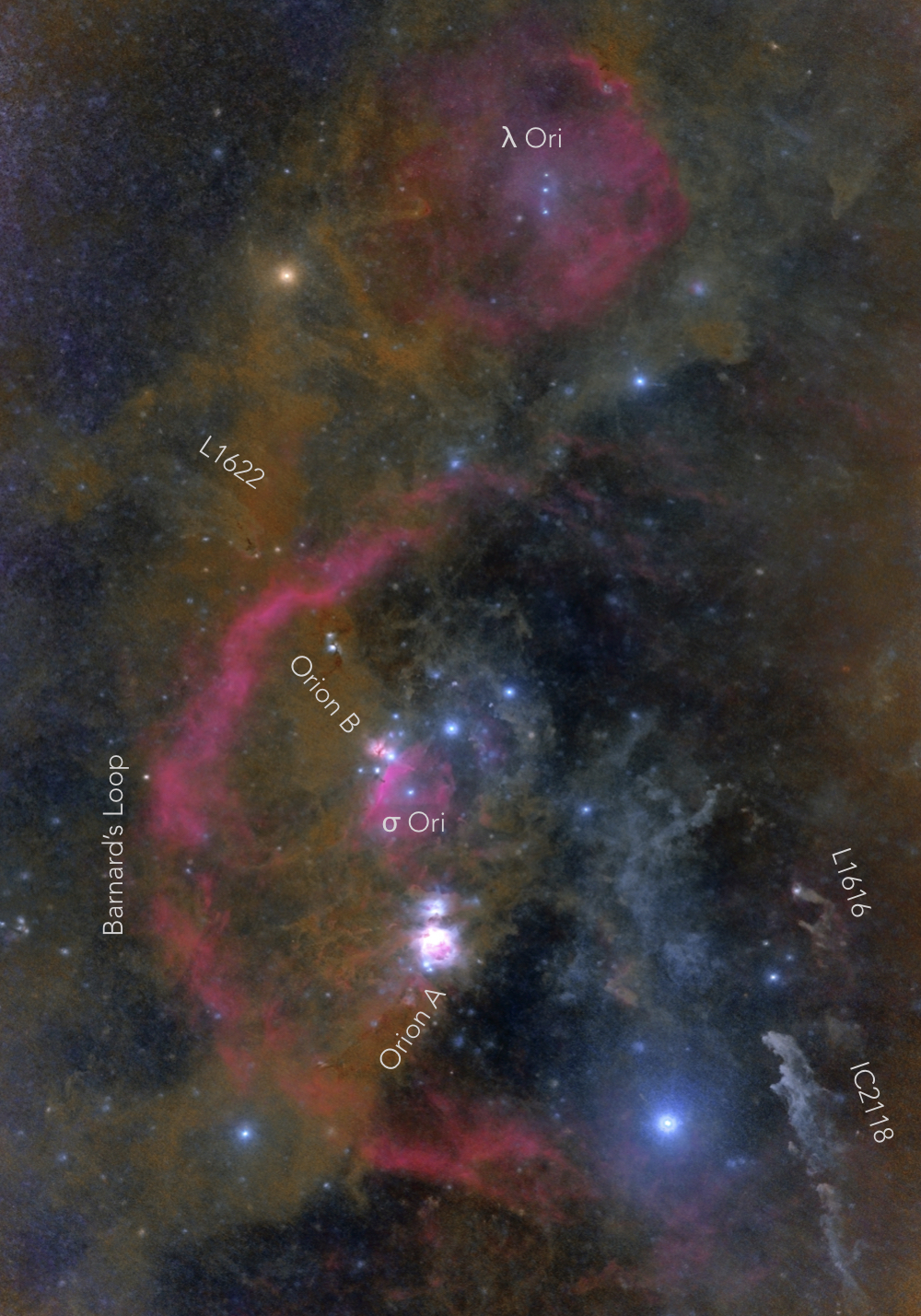}
    \end{minipage}%
    \hfill
    \begin{minipage}[t]{0.52\linewidth}
        \centering
        \includegraphics[width=\linewidth]{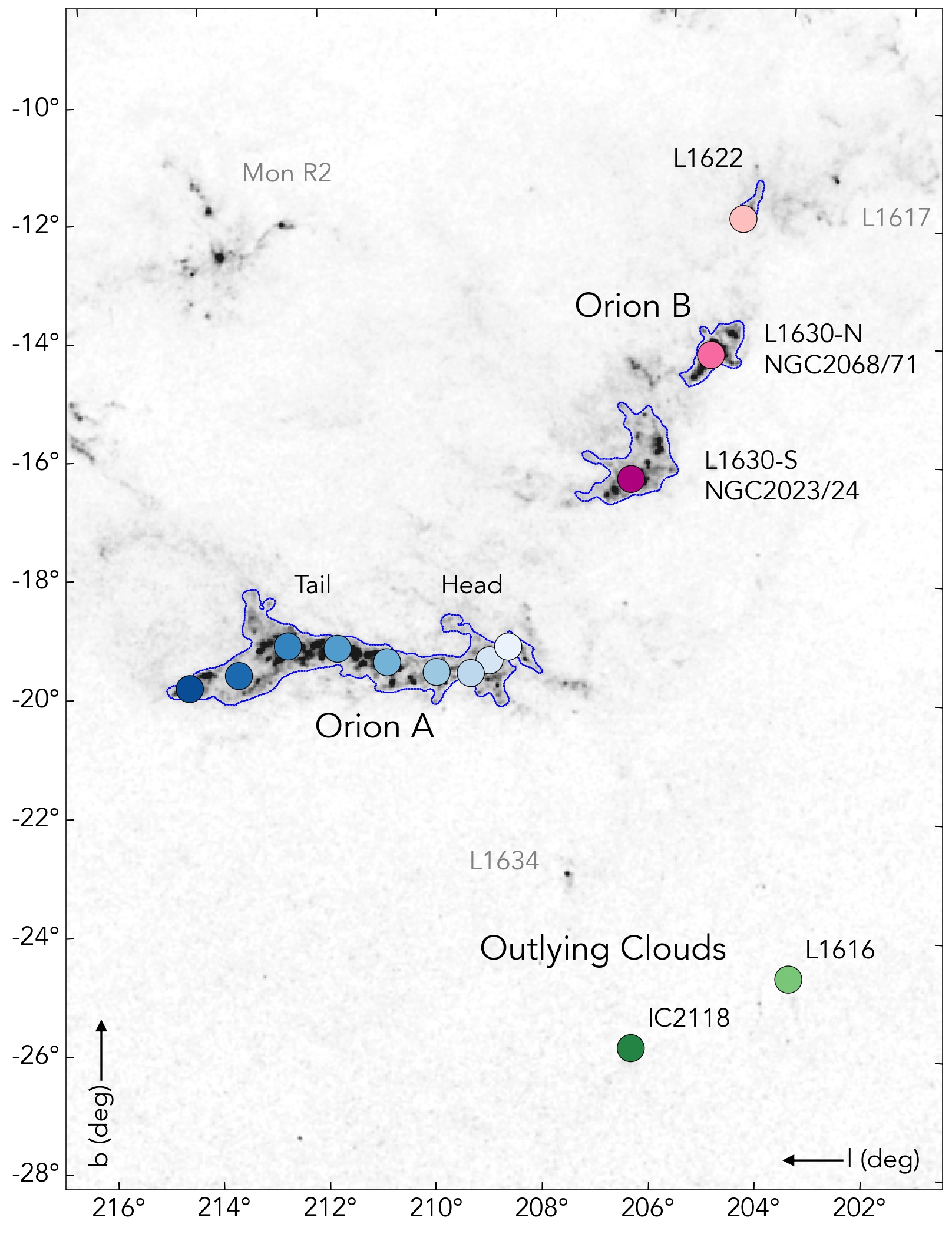}
    \end{minipage}%
    \caption{Overview of the Orion molecular cloud complex. \textit{Left:} Optical image of the Orion region (orientation roughly equatorial). Captured from deep in the heart of the Kalahari desert by Chris Beere\textsuperscript{\it a}. This image highlights well the interplay between the large-scale nebula and the dusty structures. \object{L1622} shows a large tail of scattered light that is clearly distinct from the background.
    \textit{Right:} Overview of the studied subregions in galactic orientation. The gray-scale shows regions of high extinction from a 2MASS near-infrared extinction map \citep[NICER,][]{Lombardi2011} ($A_\mathrm{K}$-range shown from 0 to 1.4\,mag). The blue contours are smoothed extinction contours (at $A_\mathrm{K} = \SI{0.42}{mag}$), only shown for the molecular clouds of interest. The colored circles are the center locations of the selected 14 subregions within the three main regions: Orion\,A (blue), Orion\,B (magenta), and the two outlying cometary clouds (green, the green circles hide the blue contours of the small cometary clouds). Additional regions are labeled for completeness in gray. The color-scaling of the filled circles is used throughout the paper to identify the subregions within the clouds. 
{\bf Notes.} \textsuperscript{({\it a})} \url{https://www.astrobin.com/users/chrisbeere}. Color in this optical image is as shot with custom white balance, processed with Starnet++ to highlight extended structures.
	}
    \label{fig:overview}
\end{figure*}


\section{Data} \label{Data}

In this section, we describe the studied subregions (Fig.~\ref{fig:overview}) and the data needed to enable an analysis of the 3D spatial motion of molecular clouds in Orion. The most prominent clouds in Orion are the GMCs \object{Orion A} and \object{Orion B}, which are both active star-forming regions, containing hundreds of YSOs. Additionally, we investigated three cometary-shaped outlying clouds, two in the southwest, \object{L1616} and \object{IC2118}, and one in the northeast, \object{L1622} \citep{Alcala2008}. 
These cometary clouds show sufficient star formation to be included in our work. 
Henceforth, we address the three studied main regions separately as Orion\,A, Orion\,B, and outlying clouds, while L1622 is described in more detail with the Orion\,B main regions (Sects.~\ref{OB-Data} and \ref{OB-Methods}) due to its projected position and partially overlapping data coverage.

To derive the average positions and velocities for each region, we used \textit{Gaia} DR2 parallaxes and proper motions\footnote{Gaia Archive: \url{https://gea.esac.esa.int/archive/}} and APOGEE-2\footnote{Apache Point Observatory Galactic Evolution Experiment, \url{https://www.sdss.org/dr16/irspec/spectro_data}} radial velocities of YSO members of the studied clouds (Sect.~\ref{yso-data}), and gas radial velocities obtained from CO emission line surveys (Sect.~\ref{gas-rv}). 
An overview of the studied Orion complex is shown in Fig.~\ref{fig:overview}, and a more detailed description of each subregion is given in Sect.~\ref{studied_regions}. 
We refer to Appendix~\ref{apx:sample-selection} for a detailed description of the quality cuts and YSO sample selections. 
For clarity, we introduce here the observed position and velocity parameters that are used throughout the paper: \\

\noindent $\alpha$, $\delta$ (deg): Right Ascension and Declination \\
$l$, $b$ (deg): Galactic longitude and latitude \\
$\varpi$\,(mas): parallax \\
$d$\,(pc): distance as derived from $1000/\varpi$ \\
$\mu_{\alpha*}\,\si{(mas/yr)}$: $\mu_{\alpha} \cos(\delta)$, proper motion along $\alpha$ \\
$\mu_\delta\,\si{(mas/yr)}$: proper motion along $\delta$ \\
$v_\alpha\,\si{(km/s)}$: tangential velocity along $\alpha$ \\
$v_\delta\,\si{(km/s)}$: tangential velocity along $\delta$ \\
\vhel\,(km/s): Heliocentric radial velocity \\
\vlsr\,(km/s): radial velocity relative to the local standard of rest (LSR)

\subsection{Collecting YSO samples} \label{yso-data}

We used YSOs with infrared-excess (Class\,II or earlier classes) for our analysis, to include only the youngest sources for each cloud, which are the most likely candidates to be located still close to their birth-sites. 
To get the best available YSO statistics we combined archival YSO catalogs with additional YSO selections (Appendix~\ref{apx:sample-selection}), while all YSO samples include a \textit{Gaia} quality criteria cut (Appendix~\ref{apx:gaia}).

First, we collected data from the literature containing YSOs and/or radial velocity measurements of young stellar members in the Orion regions of interest \citep{Alcala2004, Flaherty2008, Guieu2010, Megeath2012, Megeath2016, Kounkel2017b, Kounkel2018, Grossschedl2019A}.
Next, we added additional YSO candidates with infrared-excess by cross-matching \textit{Gaia} DR2 with the AllWISE \citep{Cutri2013} and 2MASS \citep{Skrutskie2006} catalogs using the WISE-best-neighbor and 2MASS-best-neighbor (provided in the \textit{Gaia} archive) to do photometric YSO selections using infrared colors (Appendix~\ref{apx:yso-wise-selection}). Henceforth, we call this the WISE-2MASS selection.
Such additional selections were applied for all regions, except for Orion\,A, for which an extended YSO search is already presented in \citet{Grossschedl2019A}.
Finally, to get consistent high quality radial velocities of the YSOs, we cross-matched with SDSS DR16 APOGEE-2 data within \SI{1}{\arcsec} \citep[][Majewski et al.~in prep.]{Majewski2017, Wilson2019, Joensson2020}.
The APOGEE-2 survey provides infrared spectroscopy, making it ideal to study especially young or embedded stars, and it provides overall superior radial velocities compared to \textit{Gaia}. Typical measurement errors of APOGEE-2 radial velocities that pass the applied quality criteria (Appendix~\ref{apx:apogee-cut}) are on the order of $\sim$0.05\,km/s. 
The SDSS APOGEE-2 survey is not all-sky and does not include all of our studied cloud regions, but significant regions in Orion\,A and B are included.  
For regions that were not observed by APOGEE-2 we used other resources to obtain radial velocity data. For details see Sect.\,\ref{studied_regions}.

In the following, the YSO data available for the individual subregions will be described in more detail in Sects.~\ref{OA-Data} to \ref{OC-Data}. In the Tables~\ref{tab:overview}, \ref{tab:averages_rv_pm}, and \ref{tab:averages_rv} an overview of the regions and of the respective data references is given. 
For each subregion we applied individual selections in projected coordinate space ($l$,$b$), proper motion space, distance, and for some regions in radial velocity, as described in more detail in Appendix~\ref{apx:regions}.

\subsection{Gas radial velocities} \label{gas-rv}

To get a direct estimate of a cloud's line of sight motion, we used gas radial velocities from molecular emission line surveys. Primarily, we used the \citet[][hereafter N15]{Nishimura2015} molecular emission line survey of $^{12}$CO(2-1) (at 230.54 GHz, beam size HPBW = $2.7'$, pixel scale $1'$), which covers both Orion\,A and Orion\,B (see Figs.~\ref{fig:pv-gas-ysos} and \ref{fig:pv-gas-ysos_orionb}). We compared \citetalias{Nishimura2015} to the \citet{Kong2018} CARMA-NRO Orion Survey, a high-resolution survey of the northern parts in Orion\,A, and found that \citetalias{Nishimura2015} $^{12}$CO(2-1) radial velocities agree on average well with $^{12}$CO(1-0), $^{13}$CO(1-0), and C$^{18}$O(1-0) CARMA radial velocities. Since we are only interested in average motions, the resolution of \citetalias{Nishimura2015} is sufficient for our purposes. 

To obtain gas \vlsr from the \citetalias{Nishimura2015} $^{12}$CO(2-1) emission line map, we used the \textit{Astropy} python package \texttt{spectral\_cube}. We extract subcubes for Orion\,A and Orion\,B (\object{L1630}-S/N), enclosing the regions of interest.
To extract \vlsr for each line of sight (pixel) we first smoothed the velocity channels to 1\,km/s resolution (original velocity resolution $\sim$0.08 km/s), to mitigate problems with line identification due to noise.
From the smoothed map we chose the lines by identifying local maxima and setting individual velocity ranges for each line, which excludes velocity channels outside the line-range of interest, and we calculated the 1st Moments for each selected line range (see Figs.~\ref{fig:pv-gas-ysos} and \ref{fig:pv-gas-ysos_orionb}). The result is similar to a single component Gauss-fit, where double-peaked lines are ignored, since we are only interested in the average bulk motion of larger cloud parts and not in individual line of sight properties or line substructures.
The average gas motions for subregions in Orion\,A and B were then estimated by using only pixels within a chosen extinction contour using a Herschel/Planck map \citep[][]{Lombardi2014} (smoothed outer contour at: A$_\mathrm{K} >0.5$\,mag for Orion\,A; A$_\mathrm{K} >0.4$\,mag for Orion\,B) to reduce background contamination. The average gas line of sight motions were then calculated from the Mean(\vlsr) pixel values per subregion, which delivers a measure of the bulk motion of the selected cloud parts.
To get a measure of the velocity dispersion (from the line-width) we calculated the square root of the 2nd moment of the chosen line-ranges (sigma map, see Figs.~\ref{fig:pv-sigma-oriona} and \ref{fig:pv-sigma-orionb} in Appendix\,\ref{aux:figures}). The sigma values (velocity dispersion) of the pixels within the displayed extinction contours for Orion\,A and B (see Sects.~\ref{OA-Data}, \ref{OB-Data}, and Figs.~\ref{fig:pv-gas-ysos}, \ref{fig:pv-gas-ysos_orionb}) scatter around 1.4\,km/s. This is on average on the same order as the velocity dispersion of the selected YSO samples. The velocity dispersion of individual subregions shows some variations, as can be seen in Table~\ref{tab:averages_rv_pm}, \ref{tab:averages_rv}, and Figs.~\ref{fig:pv-gas-ysos}, \ref{fig:pv-gas-ysos_orionb}, \ref{fig:pv-sigma-oriona}, \ref{fig:pv-sigma-orionb}.

The \citetalias{Nishimura2015} map does not cover all of our regions. When other molecular line observations were used we list them in Sects.~\ref{OA-Data} to \ref{OC-Data}, where we give short overviews for each of the three main regions and briefly address issues concerning data availability.
Using different molecular emission line surveys could implicate systematic differences between the studies, which are not easy to account for. Moreover, each of these studies reports the gas radial velocities relative to the local standard of rest (\vlsr). For our purposes, however, we require the heliocentric radial velocity (\vhel) as starting condition to convert the motions of all regions consistently to motions relative to LSR. 
Unfortunately, the conversions from \vhel to \vlsr, as derived from gas observations, are not mentioned explicitly in the various publications \citep[see also][and the discussion in Appendix\,\ref{apx:lsr}]{Hacar2016b} and can not be compared with each other at face value. We converted back to \vhel with the best possible guess for each data set. For \citetalias{Nishimura2015} we assumed the widely used standard solar motion of 20\,km/s as recommended by the IAU and stated in \citet[][hereafter KL86, see also Table~\ref{tab:astropy}]{Kerr1986}. We used this LSR conversion to determine \vhel for the gas, if not stated otherwise. Inaccurately converted velocities can lead to additional errors in the evaluation of the dynamical evolution of the studied regions. This does, however, not affect the main result in this work, as addressed in Appendix\,\ref{apx:lsr}.

\subsection{Studied clouds} \label{studied_regions}

Here we provide an overview for the studied molecular clouds in Orion, while we focus on data availability for individual regions. In particular, we focus on the evaluation of radial velocities, since this observable is the most inhomogeneously derived value in our study.
Detailed numbers and selection procedures are given in Tables~\ref{tab:overview} to \ref{tab:averages_rv} and in Appendix~\ref{apx:sample-selection}.

\subsubsection{Orion A} \label{OA-Data}

The GMC Orion\,A is the best studied cloud in our sample and a wealth of data is available for this region, including information on the stellar and the gaseous content. 
Orion\,A contains the following cloud parts, which are included in the analyzed subregions; these are the Orion Molecular Clouds \object{OMC1}, \object{OMC2}, \object{OMC3},  \object{OMC4}, \object{OMC5} \citep[][head of Orion\,A]{Peterson2008, Muench2008, ODell2008}, and the Lynds dark clouds\footnote{Lynds dark clouds are always abbreviated with ``L'' in-front of the individual number.} \object{L1641} and \object{L1647} \citep[][tail of Orion\,A]{Allen2008}. The clouds L1641/L1647 are further subdivided into five subregions: L1641-N, L1641-C, L1641-C/S, L1641-S, L1647-N, L1647-S (see Fig.~\ref{fig:pv-gas-ysos}). A YSO sample is taken from the YSO catalog of \citet{Grossschedl2019A}, containing 2980 YSOs with infrared-excess, a catalog based on a Spitzer\footnote{Spitzer Space Telescope \citep{Werner2004, Gehrz2007}.}, WISE, 2MASS, and VISION\footnote{VIenna Survey In OrioN, an ESO VISTA near-infrared survey by \citet{Meingast2016}.} photometric selection \citep[see also][]{Megeath2012, Megeath2016, Furlan2016}. 
To get stellar parameters we first applied \textit{Gaia} quality criteria, as given in Appendix~\ref{apx:gaia}. For the Orion\,A region we apply the following additional criteria:
$|\mu_{\alpha*}|,|\mu_{\delta}| < \SI{5}{mas/yr}$, $207^\circ<l<215^\circ$, and $-20.5^\circ<b<-18.2^\circ$. With this we retain about 31\% of the original YSO catalog. Radial velocities from APOGEE-2 are also available for a significant subsample of YSOs ($\sim$30\%) with applied radial velocity quality criteria (Appendix\,\ref{apx:apogee-cut}, additional cut, $16>v_\mathrm{HEL}\,(\mathrm{km/s})>36$). When combining \textit{Gaia} and APOGEE-2 criteria there are about $\sim$15\% of the original YSO catalog left (see Tables~\ref{tab:averages_rv_pm} and \ref{tab:averages_rv} for detailed numbers per subregion). To obtain gas radial velocities for Orion\,A we used the mentioned \citetalias{Nishimura2015} $^{12}$CO(2-1) map, which covers the whole cloud area (Fig.~\ref{fig:pv-gas-ysos}), as described in Sect.~\ref{gas-rv}.

\subsubsection{Orion B} \label{OB-Data}

The same surveys that cover Orion\,A largely cover the main clouds of Orion\,B. A Spitzer/2MASS selected YSO sample for Orion\,B and L1622 \citep{Megeath2012, Megeath2016} contains 663 YSO candidates with infrared-excess of which about 25\% pass our \textit{Gaia} quality criteria. Spitzer covered the most prominent cloud parts, which can be split up into three regions: the two main regions containing young prominent clusters in L1630 \citep[e.g.,][]{ELada1991A, ELada1991B}, \object{NGC\,2023}/\object{NGC\,2024} \citep[e.g.,][]{Meyer2008},  and \object{NGC\,2068}/\object{NGC\,2071} \citep[e.g.,][]{Gibb2008}, and the cometary cloud L1622 \citep[e.g.,][]{Reipurth2008B, Bally2009}.
There is no similar extended YSO selection available for Orion\,B as in \citet{Grossschedl2019A} for Orion\,A.
Due to the smaller sample compared to Orion\,A, we searched for additional YSO candidates in the surroundings (including regions not observed by Spitzer) using the WISE-2MASS selection, with the criteria and numbers given in Appendix~\ref{apx:yso-wise-selection}. 
Following, we discuss the Orion\,B main cloud (L1630) and L1622 separately due to different data coverage, especially concerning radial velocities. 

\textit{\sffamily L1630 South and North}: 
The Orion\,B main cloud L1630 can be split into two major components with significant active star formation, which are the clusters NGC\,2023/2024 in the south (L1630-S), and NGC\,2068/2071 in the north (L1630-N). For the Orion\,B main parts we used \citetalias{Nishimura2015} $^{12}$CO(2-1) gas radial velocities. They contain the majority of the Spitzer selected YSO candidates (635, 96\%) from the \citet{Megeath2012} survey, and we were able to extend the YSO sample with the mentioned WISE-2MASS selection (Appendix~\ref{apx:yso-wise-selection}). 
APOGEE-2 radial velocities are available for these two regions, which we compared to other radial velocity measurements in Orion\,B of young stellar members by \citet{Flaherty2008} and \citet{Kounkel2017b}. We find that they are generally in agreement with each other within the errors. If anything, there is a slight blue-shift of the \citet{Kounkel2017b} radial velocities in NGC\,2023 relative to \mbox{APOGEE-2} radial velocities, while not significant within the errors. For our analysis, we used \mbox{APOGEE-2} radial velocities due to smaller measurement errors and consistency with Orion\,A. 
After applying the \textit{Gaia} quality criteria and further individual region selections (Appendix~\ref{apx:regions}, Fig.~\ref{fig:orionb_regions}), we ended up with 57 YSOs for L1630-S and 74 YSOs for L1630-N, while 37 and 45 of these pass the additional \mbox{APOGEE-2} quality criteria (Appendix~\ref{apx:apogee-cut}). 

\textit{\sffamily The L1622 cloud}:
This is a small cometary cloud northeast to the Orion\,B main clouds and is likely located in-front of these \citep{Reipurth2008}. It is also called Orion East in \citet{Wilson2005}.
To the west lies a cluster of further cometary clouds called \object{L1617} (Fig.~\ref{fig:overview}) showing only little star formation activity, hence no YSOs were observed by \textit{Gaia}.
The \citet{Megeath2012} Spitzer catalog for L1622 contains 28 YSO candidates. After extending the sample with the WISE-2MASS YSO selection (Appendix~\ref{apx:yso-wise-selection}), applying \textit{Gaia} quality criteria and additional individual selection criteria (Appendix~\ref{apx:regions}, Fig.~\ref{fig:orionb_regions}) we ended up with 8 YSOs for L1622. 
This cloud was not covered by \mbox{APOGEE-2}, but radial velocity measurements are available from \citet{Kounkel2017b} for four young sources, which match the radial velocity and distance criteria of that region. On average these YSOs have $v_\mathrm{HEL} = 19.3 \pm 1.2\,\si{km/s}$, while the measurement errors are on average on the order of about 2\,km/s. Gas \vlsr for L1622 are reported in several independent studies. \citet{Maddalena1986} listed this cloud as CO clump Nr.\,38 and reported a value of \vlsr of 0.7\,km/s.
The region was also covered by the large-scale survey of \citet{Dame2001}, and discussed in \citet{Wilson2005}\footnote{Data from Harvard Dataverse \citep{Wilson2011_COdata}, \url{https://dataverse.harvard.edu/dataset.xhtml?persistentId=doi:10.7910/DVN/MW6HM7}}, who find on average a \vlsr of about 1\,km/s toward L1622. Moreover, \citet{Park2004} present a study of star-less cores including parts of L1622, which have on average a \vlsr of about 1.13\,km/s. Finally, \citet{Kun2008} observed the cloud with NANTEN, providing $^{12}$CO and $^{13}$CO emission line maps. They report an average \vlsr of $1.17 \pm 0.71$\,km/s toward L1622, in rough agreement with the previous studies. We adopt the value from \citet{Kun2008}, which is converted to heliocentric \vhel = 17.96\,km/s, matching within the errors with stellar radial velocities.

\subsubsection{Outlying clouds} \label{OC-Data}

We include in our study two further cometary-shaped star-forming clouds, located in the southwestern region of Orion, which are part of the outlying clouds \citep{Alcala2008}. This group includes L1616 and IC\,2118 (Witch Head Nebula). Close to these regions lies \object{L1634} (Fig.~\ref{fig:overview}), which we first intended to included in our analysis. However, the region shows an apparent overlap between two seemingly distinct clouds in reflection (optical) and dust emission, and a clear determination of YSO membership in this region was not possible. As a consequence, we did not include L1634 in our analysis.

\textit{\sffamily The L1616 cloud}: 
The cometary cloud L1616 \citep{Park2004, Alcala2004, Gandolfi2008, Alcala2008} is part of a sparse cloud structure to the southwest of the Orion GMCs. 
\citet{Alcala2004} present a list of 30 pre-main-sequence stars near L1616 of which 22 have measured radial velocities.
After adding YSOs with the WISE-2MASS selection (Appendix~\ref{apx:yso-wise-selection}) and applying \textit{Gaia} quality and individual selection criteria, we ended up with 18 YSO candidates (Appendix~\ref{apx:regions}, Fig.~\ref{fig:outlying-regions}).
The radial velocity measurements in \citet{Alcala2004} scatter around \vhel $22.3 \pm \SI{4.6}{km/s}$ \citep[see also][]{Gandolfi2008}. This average stellar radial velocity is consistent with the gas radial velocity reported in \citet[][CO clump Nr.13]{Maddalena1986} who report a value of \vlsr$\sim\SI{7.7}{km/s}$, which is converted to \vhel $\sim$ 22.6\,km/s, when using the standard solar motion from \citet{Mihalas1981}. However, we ended up with only four YSOs from \citet{Alcala2004} that are within our selection criteria, which have on average a \vhel of $24.5 \pm \SI{2.7}{km/s}$. The individual measurement errors (2 to $\SI{2.3}{km/s}$) of these four sources are on the same order as the standard deviation, hence the discrepancy to gas \vhel is likely not significant, also considering the small number statistics.
The L1616 cloud belongs to a slightly larger sparse cloud structure containing another cometary pillar, \object{[CB88]\,28} (\citealt{Clemens1988}), which, however, does not show signs of active star formation and can not be used as probe in our analysis.

\textit{\sffamily The Witch Head Nebula -- IC 2118}: 
This cometary shaped cloud is mostly known as prominent reflection nebula, located southeast of L1616 in the proximity of the supergiant Rigel \citep[spectral type B8Ia, $\varpi = 3.78$\,mas,][]{MerrillBurwell1943, vanLeeuwen2007, Shultz2014}.
\citet{Guieu2010} report 17 pre-main-sequence stars for IC\,2118, of which 10 are Spitzer selected YSOs. All of these are located in the northern part of IC\,2118, at the top of the Witch Head Nebula. We also applied our WISE-2MASS selection criteria in this region (see Appendix~\ref{apx:yso-wise-selection}). This, however, did not change the original \citet{Guieu2010} selection within the \textit{Gaia} quality criteria. Our final sample for IC\,2118 contains five YSOs (Appendix~\ref{apx:regions}, Fig.~\ref{fig:outlying-regions}). 
We extracted gas radial velocity measurements for this region from \citet{Kun2001} ($^{12}$CO(1-0) NANTEN 4m Radio Telescope). They report a \vlsr of $-2.2 \pm 1.8$\,km/s for the northern part of the could, which corresponds to the region where the small cluster of YSOs is located. This converts to \vhel of 15.4 km/s when using the standard solar motion from \citetalias{Kerr1986}.
\citet{Kun2001} report radial velocity variations across the whole Witch Head Nebula of about 10\,km/s, while in this paper we only focus on the small part at the top of the cloud containing YSOs, and we do not discuss this gradient further.
For this cloud there are no stellar radial velocities available to be compared to gas radial velocities.

\begin{table*}[!ht] 
\begin{center}
\small
\caption{Overview of the discussed 14 subregions, including the cluster Orion\,X. The table lists average cloud positions in $l,b$, as well as $\alpha,\delta$, and furthermore contains the standard deviation in $\alpha,\delta$ which was used to estimate traceback errors.}
\renewcommand{\arraystretch}{1.2}
\begin{tabular}{cllcccccc}
\hline \hline
\multicolumn{1}{l}{Label} &
\multicolumn{1}{l}{Region} &
\multicolumn{1}{l}{Subregion} &
\multicolumn{1}{c}{$l$} &
\multicolumn{1}{c}{$b$} &
\multicolumn{1}{c}{$\alpha$} &
\multicolumn{1}{c}{$\delta$} &
\multicolumn{1}{c}{$\sigma_\alpha$} &
\multicolumn{1}{c}{$\sigma_\delta$} \\
& & & 
\multicolumn{1}{c}{(deg)} & 
\multicolumn{1}{c}{(deg)} &
\multicolumn{1}{c}{(deg)} & 
\multicolumn{1}{c}{(deg)} &
\multicolumn{1}{c}{(deg)} & 
\multicolumn{1}{c}{(deg)} \\
\hline
1 & Orion\,A (tail) & L1647-S & 214.41 & -19.87 & 85.64 & -10.17 & 0.10 & 0.09 \\
2 & Orion\,A (tail) & L1647-N & 213.53 & -19.67 & 85.46 & -9.33 & 0.33 & 0.25 \\
3 & Orion\,A (tail) & L1641-S & 212.64 & -19.18 & 85.53 & -8.37 & 0.27 & 0.24 \\
4 & Orion\,A (tail) & L1641-S/C & 211.76 & -19.23 & 85.12 & -7.64 & 0.35 & 0.24 \\
5 & Orion\,A (tail) & L1641-C & 210.88 & -19.47 & 84.54 & -7.00 & 0.34 & 0.25 \\
6 & Orion\,A (head)\tablefootmark{a} & L1641-N & 209.99 & -19.63 & 84.01 & -6.33 & 0.29 & 0.25 \\
7 & Orion\,A (head) & OMC-4/5 & 209.38 & -19.65 & 83.74 & -5.82 & 0.24 & 0.16 \\
8 & Orion\,A (head) & OMC-1 & 209.05 & -19.43 & 83.79 & -5.45 & 0.21 & 0.15 \\
9 & Orion\,A (head) & OMC-2/3 & 208.72 & -19.20 & 83.86 & -5.06 & 0.26 & 0.15 \\
10 & Orion\,B (main) & L1630-S & 206.59 & -16.35 & 85.46 & -1.95 & 0.17 & 0.27 \\
11 & Orion\,B (main) & L1630-N & 205.25 & -14.22 & 86.73 & 0.19 & 0.17 & 0.21 \\
12 & Orion\,B (cometary) & L1622 & 204.77 & -11.90 & 88.56 & 1.70 & 0.07 & 0.07 \\
13 & Outlying\,Cloud (cometary) & L1616 & 203.50 & -24.70 & 76.72 & -3.33 & 0.20 & 0.12 \\
14 & Outlying\,Cloud (cometary) & IC2118 & 206.38 & -25.94 & 76.84 & -6.21 & 0.10 & 0.10 \\
15 & Cluster\tablefootmark{b} & Orion\,X & 206.04 & -21.95 & 80.25 & -4.09 & 1.29 & 1.44 \\
\hline
\end{tabular}
\renewcommand{\arraystretch}{1}
\label{tab:overview}
\tablefoot{
\tablefoottext{a}{L1641-N is connecting Orion\,A's head and tail, while we count it here to the head.}
\tablefoottext{b}{The cluster Orion\,X is included in this table for completeness. It is used to set the center position of the coordinate frame in Figs.\,\ref{fig:YZ-Orion-OX} to \ref{fig:YX-Orion-OX}.}
}
\end{center}
\end{table*}
\begin{table*}[!ht] 
\begin{center}
\small
\caption{Overview of average parameters as derived from \textit{Gaia} DR2 parallaxes and proper motions of YSOs for the 14 subregions and Orion\,X.
The listed parameters are average values as determined for each subregion with $\sigma$ referring to the standard deviation. The column $N_{\mathit{Gaia}}$ lists the number of YSOs that were used to estimate the averages.}
\renewcommand{\arraystretch}{1.2}
\begin{tabular}{lcccccccccccccc}
\hline \hline
\multicolumn{1}{l}{Subregion} &
\multicolumn{1}{c}{$N_{\mathit{Gaia}}$} &
\multicolumn{1}{c}{$\varpi$} &
\multicolumn{1}{c}{$\sigma_\varpi$} &
\multicolumn{1}{c}{$d$} &
\multicolumn{1}{c}{$\sigma_d$} &
\multicolumn{1}{c}{$\mu_{\alpha*}$} &
\multicolumn{1}{c}{$\sigma_{\mu,{\alpha*}}$} &
\multicolumn{1}{c}{$\mu_{\delta}$} & 
\multicolumn{1}{c}{$\sigma_{\mu,{\delta}}$} &
\multicolumn{1}{c}{$v_{\alpha}$} &
\multicolumn{1}{c}{$\sigma_{v,{\alpha}}$} &
\multicolumn{1}{c}{$v_{\delta}$} &
\multicolumn{1}{c}{$\sigma_{v,{\delta}}$} \\
& &
mas & mas & pc & pc & mas/yr & mas/yr & mas/yr & mas/yr &
 \multicolumn{1}{c}{km/s} & 
 \multicolumn{1}{c}{km/s} &
 \multicolumn{1}{c}{km/s} &
 \multicolumn{1}{c}{km/s} \\
\hline

L1647-S & 17 & 2.11 & 0.15 & 474 & 34 & 0.49 & 0.64 & -1.15 & 0.75 & 1.07 & 1.42 & -2.58 & 1.67 \\
L1647-N & 32 & 2.26 & 0.18 & 443 & 35 & 0.40 & 0.43 & -1.15 & 0.82 & 0.81 & 0.90 & -2.46 & 1.76 \\
L1641-S & 69 & 2.33 & 0.23 & 429 & 43 & 0.35 & 0.53 & -0.47 & 0.58 & 0.67 & 1.04 & -1.00 & 1.21 \\
L1641-S/C & 29 & 2.43 & 0.21 & 412 & 35 & 0.34 & 0.57 & -0.48 & 0.62 & 0.62 & 1.06 & -1.00 & 1.35 \\
L1641-C & 70 & 2.55 & 0.17 & 392 & 26 & 0.61 & 0.81 & -0.70 & 1.16 & 1.09 & 1.52 & -1.31 & 2.16 \\
L1641-N & 181 & 2.58 & 0.17 & 388 & 26 & 1.05 & 0.64 & 0.09 & 0.75 & 1.93 & 1.20 & 0.16 & 1.41 \\
OMC-4/5 & 116 & 2.51 & 0.16 & 399 & 25 & 1.18 & 0.67 & -0.12 & 0.77 & 2.23 & 1.29 & -0.24 & 1.49 \\
OMC-1 & 186 & 2.47 & 0.14 & 405 & 23 & 1.38 & 0.84 & -0.13 & 1.08 & 2.67 & 1.65 & -0.25 & 2.10 \\
OMC-2/3 & 115 & 2.59 & 0.16 & 387 & 24 & 1.30 & 0.70 & -0.35 & 0.95 & 2.40 & 1.32 & -0.66 & 1.79 \\
L1630-S & 57 & 2.56 & 0.20 & 391 & 31 & 0.52 & 0.80 & -0.47 & 0.94 & 0.94 & 1.47 & -0.84 & 1.73 \\
L1630-N & 74 & 2.32 & 0.17 & 430 & 31 & -0.52 & 0.82 & -0.98 & 0.41 & -1.11 & 1.74 & -2.00 & 0.84 \\
L1622 & 8 & 2.96 & 0.09 & 338 & 11 & 4.90 & 0.33 & -0.07 & 0.28 & 7.85 & 0.51 & -0.11 & 0.44 \\
L1616 & 18 & 2.59 & 0.14 & 386 & 21 & 0.75 & 0.34 & -0.94 & 0.32 & 1.39 & 0.63 & -1.73 & 0.61 \\
IC2118 & 5 & 3.41 & 0.16 & 293 & 14 & 0.70 & 0.22 & -3.44 & 0.22 & 0.97 & 0.28 & -4.79 & 0.44 \\
Orion X & 135 & 3.08 & 0.11 & 325 & 11 & 0.89 & 0.23 & -0.43 & 0.15 & 1.37 & 0.36 & -0.67 & 0.23 \\

\hline
\end{tabular}
\renewcommand{\arraystretch}{1}
\label{tab:averages_rv_pm}
\end{center}
\end{table*}
\begin{table*}[!ht] 
\begin{center}
\small
\caption{Overview of average radial velocities as determined from YSOs and gas for the 14 subregions and Orion X. The listed values are average values for each subregions with $\sigma$ referring to the velocity dispersion. $N_\mathrm{RV}$ lists the number of YSOs used to estimate the average stellar radial velocity.}
\renewcommand{\arraystretch}{1.2}
\begin{tabular}{lccccccccc}
\hline \hline
 &
\multicolumn{5}{c}{YSOs} & 
\multicolumn{4}{c}{Gas} \\
\cmidrule(lr){2-6}
\cmidrule(lr){7-10}
\multicolumn{1}{l}{Subregion} &

\multicolumn{1}{c}{$N_{\mathrm{RV}}$} &
\multicolumn{1}{c}{$v_\mathrm{HEL}$} &
\multicolumn{1}{c}{$\sigma_{v,\mathrm{yso}}$} &
\multicolumn{1}{c}{$v_\mathrm{LSR}$\tablefootmark{c}} & 
\multicolumn{1}{c}{Ref.\tablefootmark{a}} &

\multicolumn{1}{c}{$v_\mathrm{LSR}$} &
\multicolumn{1}{c}{$\sigma_{v,\mathrm{gas}}$} &
\multicolumn{1}{c}{$v_\mathrm{HEL}$\tablefootmark{d}} &
\multicolumn{1}{c}{Ref.\tablefootmark{b}} \\
& &
 \multicolumn{1}{c}{km/s} & 
 \multicolumn{1}{c}{km/s} &
 \multicolumn{1}{c}{km/s} & &
 \multicolumn{1}{c}{km/s} &
 \multicolumn{1}{c}{km/s} &  
 \multicolumn{1}{c}{km/s} & \\
\hline
L1647-S & 9 & 20.84 & 0.77 & 3.28 & 1 & 3.04 & 1.20 & 21.64 & 4 \\
L1647-N & 16 & 20.77 & 1.39 & 3.26 & 1 & 3.50 & 1.33 & 21.99 & 4 \\
L1641-S & 46 & 22.75 & 2.02 & 5.31 & 1 & 4.35 & 1.39 & 22.71 & 4 \\
L1641-S/C & 16 & 22.59 & 2.14 & 5.20 & 1 & 5.41 & 1.42 & 23.66 & 4 \\
L1641-C & 37 & 24.06 & 2.68 & 6.74 & 1 & 6.51 & 1.43 & 24.65 & 4 \\
L1641-N & 72 & 25.65 & 1.96 & 8.39 & 1 & 8.15 & 1.45 & 26.18 & 4 \\
OMC-4/5 & 48 & 26.26 & 2.41 & 9.06 & 1 & 8.54 & 1.57 & 26.48 & 4 \\
OMC-1 & 112 & 27.21 & 2.09 & 10.04 & 1 & 9.22 & 1.77 & 27.11 & 4 \\
OMC-2/3 & 38 & 28.33 & 2.60 & 11.19 & 1 & 10.46 & 1.54 & 28.30 & 4 \\
L1630-S & 37 & 27.68 & 1.61 & 10.86 & 1 & 10.12 & 1.48 & 26.82 & 4 \\
L1630-N & 45 & 27.98 & 1.37 & 11.40 & 1 & 9.46 & 1.28 & 26.51 & 4 \\
L1622\tablefootmark{e} & 4 & 19.28 & 1.17 & 2.91 & 2 & 1.17 & 0.71 & 17.96 & 5 \\
L1616\tablefootmark{f} & 4 & 24.50 & 2.70 & 7.80 & 3 & 7.70 & 1.05 & 22.59 & 6 \\
IC\,2118 & -- & -- & -- & -- & -- & -2.20 & 1.77 & 15.40 & 7 \\
Orion\,X & 4 & 19.63 & 0.28 & 2.68 & 1 & -- & -- & -- & -- \\
\hline
\end{tabular}
\renewcommand{\arraystretch}{1}
\label{tab:averages_rv}
\tablefoot{ 
\tablefoottext{a}{Reference for stellar radial velocities given in \vhel, as derived from stellar spectra.}
\tablefoottext{b}{Reference for gas radial velocities given in \vlsr, derived from $^{12}$CO emission line surveys.}
\tablefoottext{c}{The average \vhel of the YSOs is converted to \vlsr using the standard solar motion from \citet{Schoenrich2010}.}
\tablefoottext{d}{The \vlsr of the gas is converted to \vhel using the standard solar motion from \citet{Kerr1986} for all except for L1616, for which we used \citet{Mihalas1981}, since the data is from \citet{Maddalena1986}.}
\tablefoottext{e}{The four sources in L1622 were observed by \citet{Kounkel2017b} with lower resolution (observational errors between 0.8 to 1.2\,km/s) compared to APOGEE-2 (typical observational error of about 0.05\,km/s).}
\tablefoottext{e}{The four sources in L1616 were observed by \citet{Alcala2004} with lower resolution (observational errors between 2.0 to 2.3\,km/s) compared to APOGEE-2.}
}
\tablebib{
(1) APOGEE-2 SDSS-DR16;
(2) \citet{Kounkel2017b};
(3) \citet{Alcala2004};
(4) \citet{Nishimura2015} $^{12}$CO(2-1);
(5) \citet{Kun2008} $^{12}$CO(1-0);
(6) \citet{Maddalena1986} $^{12}$CO(1-0);
(7) \citet{Kun2001} $^{12}$CO(1-0).
}
\end{center}
\end{table*}
\begin{table*}[!ht] 
\begin{center}
\small
\caption{Overview of the average 6D parameters given in Galactic Cartesian coordinates and motions for the 14 subregions and Orion\,X.}
\renewcommand{\arraystretch}{1.2}
\begin{tabular}{lcccccccccccc}
\hline \hline
\multicolumn{1}{l}{Subregion} &
\multicolumn{1}{c}{$X$} &
\multicolumn{1}{c}{$Y$} &
\multicolumn{1}{c}{$Z$} &
\multicolumn{1}{c}{$X_g$} &
\multicolumn{1}{c}{$Y_g$} &
\multicolumn{1}{c}{$Z_g$} &
\multicolumn{1}{c}{$X'_\mathrm{Orion}$} &
\multicolumn{1}{c}{$Y'_\mathrm{Orion}$} &
\multicolumn{1}{c}{$Z'_\mathrm{Orion}$} &
\multicolumn{1}{c}{$U_\mathrm{LSR}$} &
\multicolumn{1}{c}{$V_\mathrm{LSR}$} &
\multicolumn{1}{c}{$W_\mathrm{LSR}$} \\
&
 \multicolumn{1}{c}{pc} & 
 \multicolumn{1}{c}{pc} &
 \multicolumn{1}{c}{pc} &
 \multicolumn{1}{c}{pc} & 
 \multicolumn{1}{c}{pc} &
 \multicolumn{1}{c}{pc} &
 \multicolumn{1}{c}{pc} & 
 \multicolumn{1}{c}{pc} &
 \multicolumn{1}{c}{pc} &
 \multicolumn{1}{c}{km/s} &
 \multicolumn{1}{c}{km/s} &  
 \multicolumn{1}{c}{km/s} \\
\hline

L1647-S & -367.90 & -251.98 & -161.13 & -8490.28 & -251.98 & -139.39 & 469.55 & 63.67 & 16.59 & -4.07 & -1.56 & -0.20 \\
L1647-N & -347.98 & -230.54 & -149.18 & -8470.33 & -230.54 & -127.49 & 439.73 & 53.22 & 17.38 & -4.68 & -1.26 & -0.43 \\
L1641-S & -340.82 & -218.31 & -140.82 & -8463.15 & -218.31 & -119.14 & 425.62 & 45.41 & 20.7 & -6.39 & -0.35 & -0.00 \\
L1641-S/C & -330.88 & -204.82 & -135.77 & -8453.20 & -204.82 & -114.12 & 409.94 & 37.69 & 19.78 & -7.35 & -0.51 & -0.37 \\
L1641-C & -317.00 & -189.54 & -130.55 & -8439.31 & -189.53 & -108.94 & 390.19 & 30.1 & 17.4 & -8.12 & -1.19 & -0.56 \\
L1641-N & -316.15 & -182.47 & -130.22 & -8438.46 & -182.47 & -108.61 & 386.47 & 24.14 & 16.24 & -10.43 & -1.01 & 0.16 \\
OMC-4/5 & -327.07 & -184.17 & -134.03 & -8449.39 & -184.17 & -112.39 & 397.67 & 20.84 & 16.68 & -10.60 & -1.35 & 0.12 \\
OMC-1 & -333.61 & -185.3 & -134.64 & -8455.93 & -185.3 & -112.99 & 403.81 & 18.97 & 18.51 & -11.22 & -1.78 & 0.35 \\
OMC-2/3 & -320.28 & -175.47 & -127.18 & -8442.58 & -175.47 & -105.56 & 385.90 & 16.04 & 19.3 & -12.07 & -2.34 & -0.33 \\
L1630-S & -335.42 & -167.88 & -110.04 & -8457.68 & -167.88 & -88.38 & 388.94 & 2.54 & 39.02 & -11.48 & -0.43 & 0.13 \\
L1630-N & -377.26 & -177.93 & -105.73 & -8499.51 & -177.93 & -83.97 & 426.22 & -6.92 & 58.79 & -11.19 & 0.34 & -1.07 \\
L1622 & -300.02 & -138.41 & -69.65 & -8422.17 & -138.41 & -48.08 & 332.26 & -8.27 & 59.62 & -4.55 & 0.83 & 10.26 \\
L1616 & -321.88 & -139.96 & -161.44 & -8444.27 & -139.96 & -139.81 & 385.58 & -16.53 & -17.81 & -7.01 & 2.01 & -1.83 \\
IC\,2118 & -236.15 & -117.12 & -128.23 & -8358.45 & -117.12 & -106.82 & 292.47 & 0.83 & -19.73 & 1.30 & 2.14 & -0.67 \\
Orion\,X & -270.80 & -132.29 & -121.47 & -8393.08 & -132.29 & -99.98 & 324.94 & 0.0 & 0.0 & -5.02 & 3.00 & 0.75 \\

\hline
\end{tabular}
\renewcommand{\arraystretch}{1}
\label{tab:averages_xyz}
\tablefoot{$X,Y,Z$ are Heliocentric Galactic Cartesian coordinates; $X_g, Y_g, Z_g$ are Galactocentric Cartesian coordinates; and $X'_\mathrm{Orion}, Y'_\mathrm{Orion}, Z'_\mathrm{Orion}$ are transformed Cartesian coordinates with the $X$-axis pointing toward a central position in Orion (toward $l,b = \SI{206.038}{\degree}, \SI{-21.945}{\degree}$). The last three columns are Galactic Cartesian velocities relative to the LSR.}
\end{center}
\end{table*}

\section{Methods} \label{Methods}

In this section we describe our methods to evaluate the 3D space motions of molecular clouds in Orion. First, we demonstrate the validity of using YSOs as proxy for cloud proper motions and parallaxes. Next, we present the methods used to obtain the average positions and motions for the individually discussed clouds. Finally, we introduce our approach to estimate the orbits of the clouds in the Milky Way and their 3D space motions, and we use these motions to estimate momenta for a subsample of the studied regions.

\subsection{YSOs as proxies for cloud parameters}

We used YSOs with infrared excess to indirectly determine the proper motions and distances of the studied star-forming molecular clouds. These young stars \citep[$\lesssim3$\,Myr, e.g.,][]{Dunham2015} are still close to their birth sites \citep[e.g.,][]{Heiderman2010, Gutermuth2011, Grossschedl2019A, Pokhrel2020}, and it is well established in the literature that young stars in general share the radial velocities of their parental molecular clouds \citep[e.g.,][]{Furesz2008, Tobin2009, Hacar2016b}. 
The observed agreement suggests that, on average, also the proper motions and distances of gas and YSOs should be approximately the same.

For our purposes, we used Class\,II (or earlier class) YSOs in order to retain only the youngest possible selection to maximize the chances that the young stars share the same space motion as the gas. We did not include Class\,III sources \citep[e.g.,][]{Pillitteri2013}, but a first analysis indicates that many Class\,III candidates still share the same overall motions and distances as the Class\,II candidates. This suggests that, in the future, also Class\,III samples could provide important insight on the dynamics of molecular cloud complexes, when no or too little Class\,II members are available. 

\subsection{Evaluating average positions and motions of the subregions}
\label{6d-parameters}

The 6D parameters (3D position, 3D motion) of the subregions were determined as the mean of the observed parameters. We parameterized the scatter with the standard deviation of the mean ($\sigma$). The resulting 6D average parameters are given in Tables\,\ref{tab:overview} to \ref{tab:averages_rv}.
The 3D positions were derived from projected 2D average $l,b$ or $\alpha,\delta$ positions, and from average \textit{Gaia} DR2 parallaxes $\varpi$ (mas/yr). The distances are determined by inverting the mean parallaxes, $d(pc) = 1000/\varpi$\,(mas). This approach does not include any systematic correction \citep{Luri2018, Lindegren2018, Stassun2018}, since systematic errors in the parallax measurements are unknown for the Orion region. Moreover, we did not use an inference procedure to account for the nonlinearity of the transformation or the asymmetry of the resulting probability distribution as given by \citet{BailerJones2018}, since these distances do not represent individual young stellar populations.

The 3D motions were obtained from \textit{Gaia} DR2 proper motions of YSOs ($\mu_{\alpha*}$, $\mu_\delta$), combined with average gas radial velocities as determined from Mean(\vlsr) (converted to \vhel, see Sect.~\ref{gas-rv}). To compare YSO and gas radial velocities we also calculated stellar average \vhel values from YSO samples. Additionally, we calculated the YSO tangential velocities ($v_\alpha$, $v_\delta$), given in km/s in Table\,\ref{tab:averages_rv_pm}. The values $\sigma_{v,\mathrm{gas}}$, $\sigma_{v,\mathrm{yso}}$, $\sigma_{v,\alpha}$, and $\sigma_{v,\delta}$ are a measure of the velocity dispersion.
To test the significance of these parameters we compare with typical measurement errors of the stellar astrometry.
The stellar radial velocities taken from \mbox{APOGEE-2} have typical errors of about 0.04 to 0.07\,km/s, and the tangential velocities as derived via proper motion and parallax (Equ.~\ref{equ:tangential}) have typical errors of about 0.2 to 0.5\,km/s, hence the errors in tangential direction are almost an order of magnitude larger compared to the line of sight direction. For most regions, the typical errors are lower than the determined velocity dispersions ($\sigma\sim0.5$ to 3\,km/s, see Tables~\ref{tab:averages_rv_pm} and \ref{tab:averages_rv}), while velocity dispersions near 0.5\,km/s in tangential direction could be dominated by errors.
Generally, the $\sigma$ values should be seen as an approximation of the true velocity dispersion of the stellar samples. This is because the listed values were derived from YSO subsamples (i.e., high quality observation in \textit{Gaia} and/or \mbox{APOGEE-2} required) within subregions defined with rigid boundaries, therefore the used YSO samples are not complete and could suffer from a selection bias. In addition, measurement errors could inflate the measured $\sigma$ values.

Due to differences of the three chosen regions (Orion\,A, Orion\,B, outlying clouds), we shortly discuss them individually in the next subsections. For each region, we check the validity of using YSOs as a proxy for cloud parameters by comparing YSO and gas radial velocities.

\subsubsection{6D parameter determination for Orion A} \label{OA-Methods}

Since Orion\,A covers a quite large area in the sky (almost $\SI{20}{deg^2}$), shows gradients in both distance and velocity, and most importantly, since the 3D shape shows a bent structure, we decided to split the region into nine subregions (see also Sect.~\ref{OA-Data} and Fig.\,\ref{fig:pv-gas-ysos}). First, to get the cloud's line of sight motions, we extracted only those radial velocity measurements from the $^{12}$CO map (Sect.~\ref{gas-rv}) that fall within a specific extinction contour (smoothed outer contour at A$_K = \SI{0.5}{mag}$) to eliminate possible background contamination. This approach reduced the velocity scatter in the Position-Velocity-Diagram  (PV-diagram) for Orion\,A significantly \citep[Fig.~\ref{fig:pv-gas-ysos}, compare to Fig.\,1 in][]{Hacar2016b}. Additionally, we excluded the northeast part of Orion\,A's tail (see Fig.~\ref{fig:pv-gas-ysos}, empty contour), due to lower intensity CO measurements, low YSO statistics, and uncertain distance estimates.

Next, we split the cloud based on known subregions near the head \citep[see also][]{Getman2019}, and based on radial velocities at the tail, since there are regions with almost constant velocities, interrupted by velocity-jumps of about 1 to 2\,km/s. The cloud-separations were applied at the following positions along $l$\,(deg): 214.408, 213.525, 212.642, 211.758, 210.875, 209.992, 209.383, 209.050, 208.717. The first six bins have a width of \SI{53}{\arcmin} and the last three bins near the ONC have a width of \SI{20}{\arcmin}. These subregions correspond to known cloud parts as introduced in Sect.~\ref{OA-Data}. The PV-diagram in Fig.~\ref{fig:pv-gas-ysos} illustrates this approach.

The average positions and motions were then determined from YSO and gas parameters per bin. Average $l$ values for each subregion were determined from the mid bin positions, and average $b$ values were chosen manually to match with regions of high column-density (these positions match well with average YSO $b$ positions). 
To determine \vhel we used the average \vlsr measurements of the gas from \citetalias{Nishimura2015} per bin, and converted it to \vhel using the \citetalias{Kerr1986} standard solar motion. 
We then compare gas to YSO radial velocities (\vhel of YSOs was converted with \citetalias{Kerr1986} to \vlsr only for Figs.~\ref{fig:pv-gas-ysos} and \ref{fig:pv-gas-ysos_orionb}), which follow on average the same trend as the gas (within the $\sigma$).

There are shifts between gas and YSO \vlsr, with YSOs near the tail being slightly blue-shifted ($\Delta v_\mathrm{LSR} \sim 1$ to 2\,km/s), and with a reversed trend closer to the head. A slight blue-shift of YSO radial velocities could be caused by the general dispersion of stars away from their birth-sites, considering that the YSOs in the front moving toward us will be less extincted by dust and are more likely observed in the optical and with higher quality. On the other hand, the reversed trend toward the head does not fit within this picture (also not visible in Orion\,B, see Fig.~\ref{fig:pv-gas-ysos_orionb}). Therefore, an unambiguous interpretation of any shift is not possible, since the deviations are within the measured velocity dispersion and can be caused by statistical errors, systematic errors, or inconsistent LSR conversion.
In general, the agreement of gas and YSO radial velocities within the scatter validates our assumption to use YSOs as proxy for cloud parameters, and we averaged the YSO \textit{Gaia} DR2 proper motions and parallaxes within the same bins to complete the 6D parameter space.

We excluded the northwestern tip of the head that overlaps with NGC\,1977, since this cluster appears to be decoupled from the gas in projection, and the velocities show a significant deviation of YSOs versus gas, as visible in the top and bottom panels of Fig.~\ref{fig:pv-gas-ysos}.
\object{NGC\,1977} may be located behind the cloud, since there is no or little gas that would cause extinction toward cluster members. Indeed, there is evidence that the B1V star \object{42\,Ori} has blown out the gas in this region via radiative feedback \citep[e.g.,][]{Bouy2014, Pabst2020} and may be even pushing the gas at the tip of the cloud back toward the Sun. This would be in agreement with the red-shifted velocities of the cluster compared to the gas. The young compact cluster NGC\,1977 is discussed further in Sect.~\ref{sec:groups} and in Appendix~\ref{apx:ycc}.

\subsubsection{6D parameter determination for Orion B} \label{OB-Methods}
Orion\,B is split into three main components, as also described in Sect.~\ref{OB-Data}. For the two subregions in L1630, we used the \citetalias{Nishimura2015} map to determine the average \vhel from the gas, in the same manner as for Orion\,A. The corresponding PV-diagram is shown in Fig.~\ref{fig:pv-gas-ysos_orionb}, which shows YSO and gas \vlsr for the subregions L1630-S/N. Within the errors, the \vlsr of YSOs and gas agree with each other, while in L1630-N the YSO radial velocities seem to be slightly red-shifted (on average about 1.5\,km/s). It is not clear if this is a significant shift. For example, \citet{Kounkel2017b} do not find a shift of YSO to gas radial velocities in this region. If anything, they find a slight blue-shift of YSOs in L1630-S suggesting that the shift in Figure~\ref{fig:pv-gas-ysos_orionb} is not significant and is likely caused by systematics or erroneous LSR conversion (Appendix\,\ref{apx:lsr}). The other five parameters were then determined from averaging the parameters from the chosen YSO samples. For further details on these regions and sample selection see Appendix~\ref{apx:regions} and Fig.~\ref{fig:orionb_regions}.

For L1622 we used the \vlsr of 1.17\,km/s as reported in \citet{Kun2008}, which is converted to \vhel $\sim$ 17.96 km/s, using the \citetalias{Kerr1986} standard solar motion. Compared to YSO velocities (average $v_\mathrm{HEL} = 19.3 \pm 1.2\,\si{km/s}$, \citealp{Kounkel2017b}), we find that the latter are relatively red-shifted ($\sim$1.3\,km/s), while not significant within the errors (typical stellar radial velocity error about 2\,km/s). Moreover, inherent systematics of the observations, small number statistics, or LSR conversion errors could be responsible for this shift. For our analysis, we used the given gas radial velocity to obtain the line of sight motion, while the other parameters were obtained from averages of the YSO sample. The position ($l,b$) was adjusted manually according to high column-density in L1622 (Appendix~\ref{apx:regions} and Fig.~\ref{fig:orionb_regions}).

\subsubsection{6D parameter determination for the outlying clouds} \label{OC-Methods}

For the two star-forming cometary clouds, L1616 and IC\,2118, we used the YSO samples as defined in Sect.~\ref{OC-Data}. To obtain radial velocities we used the gas velocities from CO observations as reported by \citet{Maddalena1986} for L1616 and \citet{Kun2001} for IC\,2118, given in Sect.~\ref{OC-Data} and Table~\ref{tab:averages_rv_pm}.
For L1616 the radial velocities of YSOs as reported by \citet{Alcala2004} are consistent with the CO velocities by \citet{Maddalena1986} within the errors, and we used the gas \vlsr = 7.7\,km/s, which is \vhel = 22.6\,km/s.
For IC\,2118 only gas radial velocities are available. Based on the findings for the other clouds in our sample, we assumed that YSOs also share on average similar motions as the gas of the associated molecular cloud. Future investigations are needed to confirm this assumption. The other parameters were again determined from average YSO parameters. A more detailed description is given in Appendix~\ref{apx:regions} and Fig.~\ref{fig:outlying-regions}.

\begin{figure*}[!ht]
    \centering
    \includegraphics[width=1\linewidth]{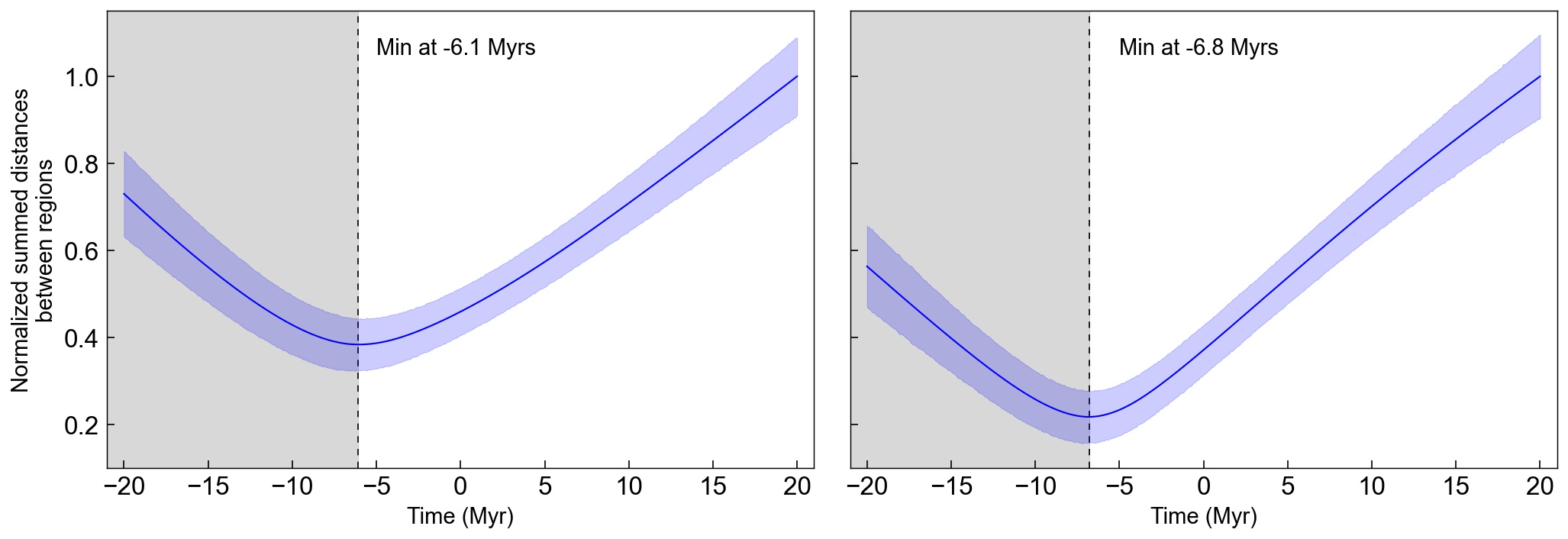}
    \caption{
    Estimated time when the subregions show the most compact configuration (minimum distances). Shown is the summed distance from each region to every other region at each time step (for $\pm$20\,Myr with 0.1\,Myr timesteps), normalized to the maximum. The blue shaded area shows 1$\sigma$ uncertainties, calculated by sampling the standard deviations around the averages. 
    \text{Left:} The summed distances were calculated for all 14 investigated regions, which leads to a minimum at $-6.1$\,Myr. 
    \textit{Right:} The summed distances were only calculated for nine regions by excluding the tail of Orion\,A. When excluding the tail the minimum gets more pronounced and shifts to $-6.8$\,Myr. 
    Beyond these minima the distances start to rise again (gray shaded areas), which means that any motions of the clouds before about $-6$\,Myr should be omitted. 
    }
    \label{fig:min-distance}
\end{figure*}

\subsection{Galactic Cartesian coordinates and Galactic orbit estimation} \label{galpy}

\begin{figure*}[!ht]
    \centering
    \includegraphics[width=1\linewidth]{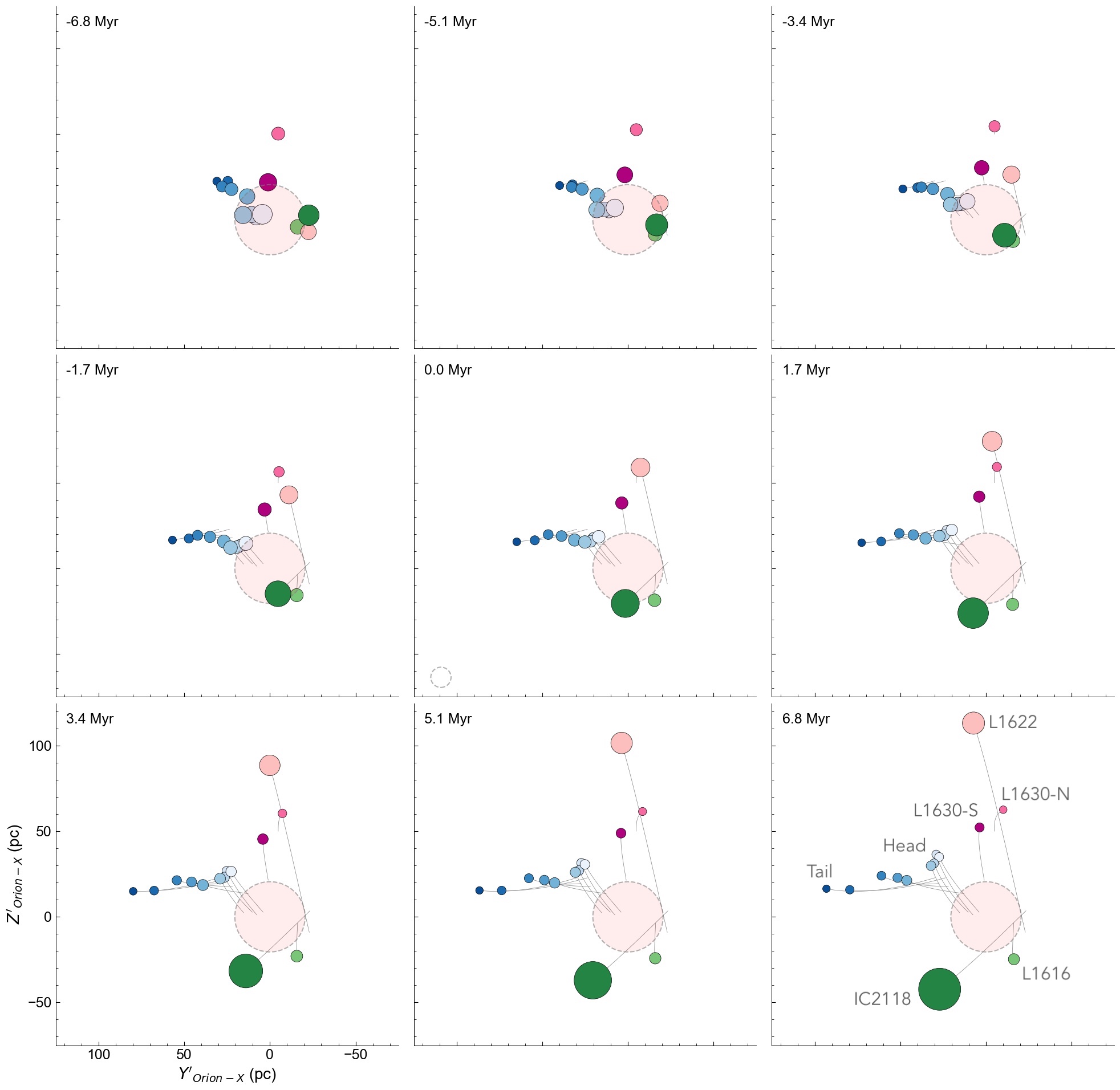}
    \caption{Nine time-snapshots showing the relative motions of the subregions in $Z'_\mathrm{Orion-X}$ vs $Y'_\mathrm{Orion-X}$ from -6.8 to 6.8\,Myr. This projection represents a face-on view of Orion as viewed from the Sun. The x-axis of the coordinate system points toward $(l,b) = (\SI{206.038}{\degree}, \SI{-21.945}{\degree})$. This is the average position of the cluster Orion\,X \citep{Chen2020}. Its extent is shown by the red disk in the center with a gray dashed outline.  
    The 14 subregions are shown as filled circles (colored as in Fig.~\ref{fig:overview}) and labeled in the last panel (for Orion\,A only head and tail are indicated). The symbol sizes are scaled with distance (along the 3rd axis) and are normalized relative to the gray-dashed open circle given in the lower-left corner of central panel (circle size normalized for Orion\,X); larger points are in-front and smaller points in the back of Orion\,X, to give an impression of depth. The symbol sizes are scaled for each projection individually and are not comparable between Figs.~\ref{fig:YZ-Orion-OX} to \ref{fig:YX-Orion-OX}.
    See text for more information. A movie version is available online at \url{https://homepage.univie.ac.at/josefa.elisabeth.grossschedl/figure7-YZ-orion-timelapse.mp4}.
    See also the \href{https://homepage.univie.ac.at/josefa.elisabeth.grossschedl/orion-bb.html}{link} in Fig.~\ref{fig:orion-bb} for an interactive version.
    }
    \label{fig:YZ-Orion-OX}
\end{figure*}

\begin{figure*}[!ht]
    \centering
    \includegraphics[width=1\linewidth]{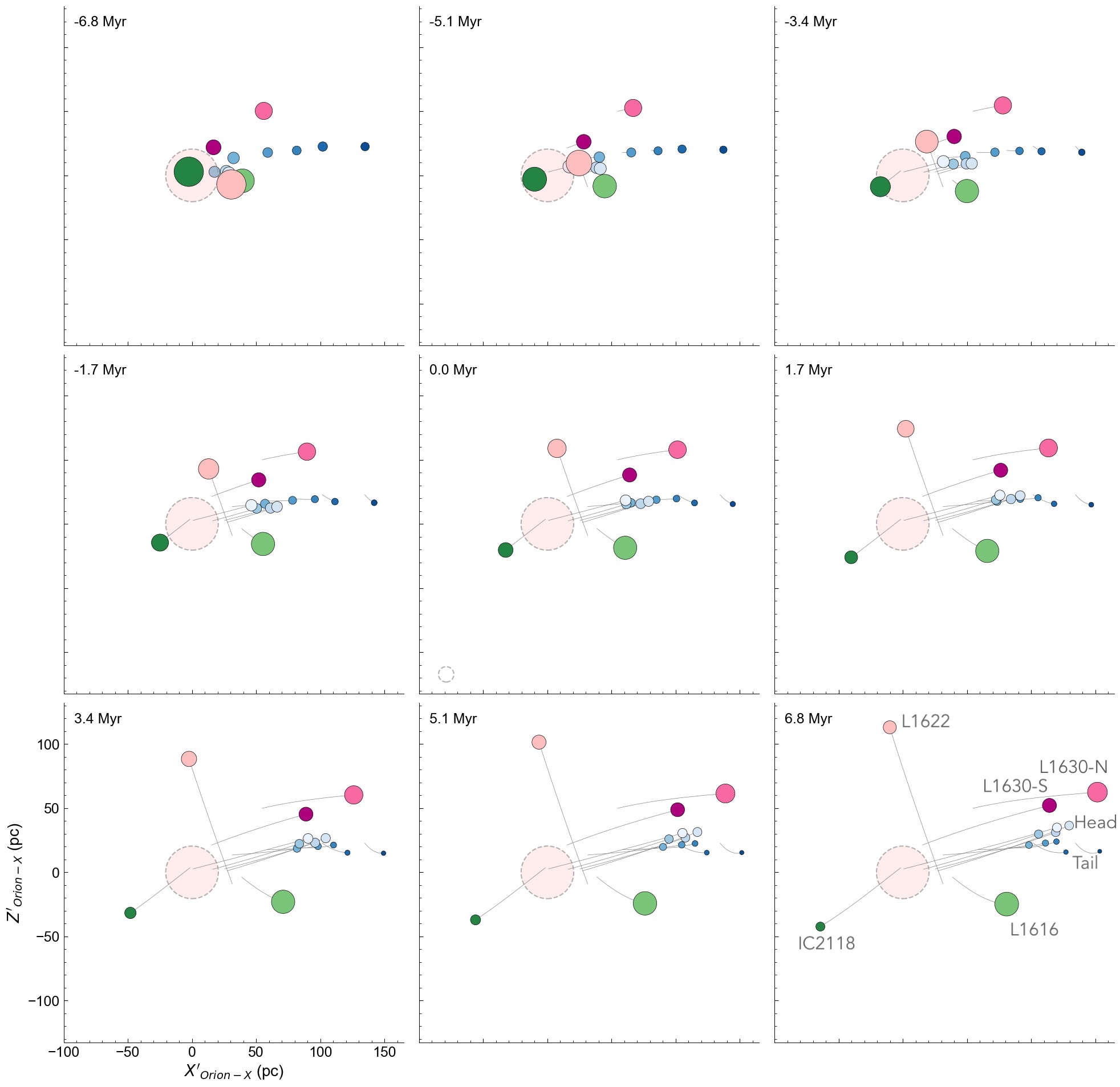}
    \caption{Nine time snapshots showing the relative motions of the subregions in $Z'_\mathrm{Orion-X}$ vs $X'_\mathrm{Orion-X}$ from -6.8 to 6.8\,Myr. This projection represents a side view of Orion. See Fig.~\ref{fig:YZ-Orion-OX} and text for more information. A movie version is available online at \url{https://homepage.univie.ac.at/josefa.elisabeth.grossschedl/figure8-XZ-orion-timelapse.mp4}.
    See also the \href{https://homepage.univie.ac.at/josefa.elisabeth.grossschedl/orion-bb.html}{link} in Fig.~\ref{fig:orion-bb} for an interactive version.
    }
    \label{fig:XZ-Orion-OX}
\end{figure*}

\begin{figure*}[!ht]
    \centering
    \includegraphics[width=1\linewidth]{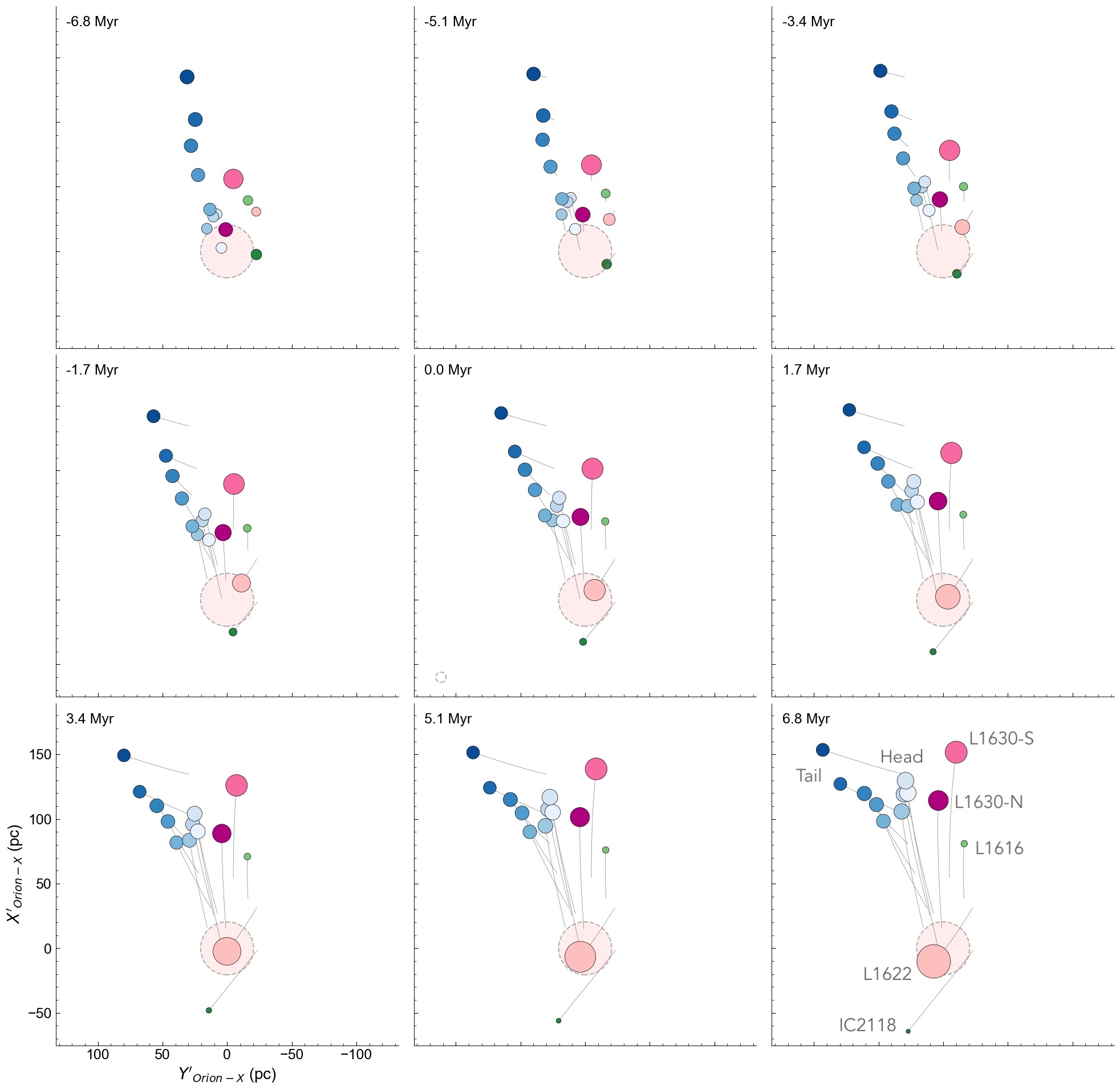}
    \caption{Nine time snapshots showing the relative motions of the subregions in $X'_\mathrm{Orion-X}$ vs $Y'_\mathrm{Orion-X}$ from -6.8 to 6.8\,Myr. This projection represents a top-down view of Orion. See Fig.~\ref{fig:YZ-Orion-OX} and text for more information. A movie version is available online at \url{https://homepage.univie.ac.at/josefa.elisabeth.grossschedl/figure9-YX-orion-timelapse.mp4}.
    See also the \href{https://homepage.univie.ac.at/josefa.elisabeth.grossschedl/orion-bb.html}{link} in Fig.~\ref{fig:orion-bb} for an interactive version.
    }
    \label{fig:YX-Orion-OX}
\end{figure*}

We used the Python package \textit{Astropy} \citep{Astropy2013, Astropy2018} to calculate Galactic Cartesian coordinates, which were used to visualize our results in 3D. In Table~\ref{tab:averages_xyz} we show the resulting coordinates given Heliocentric ($X$, $Y$, $Z$), Galactocentric ($X_{g}$, $Y_g$, $Z_g$), and also transformed into a coordinate systems that points toward Orion. For the latter, the x-axis ($X'_\mathrm{Orion}$) points toward a central position in Orion (at $l,b = \SI{206.038}{\degree}, \SI{-22.945}{\degree}$). The choice of this position is elaborated below in the results (Sect.~\ref{results_relative_motions}). 
The coordinate $X'_\mathrm{Orion}$ is equal to the distance from the Sun for the chosen central position, while $Y'_\mathrm{Orion}$ and $Z'_\mathrm{Orion}$ are roughly parallel to $l$ and $b$, respectively. 
These coordinates are used in Figs.~\ref{fig:YZ-Orion-OX} to \ref{fig:YX-Orion-OX}, where they are given relative to the central position (cluster Orion\,X at zero, see Sect.~\ref{results_relative_motions}). For clarity the coordinates are then written as $X'_\mathrm{Orion-X}$, $Y'_\mathrm{Orion-X}$, $Z'_\mathrm{Orion-X}$.
Finally, we list the Galactic Cartesian velocities relative to LSR ($U_{\mathrm{LSR}}$, $V_{\mathrm{LSR}}$, $W_{\mathrm{LSR}}$), which are the time derivatives along $X$, $Y$, $Z$;
$U_{\mathrm{LSR}}$ is positive toward the Galactic center in the solar neighborhood,
$V_{\mathrm{LSR}}$ is positive in the direction of Galactic rotation, and 
$W_{\mathrm{LSR}}$ is positive toward the Galactic north pole.

Next, we derived the Galactic orbits of the selected star-forming regions and their relative motions by using the average 6D parameters of the selected subregions as starting conditions.
To get the orbital motion of each subregion, we used the Python package \textit{Galpy} by \citet{Bovy2015} in combination with \textit{Astropy}. \textit{Galpy} enables us to estimate orbits on a series of predefined potentials, including potentials, which approximate the Milky Way. We used a Milky Way potential that includes a disk, bulge, and halo component \citep[\texttt{galpy.potential.MWPotential2014},][]{Bovy2015}. This approach ignores the gravitational field of the gas clouds or any other acceleration or damping mechanisms acting within a region, and consequently should be considered an approximation of the real dynamics. However, the gravitational potential of the Milky Way dominates over that of single GMCs and should dominate the overall dynamics. \textit{Galpy} allows us to trace back the orbits of the selected clouds in Orion with some confidence for the last few million years. 
To properly estimate the orbits of objects in the Milky Way, one needs to know the Sun's position and its Galactic motion. We use the default values from \texttt{Astropy\,4.0} (see Table~\ref{tab:astropy}).

\begin{figure}[!t]
\small
    \centering
        \includegraphics[width=1.02\linewidth]{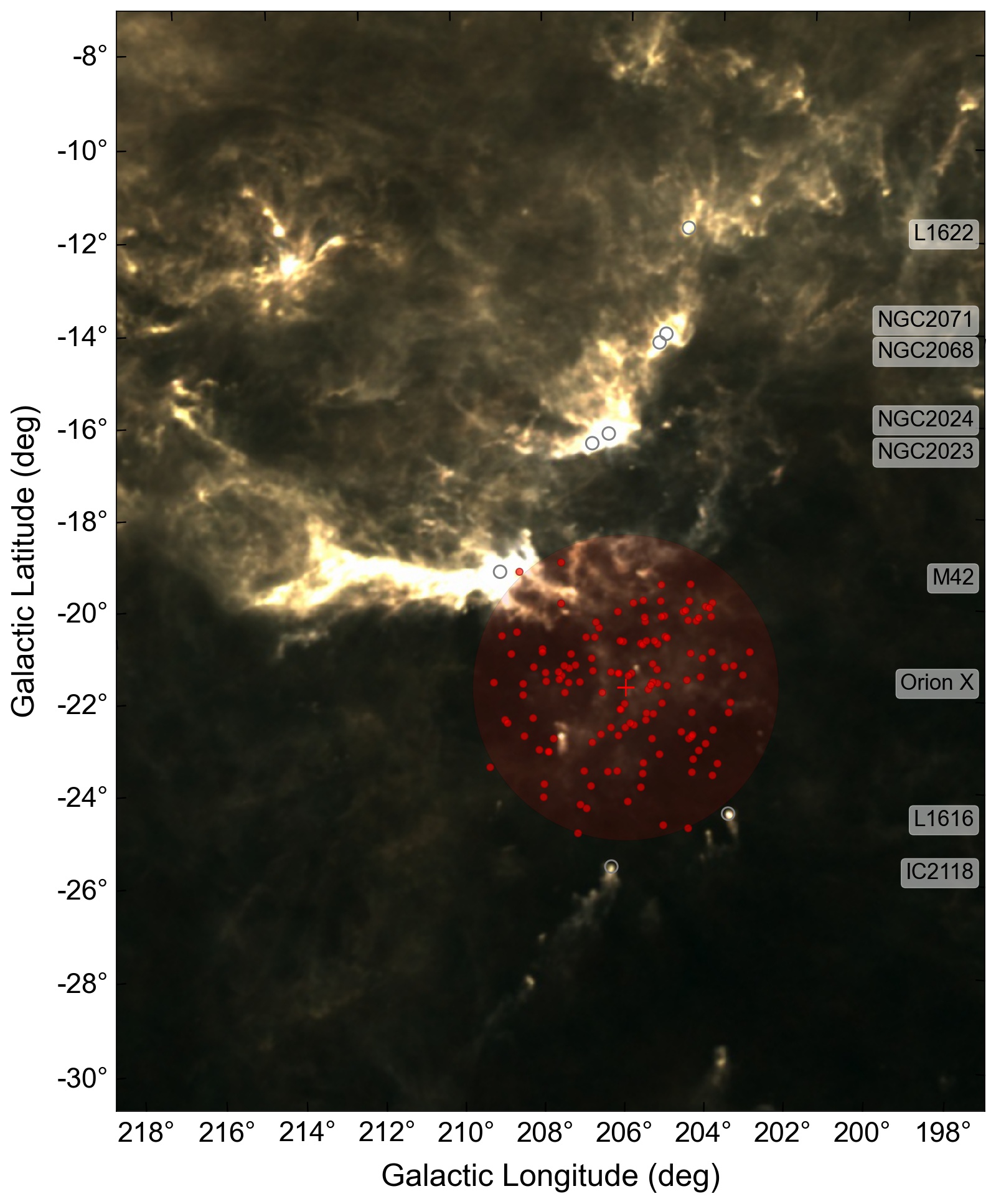}
    \caption{Planck HFI composite \citep{Planck2014} showing an overview of the Orion molecular cloud complex. Because of its sensitivity and dynamic range, the Planck image shows clearly the wind-blown appearance of the molecular gas in this whole region, shaped by the feedback of massive stars. The small gray open circles mark the investigated regions, with corresponding labels on the right. In Orion\,A only the position of M42 is indicated, and in Orion\,B the clusters are labeled separately.
    The large red filled circle encloses the position of the cluster Orion\,X \citep{Bouy2015}, with a diameter of about $\SI{7}{\degree}$ ($\sim \SI{40}{pc}$ at the cluster's average distance of $\sim \SI{325}{pc}$). The individual stellar members of the cluster \citep[from][]{Chen2020} are represented as red dots.
    }
    \label{fig:planck-orionx}
\end{figure}

\begin{figure}[!t]
\small
    \centering
        \includegraphics[width=1\linewidth]{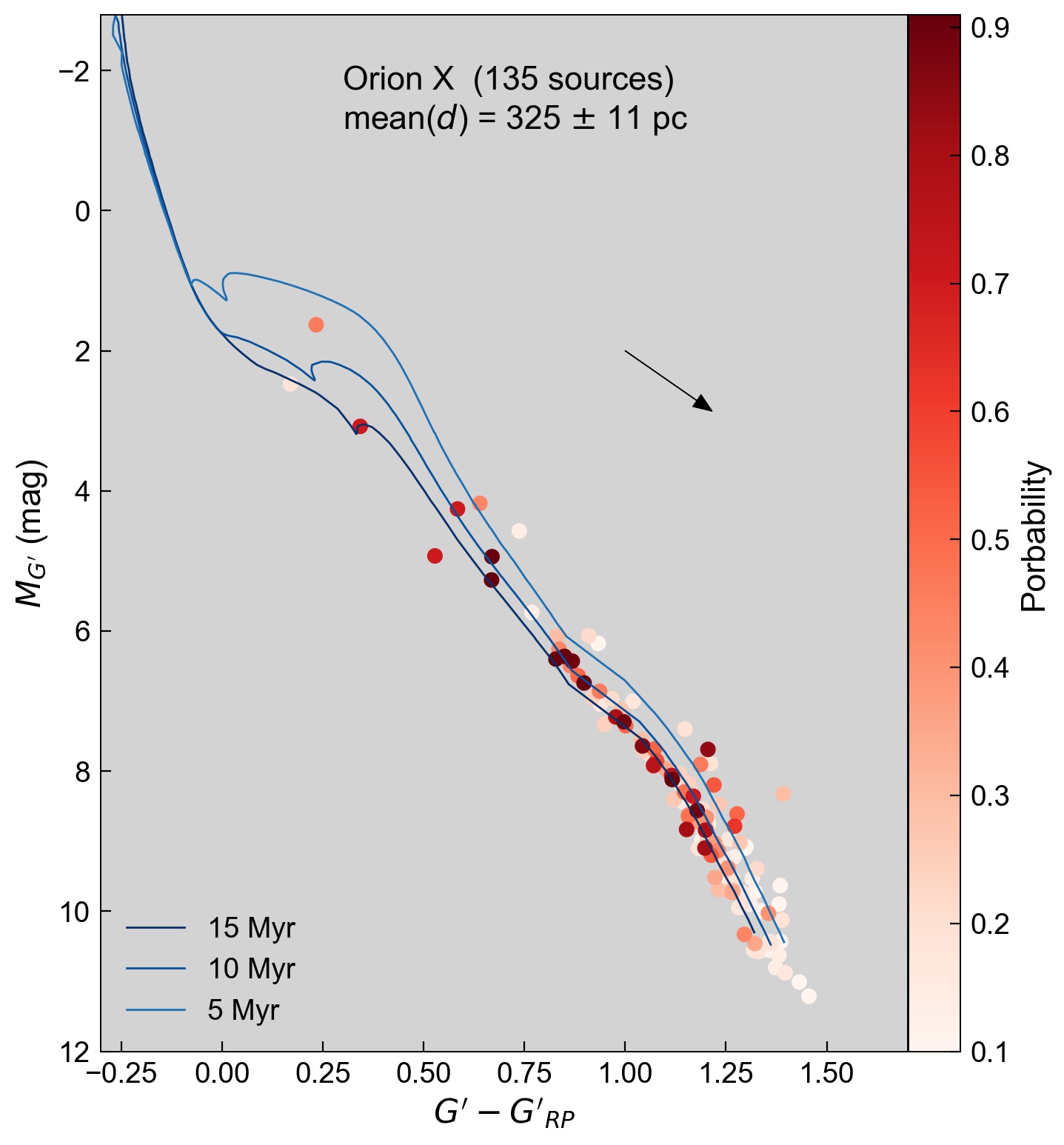}
    \caption{Color-absolute-magnitude diagram for the \textit{Gaia} passbands $G'$ and $G'_\mathrm{RP}$ (corrected DR2 photometry, see Appendix~\ref{apx:gaia}). The red dots show stellar members of Orion\,X \citep[][]{Chen2020}, color-coded for membership probability. The blue lines are PARSEC isochrones for 5, 10, and 15\,Myr (see legend). The black arrow shows the direction of extinction with a length of $A_V = 1$\,mag.
    }
    \label{fig:hrd-orionx}
\end{figure}

\begin{figure}[!ht]
    \centering
    \includegraphics[width=1\linewidth]{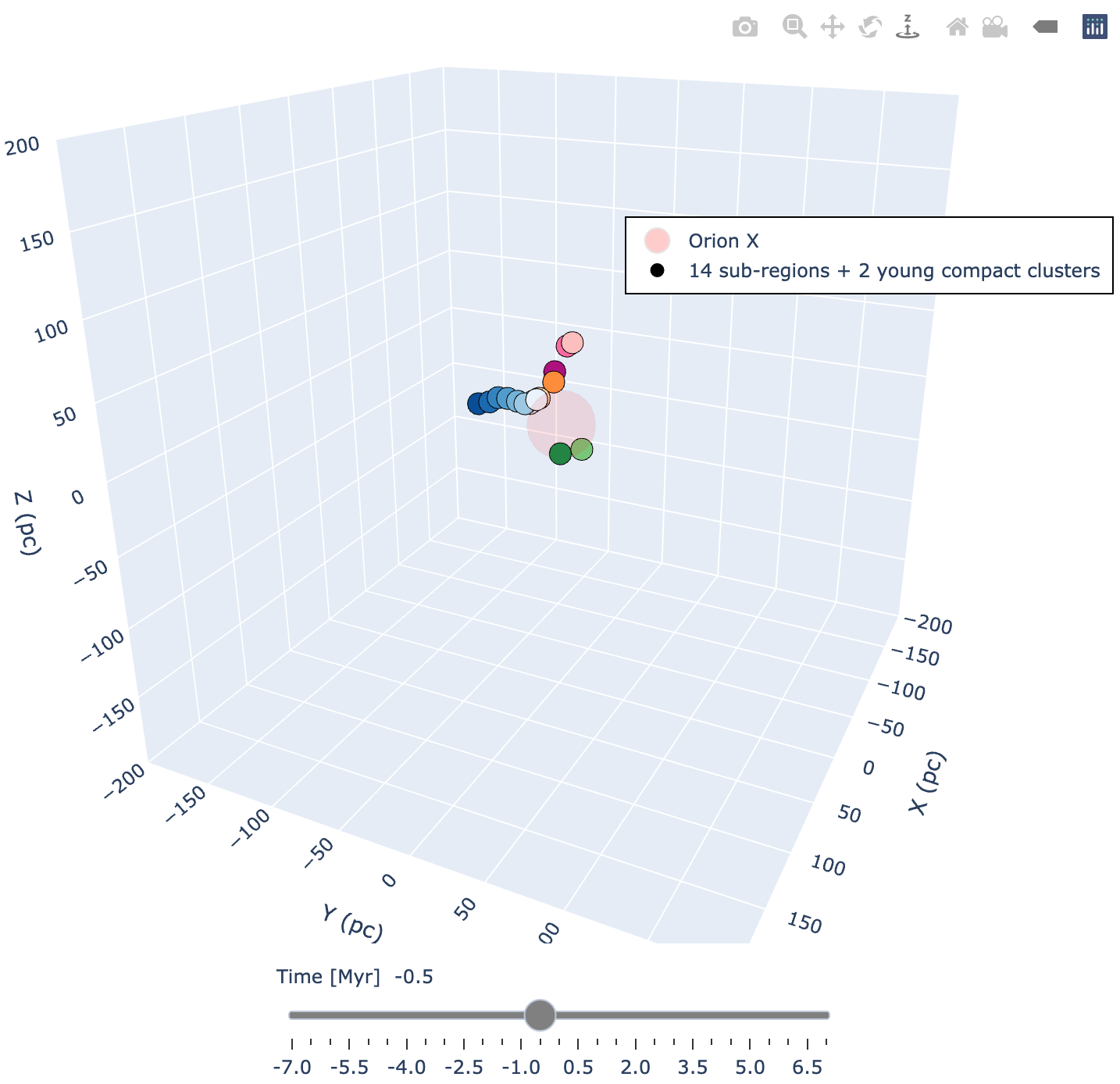}
    \caption{Interactive 3D scatter plot in Galactic Cartesian coordinates centered on Orion\,X. Shown is a snapshot at time = 0\,Myr positioned to represent the view from the Sun toward Orion.
    The interactive plot includes a timeslider for -7 to +7\,Myr and shows the relative positions and motions of the Orion subregions over this time span, allowing a 3D investigation of the evolution of the region. Symbols and colors are the same as in Figs.~\ref{fig:overview}, \ref{fig:YZ-Orion-OX}, \ref{fig:XZ-Orion-OX}, \ref{fig:YX-Orion-OX}, and \ref{fig:ycc}. The figure includes the positions and motions of two young compact clusters, NGC\,1977 and $\sigma$\,Ori, as further discussed in Sect.~\ref{sec:groups} and Appendix~\ref{apx:ycc}.
    The interactive 3D version is available online at  \url{https://homepage.univie.ac.at/josefa.elisabeth.grossschedl/orion-bb.html}.
    }
    \label{fig:orion-bb}
\end{figure}

\section{Results} \label{Results}

In this section we present the resulting 6D parameters for the selected molecular clouds in Orion and the 3D space motions of these clouds. This is followed by a momentum analysis using the 6D information.

\subsection{Galactic Cartesian representation of the clouds 3D orientation and motions} \label{results_cartesian_3d}

In Tables~\ref{tab:averages_rv_pm} and \ref{tab:averages_rv} we present the resulting 6D parameters ($\alpha$, $\delta$, $\varpi$, $\mu_{\alpha*}$, $\mu_\delta$, \vhel) for each subregion, as determined from average properties of YSOs and gas. 
The distances $d$ to the clouds, as determined from YSOs, mostly agree with other studies based on other methods, for example, by \citet{Zucker2019a,Zucker2020}. 
We find the largest discrepancy for L1622, where we get a closer distance ($d = 338\pm11$\,pc) compared to \citet{Zucker2020} ($d = 418\pm20$\,pc) by 80\,pc. This could be because L1622 covers a rather small solid angle in the sky, and is projected on top of more distant clouds likely associated with the Orion\,B main clouds. This overlapping-cloud scenario is consistent with gas radial velocity measurements, where L1622 shows a blue-shifted motion relative to its surroundings (by $\Delta v_\mathrm{LSR} \sim 9$\,km/s), suggesting it is a different cloud \citep[see also][]{Reipurth2008}. Moreover, the outline of the L1622 cloud can be seen distinct form the background in the optical in Fig.\,\ref{fig:overview} (left panel).
Toward IC\,2118 \citet{Zucker2020} give distances for three lines of sight ($328\substack{+15\\-20}$\,pc, $273\substack{+8\\-11}$\,pc, $283\substack{+16\\-30}$\,pc), which scatter around our distance determination of $d=293\pm14$\,pc. On the other hand, their distance for L1616 ($392\substack{+8\\-7}$\,pc) fits well with ours ($386\pm21$\,pc).
For Orion\,A and B a comparison is not straightforward since they report several positions, which deviate from the projected high column-density regions of the clouds, while the \citet{Zucker2020} distances in these regions scatter around our findings. In conclusion, we find that estimating distances to molecular clouds based on YSO distances delivers consistent results within the errors when compared to other methods. This was already demonstrated in \citetalias{Grossschedl2018}. 

In Table~\ref{tab:averages_xyz} we present the Galactic Cartesian representation of the average cloud parameters, as introduced in Sect.~\ref{galpy}. The Cartesian LSR velocities in the table deliver some first results for Orion. First, the current dominating motion is in the $X$-direction ($U_{\mathrm{LSR}}$), which is mostly negative, signifying that the clouds move away from the Galactic center, probably toward their apogalacticon, except for IC\,2118.
Second, all the motions in $Z$-direction ($W_{\mathrm{LSR}}$) are close to zero, except for L1622, which moves toward the Galactic plane with relatively high velocity. $W_{\mathrm{LSR}} \sim 0$\,km/s means that most of the clouds in Orion have reached their maximum distance to the Galactic mid-plain (with distances between 80 to 140\,pc below the plane), where they now have slowed down to zero vertical velocity and will consequently not move farther away but start to fall back toward the plane.

When investigating the 3D positions of L1622, L1616, and IC\,2118 in more detail, we find that all three clumps are almost at an equal distance to each other ($\sim 90$\,pc to 95\,pc), forming a quasi-equilateral triangle, which can not be seen in projection. This is intriguing and deserves further analysis.
Considering IC\,2118, one can see in the optical (Fig.\,\ref{fig:overview}) that the whole Witch Head Nebula is a prominent reflection nebula, likely illuminated (and maybe partially shaped) by the supergiant Rigel \citep{Kun2001}. With the here estimated distance to the head of IC\,2118 we determine a rough separation between IC\,2118 and \object{Rigel} \citep[$d \sim 265$\,pc,][]{vanLeeuwen2007} of about 32\,pc. The radial velocity of Rigel is given with \vhel$\sim$17.8 \citep{Gontcharov2006}, which means that its line of sight motion is about 2\,km/s red-shifted relative to IC\,2118, indicating that IC\,2118 might only pass by the supergiant and not move together with it. 
Generally, the improved estimate of the 3D separation of the two objects (compared to previous 2D estimates) allows a better quantification of stellar feedback from supergiants such as Rigel in the future.

The clouds, which clearly show peculiar motions, especially L1622 and also IC\,2118, could be a result of external perturbations that accelerated these parts of Orion away from the bulk motion of the region. This fits to the scenario that we introduced in this paper and suggested in \citetalias{Grossschedl2018}, attributing the bent structure of the Orion\,A cloud to be shaped by feedback of massive stars. The diverting motions suggest that external perturbations are very likely and have influenced other parts in addition to Orion\,A. We investigate this idea further in the following section, where we look at the relative motions of the molecular clouds in Orion in more detail.

\subsection{Relative space motions of molecular clouds in Orion} \label{results_relative_motions}

To test the assumption that some of the clouds in Orion were potentially pushed by some feedback event that took place roughly between the studied clouds, we first determine the point in time when the subregions were closest. To this end, we traced the orbits back and forth in time (from $-20$\,Myr to $+20$\,Myr) to determine the moment where the regions show the most compact configuration. In particular, we calculated the sum of all Cartesian distances between all subregions at timesteps of 0.1\,Myr to find the minimum. The result is shown in Fig.~\ref{fig:min-distance}, where we plot the normalized summed distances versus time, first using all 14 subregions and second using only nine subregions. The second version excludes the last five regions in the tail of Orion\,A, since the tail is, as shown in \citetalias{Grossschedl2018}, at larger distances than the head, and is likely unperturbed by the feedback event. 
We find that the minimum lies at $-6.1$\,Myr when using 14 regions and at $-6.8$\,Myr when using nine regions, while it is more pronounced for the second version without Orion\,A's tail. We conclude that the subregions were closest about $5$ to $7$\,Myr ago, which roughly sets the dynamical age of an expansion event. Due to the various uncertainties involved, a more precise estimate is not feasible at the moment. The uncertainties of the tracebacks include measurement errors, statistical errors, LSR conversion inconsistencies, systematic offsets in the different data sets, and the ignored gravitational field of the gas clouds.

To analyze the situation in more detail we further investigated the relative motions of the clouds. To this end, we need to define a central position and rest velocity to derive the relative motions. However, a clear determination of such a central point of origin is not straightforward due to different reasons: 
(a) There is likely not a single point of origin in the first place. Several massive stars formed in the region and produced feedback (radiation, ionization, winds, supernovae), as indicated by the nested shells in the Orion-Eridanus superbubble \citep[e.g.,][see also the discussion in Sect.~\ref{sec:superbubble}]{Ochsendorf2015, Joubaud2019}. More likely, the origin of such feedback could have resided in one ore several relatively older stellar group(s) (see Sect.\,\ref{sec:groups}), which are located throughout the larger Orion complex, such as the OB association called Orion OB1 \citep[e.g.,][]{Blaauw1964,Brown1994};
(b) Taking simply the average position and motion from the 14 studied cloud parts (as observed today), or their center of mass, would be biased by the chosen cloud sample. Moreover, the studied molecular clouds had different initial masses and densities and therefore were likely influenced differently by a feedback event. A momentum analysis could help, even if it brings further significant uncertainties, especially due to the unknown initial masses and densities, as discussed further in Sects.~\ref{momentum} and \ref{discussion:mom}; 
(c) If choosing one of the older stellar clusters in the region, then also the age of this progenitor cluster needs to fit into the scenario with about 10\,Myr or older to allow for stellar feedback in the form of supernovae, to fit the ages of the presented YSO samples, which are all younger than about 5\,Myr. Determining cluster ages is again not free of uncertainties, as elaborated below; 
(d) Finally, the uncertainties in the determined 6D cloud parameters do not allow a perfect analysis, since the errors will grow with each timestep.

To set a point of origin and rest frame we decided to look for possible progenitor clusters, which could have been the hosts of massive stellar feedback in the form of radiation, ionization, winds, and supernovae. Considering the above mentioned caveats, any chosen point of origin should only be seen as an approximation.
There are several studies who did a cluster analysis in the Orion region, including \citet{Kounkel2018, Kounkel2019}, \citet{Zari2019}, \citet{Kos2019}, and \citet{Chen2020}. All of these studies deliver partially overlapping results and overall rather complex stellar groups and subgroups.  
For our purposes, we investigated the 25 comoving groups of young stars in Orion that were recently identified by \citet{Chen2020}, to get the most reasonable central position and subsequently relative cloud motions. The authors selected the individual stellar populations by applying machine learning methods to \textit{Gaia}\,DR2 astrometry, while many of these represent well-known clusters. By comparing the positions and motions of the individual populations to the cloud ensemble, we identified the comoving stellar group \object{Orion X} \citep{Bouy2015} as a possible point of origin for the feedback scenario. It is located roughly between the ONC and the outlying clouds, extending for about $\SI{7}{\degree}$ ($\sim\SI{40}{pc}$), as shown in Fig.~\ref{fig:planck-orionx}. 

We investigated the age of Orion\,X by comparing with isochrones in an \textit{Gaia} color-absolute-magnitude diagram (CMD, equivalent to an Hertzsprung-Russel Diagram, HRD), as shown in Fig.~\ref{fig:hrd-orionx}. We used PARSEC isochrones from \citet{Bressan2012} with the \citet{Weiler2018} \textit{Gaia} DR2 passbands, solar metallicity (metal fraction $z = 0.0152$), and no extinction. 
An extinction vector of length A$_\mathrm{V} = 1$\,mag is shown in Fig.~\ref{fig:hrd-orionx}, indicating how extinction would influence the colors and magnitudes. The extinction law is derived from an extinction curve with R$_\mathrm{V}$=3.1 taken from \citealt{Cardelli1989} and \citealt{ODonnell1994}.
The Orion\,X group members are color-coded in Fig.~\ref{fig:hrd-orionx} for frequentist membership probability, giving the percentage of times a star appears in an assigned group, according to the method described in \citet{Chen2020}. Sources with higher membership probability appear less scattered in the CMD, confirming their group membership due to similar ages.
From the investigated isochrones we determined that the cluster age is likely between 8 and 15\,Myr. It is not feasible to state a more precise age due to the scatter of the cluster members in the CMD-space, the uncertain metallicity, systematic errors intrinsic to theoretical isochrones, and possible foreground extinction\footnote{Unheeded extinction would make the stars seem slightly too young.}. The given lower limit could signify a problem for the proposed scenario, while an older age than 8\,Myr seems more likely, especially when considering high-probability members.

For above mentioned reasons we chose the average position and motion of Orion\,X (see last row in Tables~\ref{tab:overview} to \ref{tab:averages_xyz}) to determine the relative motions for our cloud sample. To this end, we put the average position of Orion\,X in the center of our Cartesian coordinate frame, and we individually calculate the average orbit of the cluster the same way as for the subregions. The x-axis ($X'_\mathrm{Orion-X}$) of this frame is oriented toward the average Orion\,X position ($l,b = \SI{206.038}{\degree}, \SI{-21.945}{\degree}$), as given in Sect.~\ref{galpy}, which allows a better orientation and interpretation of the situation.

The results are presented in Figs.\,\ref{fig:YZ-Orion-OX} to \ref{fig:YX-Orion-OX}, which show the positions of the subregions at several snapshots in time from three different viewing angels, while the central panels in each figure represent the situation today.
The subregions are colored as in Fig.~\ref{fig:overview}. The point-sizes are scaled with distance within the current projection (i.e., along the third axis), and they are normalized relative to Orion\,X (gray dashed circles in bottom-left corner of middle panels). The larger red ``disk'' at (0,0) of each panel represents the location and extent of Orion\,X, which we determined to be about 40\,pc, from known members and for the situation today (see Fig.~\ref{fig:planck-orionx}). The relative trails of the regions are shown as gray lines. Each figure shows nine snapshots in time between -6.8 and 6.8\,Myr. The starting time was chosen based on the minimum distances between the regions as demonstrated in Fig.~\ref{fig:min-distance}.
Figure~\ref{fig:YZ-Orion-OX} shows the positions of the subregions in the $Y'_\mathrm{Orion-X}$/$Z'_\mathrm{Orion-X}$ plane that approximates their projected positions as viewed from the Sun at $t=0$\,Myr\footnote{The Sun moves away from that point of view with time.}, hence it presents a face-on view. 
Figure~\ref{fig:XZ-Orion-OX} ($X'_\mathrm{Orion-X}$/$Z'_\mathrm{Orion-X}$) shows a side-view within this coordinate frame, revealing the different distances of the subregions relative to the Sun. Especially the prominent tail of Orion\,A is clearly visible. 
Figure~\ref{fig:YX-Orion-OX} ($Y'_\mathrm{Orion-X}$/$X'_\mathrm{Orion-X}$) shows a top-down view and highlights again the bent structure of Orion\,A's head. This last orientation is similar to Fig.\,4 in \citetalias{Grossschedl2018}. 
Additionally, an interactive 3D plot is available in Fig.~\ref{fig:orion-bb}, allowing a more detailed investigation of the positions and motions of the subregions from all viewing angles and at various timesteps.

The analysis of the cloud's 3D space motions, as presented in Figs.\,\ref{fig:YZ-Orion-OX} to \ref{fig:orion-bb}, reveals that the clouds indeed were closest about 6\,Myr ago. This supports the idea that some feedback event(s) took place located near the head of Orion\,A and between the studied regions, since all subregions move radially away from a rough common center.
However, they do not seem to converge to a central point. This could be due to uncertainties or indicate that there was not a single event that accelerated the regions. Additionally, the paths of the small clumps (L1622, L1616, and IC\,2118) seem to cross each other when starting at -6.8\,Myr, which puts an upper limit to the age of a possible feedback event. This is further discussed in Sect.~\ref{sec:superbubble}.

For individual subregions we see that especially the small cometary clouds show quite high relative motions. This is feasible, since the lower mass clouds were likely affected differently compared to their high-mass counterparts in the region. L1622 is located closest to the Galactic plane and at the same time continues to move toward the plain at high speed. In the tracebacks around $-6$\,Myr we see that L1622 originates in the same region as IC\,2118 and L1616, even though it can be found today in a completely different environment, and actually was introduced in this paper with the Orion\,B main clouds based on its projected position.

For Orion\,A we see that the head of the cloud indeed seems to relatively approach the tail, supporting the scenario in \citetalias{Grossschedl2018}, where we already argue that the head of the cloud was pushed (see also Figs.\,\ref{fig:ysos_uvw_glon} and \ref{fig:ysos_uvw_relative}).
The relative motions of Orion\,A's tail in Figs.\,\ref{fig:YZ-Orion-OX} to \ref{fig:YX-Orion-OX} also show a motion departing from Orion\,X, opposing our argument that the tail is largely unperturbed, which we made based on the facts that it is a more distant and a more quiescent star-forming region.
Likely the Orion\,X cluster has some relative motion on its own and did not influence the tail as strongly as indicated in Figs.~\ref{fig:YZ-Orion-OX} to \ref{fig:YX-Orion-OX} (see also Sect.~\ref{results-PVD} and Fig.~\ref{fig:PVD-OrionX}).

The Orion\,B main clouds also move radially away from the central position, while L1630-S fits better in this scenario with Orion\,X as center compared to L1630-N. It could be that Orion\,B was instead influenced by feedback originating from different clusters besides Orion\,X. 
Other possible progenitor clusters include groups with ages of about 10\,Myr or older, for example, OB1a in the northeast \citep{Warren1977, Brown1994, Briceno2001, Briceno2008}, or the stellar groups found near the Belt stars of Orion, also named Orion Belt population \citep[OBP,][]{Kubiak2017} in \citet{Chen2020} or Orion\,D in \citet{Kounkel2018}.

To better understand the role of feedback in the whole region, all groups in Orion need to be studied in more detail in that context, implying first a more robust and more statistically significant membership analysis followed by better age determinations. In Sect.~\ref{sec:groups} we briefly discuss some groups in the context of the feedback scenario. A more detailed study of all groups in Orion goes beyond the scope of this paper.

\begin{table*}[!ht] 
\begin{center}
\small
\caption{
Estimated relative velocities ($v_\mathrm{rest}$), momenta ($p$), and kinetic energies (E$_\mathrm{kin}$) for nine subregions. The values were determined using three different rest velocities $v_\mathrm{rest}$ to get $v_\mathrm{rel}$. The total cloud masses $M_\mathrm{cloud}$ are a combination of gas mass and stellar masses from YSOs. Col.\,3 gives the projected surface area $a_\mathrm{today}$ of each subregion.
}
\renewcommand{\arraystretch}{1.2}
\begin{tabular}{lrcccccccccc}
\hline \hline
 & & & 
\multicolumn{3}{c}{relative velocities} & 
\multicolumn{3}{c}{momenta} &
\multicolumn{3}{c}{kinetic energies} \\
\cmidrule(lr){4-6}
\cmidrule(lr){7-9}
\cmidrule(lr){10-12}
\multicolumn{1}{l}{Subregion} &
\multicolumn{1}{c}{$M_\mathrm{cloud}$} &
\multicolumn{1}{c}{$a_\mathrm{today}$} &
\multicolumn{1}{c}{$v_\mathrm{rel,ox}$} &
\multicolumn{1}{c}{$v_\mathrm{rel,tail}$} &
\multicolumn{1}{c}{$v_\mathrm{rel,obpn}$} &
\multicolumn{1}{c}{$p_\mathrm{ox}$} &
\multicolumn{1}{c}{$p_\mathrm{tail}$} &
\multicolumn{1}{c}{$p_\mathrm{obpn}$} &
\multicolumn{1}{c}{$E_\mathrm{kin,ox}$} &
\multicolumn{1}{c}{$E_\mathrm{kin,tail}$} &
\multicolumn{1}{c}{$E_\mathrm{kin,obpn}$} \\
 &
 \multicolumn{1}{c}{M$_\odot$} & 
 \multicolumn{1}{c}{pc$^2$} &
 \multicolumn{1}{c}{km/s} &
 \multicolumn{1}{c}{km/s} & 
 \multicolumn{1}{c}{km/s} &
 \multicolumn{1}{c}{M$_\odot$\,km\,s$^{-1}$} &
 \multicolumn{1}{c}{M$_\odot$\,km\,s$^{-1}$} & 
 \multicolumn{1}{c}{M$_\odot$\,km\,s$^{-1}$} &
 \multicolumn{1}{c}{$10^{47}$\,erg} &
 \multicolumn{1}{c}{$10^{47}$\,erg} &
 \multicolumn{1}{c}{$10^{47}$\,erg} \\
\hline
L1641-N & 3674 & 21 & 6.76 & 3.16 & 3.83 & 24818 & 11594 & 14084 & 16.67 & 3.64 & 5.37 \\
OMC-4/5 & 1417 & 10 & 7.10 & 3.39 & 4.10 & 10065 & 4797 & 5817 & 7.11 & 1.61 & 2.37 \\
OMC-1 & 2765 & 13 & 7.84 & 4.13 & 4.71 & 21665 & 11409 & 13035 & 16.88 & 4.68 & 6.11 \\
OMC-2/3 & 1600 & 10 & 8.91 & 5.06 & 5.89 & 14265 & 8097 & 9434 & 12.64 & 4.07 & 5.53 \\
L1630-S & 2918 & 35 & 7.34 & 4.16 & 4.69 & 21410 & 12123 & 13676 & 15.62 & 5.01 & 6.37 \\
L1630-N & 4178 & 29 & 6.96 & 3.99 & 4.97 & 29091 & 16672 & 20764 & 20.14 & 6.61 & 10.26 \\
L1622 & 637 & 7 & 9.77 & 11.08 & 8.92 & 6227 & 7057 & 5685 & 6.05 & 7.77 & 5.04 \\
L1616 & 180 & 3 & 3.41 & 2.94 & 4.08 & 614 & 530 & 735 & 0.21 & 0.15 & 0.30 \\
IC\,2118 & 123 & 3 & 6.53 & 9.05 & 8.99 & 800 & 1109 & 1101 & 0.52 & 1.00 & 0.98 \\
\hline
\end{tabular}
\renewcommand{\arraystretch}{1}
\label{tab:momentum}
\tablefoot{The abbreviations indicate which rest velocity was used to calculate $v_\mathrm{rel}$: Orion\,X (ox), L1641-S/C in Orion\,A's tail (tail), OBP-Near (obpn).}
\end{center}
\end{table*}

\subsection{Estimating momenta and kinetic energies} \label{momentum}

Because we are in the unique position of having in hand 3D cloud motions, we can estimate the momenta and kinetic energies of the cloud subregions. Assuming that an external feedback event (or events) caused the observed 3D motions, one should be able to see a signature of this event in the clouds' momenta and kinetic energies. Such an analysis is not straightforward and will necessarily have significant uncertainties, hence we discuss these shortly at the end of this section.

\begin{figure*}[!ht]
    \centering
    \includegraphics[width=1\linewidth]{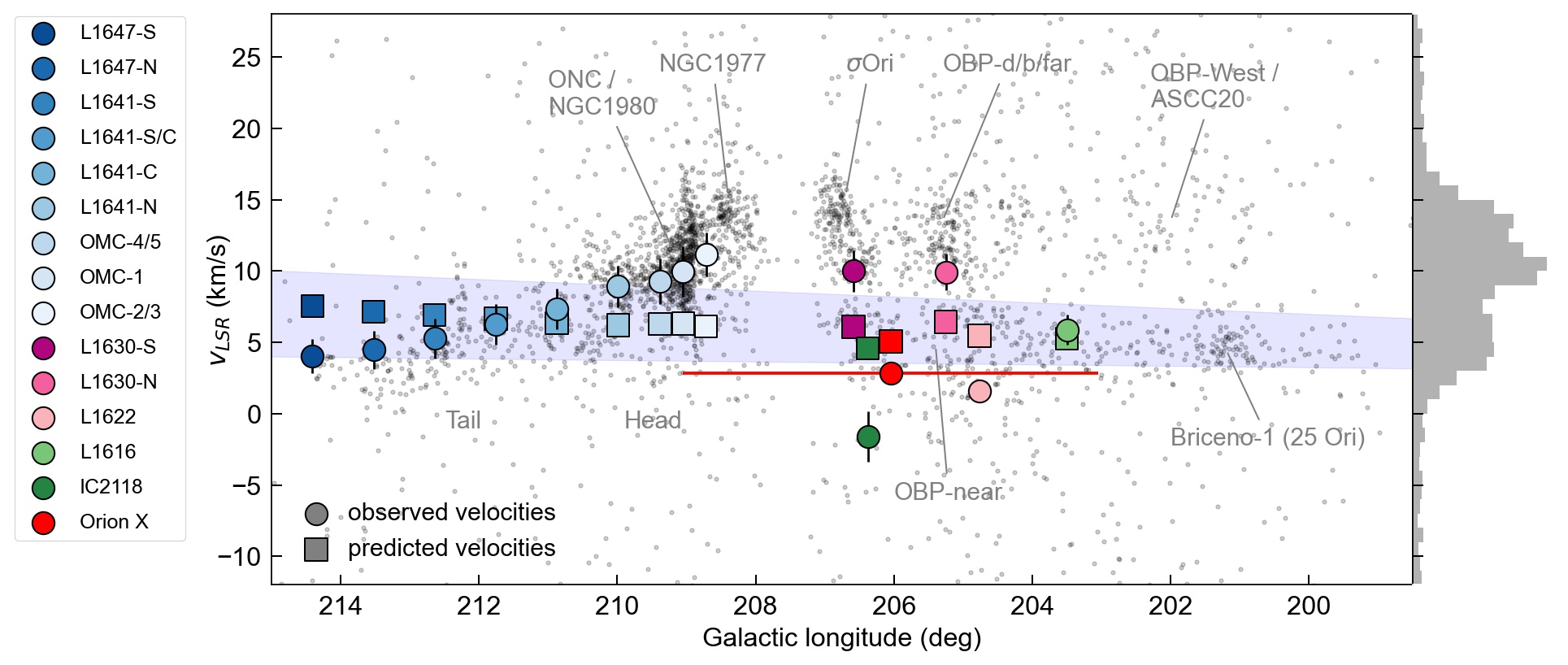}
    \caption{%
    Position-velocity diagram (\vlsr vs $l$) for \mbox{APOGEE-2} observed sources in the southern Orion region. The 14 subregions are shown as colored circles (see legend and Fig.~\ref{fig:overview}), with the error bars representing the velocity dispersion (Table~\ref{tab:averages_rv}). 
    The red circle indicates the average position and velocity of Orion\,X in this PV-space (error bar smaller than the symbol size), with the red horizontal bar showing the projected $l$-extent of the scatter of known group members \citep{Chen2020}. 
    The box-symbols represent the predicted PV-position of each corresponding cloud, if they were following the Galactic rotation curve without perturbation. The blue shaded area is shown as reference and encloses predicted motions for objects that are located between $\SI{200}{pc}<d<\SI{600}{pc}$, $\SI{215}{\degree}>l>\SI{198.5}{\degree}$, and $\SI{-27}{\degree}<b<\SI{-11}{\degree}$.
    For orientation we labeled the head and tail of Orion\,A, as well as other well-known stellar groups visible as over-densities in the PV-scatter (see Sect.~\ref{sec:groups}).
    The distribution of \vlsr motions is presented in the histogram on the right, which is bimodal with one peak centered at about 5\,km/s and another between 10 and 15\,km/s. The latter is dominated by the ONC and other young stellar groups. 
    The gap in the middle ($l \sim {207.5}$) is largely caused by data incompleteness in \mbox{APOGEE-2}, in combination with a lack of stellar clusterings in this region.}
    \label{fig:PVD-OrionX}
\end{figure*}

To get an estimate of the momenta and kinetic energies we first need to determine the cloud masses and relative velocities. We estimate the cloud masses from dust emission \citep[Herschel-Planck,][]{Lombardi2014} and dust extinction \citep[2MASS,][]{Lombardi2011} maps. The detailed procedure is described in Appendix~\ref{apx:masses}.
To get relative cloud motions we take a conservative approach and assume three different rest velocities to obtain a measure of the systematic error. First, we used the Orion\,X 3D space motions as rest velocity ($v_\mathrm{rest,ox}$), which is also used in Figs.~\ref{fig:YZ-Orion-OX} to \ref{fig:YX-Orion-OX}, second, we used the space motions of L1641-S/C in Orion\,A's tail as rest velocity ($v_\mathrm{rest,tail}$), and third, we used the space motions of OBP-Near ($v_\mathrm{rest,obpn}$) \citep[][see Sect.~\ref{sec:groups}]{Kubiak2017, Chen2020}. L1641-S/C was chosen, since this region seems to be unperturbed and does not show any deviating motion compared to predicted Galactic rotation (see Fig.~\ref{fig:PVD-OrionX}). OBP-Near also shows average motions similar to predicted motions and is therefore added here. We calculated the relative velocities ($v_\mathrm{rel}$) of the subregions in the $UVW$ frame (see Table~\ref{tab:averages_xyz}) to either Orion\,X, L1641-S/C, or OBP-Near velocities, for which we determined the following rest velocities in units of km/s: \\

\noindent $v_\mathrm{rest,ox}$: ($U$, $V$, $W$)$_\mathrm{LSR}$ = (-5.019, 3.000, 0.747), \\
$v_\mathrm{rest,tail}$: ($U$, $V$, $W$)$_\mathrm{LSR}$ = (-7.355, -0.513, -0.366), \\
$v_\mathrm{rest,obpn}$: ($U$, $V$, $W$)$_\mathrm{LSR}$ = (-7.109, 0.066, 1.751). \\

\noindent The momenta ($p$) and kinetic energies ($E_\mathrm{kin}$) of the cloud parts were then estimated with:
\begin{equation} 
\label{equ:mom_ekin}
\begin{aligned}
  & v_\mathrm{rel} = \sqrt{(U - U_\mathrm{rest})^2 + (V - V_\mathrm{rest})^2 + (W - W_\mathrm{rest})^2}, \\
  & p_\mathrm{cloud} = M_\mathrm{cloud} \cdot v_\mathrm{rel}, \\
  & E_\mathrm{kin,cloud} = M_\mathrm{cloud} \cdot v_\mathrm{rel}^2 \cdot 0.5.
\end{aligned}
\end{equation}
As pointed out above, we have to consider several uncertainties that could influence the resulting momenta and energies. These include (1) cloud masses, which can be estimated roughly within a factor of two (see Appendix \ref{apx:masses}); (2) the number of stars that have formed since the event (the total stellar mass); and (3) the relative cloud velocities, which requires the knowledge of the rest velocity $v_\mathrm{rest}$ (point of origin). Further uncertainties arise from (4) splitting clouds into subparts with sharp borders, especially for Orion\,A and the two main clouds parts in Orion\,B. 
Moreover, (5) we ignore any gravitational interactions between the clouds in our analysis. The here presented relative cloud motions are derived from today's observed velocities, while the observed trajectories could be slightly altered from their tracks (which they would have from a single push) due to gravitational interactions. Especially in the giant molecular clouds --- Orion\,A and Orion\,B --- internal gravitational interactions between the regions or self-gravity could influence the 3D motions. 
Finally, (6) the determined average 6D parameters include measurement and statistical uncertainties, propagating into these estimates.
These uncertainties have to be kept in mind while interpreting the results, which are listed in Table~\ref{tab:momentum} for nine subregions. We excluded Orion\,A's tail, which is likely not affected by external feedback. 

We find that the total momentum of the nine subregions is about one order of magnitude lower compared to the total radial momentum output of one supernova, which is about 2 to $4 \times 10^{5}$\,M$_\odot$\,km\,s$^{-1}$ \citep[e.g.,][see Sect.~\ref{discussion:mom}]{Kim2015, Walch2015, Haid2016}, and the sum of the observed kinetic energies is about two to three orders of magnitude lower compared to the total energy output of a supernova of about $1 \times 10^{51}$\,erg, while kinetic energy transfer is difficult to quantify, since it gets partially lost in thermal energy and can evaporate. 
We further analyze and discuss the implications of these results in Sect.~\ref{discussion:mom} and Fig.~\ref{fig:momentum}, where we focus on the momentum analysis.

\section{Discussion} \label{Discussion}

In this section we discuss the implications of the derived radial motions in the studied cloud sample. In our study of the 3D shape of Orion\,A \citepalias{Grossschedl2018} we found that this cloud is twice as long as previously assumed with a peculiar bent head. Very recently, \citet{Rezaei2020} using a different approach (Gaussian processes-based) confirmed the overall shape of Orion\,A, further motivating the analysis in this paper. We suggested in \citetalias{Grossschedl2018} that the cloud was perturbed by external forces, likely feedback forces from massive stars. We argued that if such a feedback event occurred in the recent past in Orion, which is likely given the number of massive stars in the region, it must have left a signature in the observed motions of the youngest stars and the gas from where they are emerging. Here we attempt to put Orion\,A into a larger context by exploring, for the first time, the 3D dynamics of the Orion star-forming complex by combining gas line of sight motions with space motions of YSOs. 


A main result of this paper is the discovery of coherent radial cloud motions on 100-pc scales in the Orion complex. We argue that the best explanation for the observables today calls for a major feedback event that took place in Orion about 6\,Myr ago that we name the Orion Big Blast event (Orion-BB event). This feedback event shaped the distribution and kinematics of the gas we observe today in two fundamental ways. First, it accelerated gas clouds radially away from a region we tentatively associate with the Orion\,X stellar population, and second, it compressed these clouds, increasing their star formation rate. The head of Orion\,A, containing the Orion Nebula Cluster, is a good example of the latter, as it displays today a star formation rate about an order of magnitude higher than the unperturbed tail region \citep{Grossschedl2018B}.

The seminal discovery of expanding associations of OB stars by  Viktor Ambartsumian \citep[e.g.,][]{Ambartsumian1947, Ambartsumian1958}, followed by the work of Adriaan Blaauw \citep[][]{Blaauw1964, Blaauw1991}, gave rise to a feedback-driven star formation model proposed by \citet{Elmegreen1977}, now the classical feedback-driven star formation model. The star formation scenario proposed in this paper offers strong support for the broad premise of the classical model, namely, that a previous generation of stars has a significant impact on the formation of the next. Unlike the classical model, and because we know today that the entire Orion complex is part of the much larger pre-existing gas structure, the Radcliffe Wave \citep{Alves2020}, there is no implicit need in our view to ``collect-and-collapse'' gas as in the classical model, but only to ``compress-and-collapse'' the pre-existing gas in the complex.

\subsection{Signatures of feedback in the large scale radial velocity structure} \label{results-PVD}

Our results indicate that a major feedback event took place in Orion about 6 Myr ago. If the regions investigated in this paper were indeed perturbed by a large feedback event, one could expect, to zero order, a roughly bimodal velocity distribution for the gas and young stars in the complex: (a) stars and gas not affected by the feedback event moving at a primordial radial velocity, and (b) the perturbed gas and the young stars associated with it moving at a different radial velocity. 

In Fig.\,\ref{fig:PVD-OrionX} we present a PV-diagram for stars observed by \mbox{APOGEE-2} in the southern Orion region (gray dots), with applied quality criteria. The 14 subregions are shown as filled circles and colored as in Fig.\,\ref{fig:overview}, and the average PV-position of Orion\,X is shown in red. For individual sources (observed by \mbox{APOGEE-2}) stellar radial velocities were used, while for the 14 subregions the gas radial velocities were used. The predicted velocities (box symbols) are the expected velocities from Galactic rotation alone, without external pressure from a feedback event, calculated for each subregion individually using a Milky Way potential \citep[][]{Bovy2015}, with the help of \textit{Astropy} and \textit{Galpy}\footnote{\texttt{galpy.potential.MWPotential2014}, \texttt{galpy.potential.vcirc}}.
The predicted velocities are a function of position ($d$, $l$, $b$) and they fall at \vlsr about 5 to 7\,km/s for stars in Orion.

It is clear from Fig.\,\ref{fig:PVD-OrionX} that the observed radial velocities present a bimodal distribution (see histogram on the right y-axis) and that most stars (and subregions) in Orion are located above the predicted velocities, forming an arc-like structure above \vlsr$\sim$5\,km/s, except for IC\,2118 and L1622. This shows that most young stars in Orion have relatively red-shifted line of sight motions with respect to predicted motions. 
This supports the feedback-driven ``push scenario'', where an external feedback event took place largely in-front of pre-existing gas, from the Sun's point of view. The clouds in the region also largely follow this arc. Only the cometary clouds L1622 and IC\,2118 have blue-shifted velocities and are moving relatively to the front, indicating that they were located between the feedback event and the Sun and that only a small fraction of the gas was on ``our'' side of the event. L1616 seems unperturbed in this PV-space, since it has only a minor component of motion along the line of sight direction. 

The position of Orion\,X in this PV-diagram indicates that the cluster roughly shares the large scale motion of unperturbed regions in Orion, showing slightly blue-shifted velocities. We note that only four stars in Orion\,X have been observed by \mbox{APOGEE-2} with sufficient quality, hence the average \vlsr of the cluster is likely more uncertain than the scatter of the observed data points (error bar fits within the red circle), and the resulting velocity dispersion (standard deviation) is underestimated due to low statistics. 

The predicted velocities in the PV-diagram are an approximation, as Orion is part of the Radcliffe Wave, itself deviating slightly from pure Galactic rotation \citep{Alves2020}.
In this context it is interesting that the tail of Orion\,A, unperturbed by feedback, appears slightly blue-shifted in comparison to the average rotation, which cloud be a signature of the Radcliffe Wave motion.
In general, assuming simple Galactic rotation for any single cloud is always an approximation, since various forms of gravitational or feedback forces constantly act within the Galaxy, as shown recently, for example, in \citet{Smith2020} or \citet{Jeffreson2020} via numerical simulations. 
On average GMCs have a cloud-to-cloud velocity dispersion of about 6 to 8\,km/s \citep{Stark1984,Stark1989}, while this was estimated for a larger GMC sample and not within a single star-forming region.
The point we want to make here is that the deviation caused by feedback (on the order of 10 km/s, +5 and -5 km/s from the average) dominates the velocity distribution in Orion, revealing that feedback has a major impact on the gas dynamics of the entire complex, and subsequently on the velocities of most of the young stars that formed inside the perturbed gas in Orion. 

Additional evidence of feedback can be seen in the large-scale structure of gas and dust in the whole Orion region, which largely shows a wind-blown appearance, as can be seen, for example, in dust emission (Fig.~\ref{fig:planck-orionx}) or even in the optical (Fig.~\ref{fig:overview}, left panel). Both Orion\,A's head and Orion\,B show evidence of cometary-like tails and streamers blown away from a central region near Orion\,X, while being compressed on the southwestern rims, where YSOs are predominantly formed.
The cometary clouds (L1622, L1616, IC\,2118) have orientations with their tails pointing away from a central location in front of Orion A and B. Further cometary clouds in this region (with less or no star formation) show a similar structure and orientation, such as the L1617 cometary clouds to the northwest or L1634 in the south (Fig.\,\ref{fig:overview}). 

We have now several lines of evidence (spatial and dynamical) that Orion\,A's head, also known as the Integral Shape Filament, clearly seems to have been pushed backwards while the tail seems dynamically unperturbed, confirming our assumption of a compressed head \citepalias{Grossschedl2018} and a largely unperturbed tail. We also know that the star formation rate in the head of Orion\,A is about an order of magnitude higher compared to that of the tail \citep{Grossschedl2018B}. Together, these facts naturally lead to the conclusion that practically all of the very young stars (Class I and II) in the star formation rich head of Orion\,A, but also in Orion\,B, and possibly some of the young but dust-free populations such as the OBP \citep[][see also Sect.~\ref{sec:groups}]{Kubiak2017, Chen2020}, are a product of large-scale triggering. This conclusion implies that at the genesis of the Orion Nebula Cluster (and NGC\,2023/2024 in Orion\,B) lies a feedback, compression, and triggering process, which will need to be taken into account in cluster formation models. 

We caution against a ``one-event-fits-all'' feedback event scenario, and address this further below. For example, the situation in the Orion\,B main cloud seems more complex than Orion\,A.  
Our determined relative motions of the two main parts in Orion\,B (especially L1630-N) do not fit perfectly in the picture with Orion\,X as progenitor. There are clearly perturbations visible in the radial velocities of the gas in Orion\,B going beyond the shown PV-diagram in Fig.~\ref{fig:pv-gas-ysos_orionb}, where the middle cloud part was excluded \citep[see, e.g.,][]{Nishimura2015, Bron2018, Orkisz2019}. Additionally, the proper motions of the studied YSO samples in L1630-S/N do not show a clear single peak in proper motion space (Fig.\,\ref{fig:orionb_regions}) suggesting further perturbations and indicating a more complex dynamical status than a ``simple push from the front'' for these regions. 

Finally, the HII-region \object{IC 434}, which illuminates the Horse Head Nebula, is likely interacting with the southern part of the cloud \citep[see also][]{Bally2018, Orkisz2019}. The driver of this HII-region is the $\sigma$\,Orionis cluster \citep{Walter2008, Caballero2010a, Caballero2010b, Hernandez2007, Ochsendorf2014, Caballero2019}, which is located directly to the southwest of L1630-S, with a massive multiple system at its center, containing the O and B stars $\sigma$ Ori AB. 
\citet{Ochsendorf2014} propose that the illuminated dust structure IC\,434 is a first example of a dust (or bow) wave created by massive stars moving through the interstellar medium. They argue that dust and bow waves are most efficiently formed around weak-wind stars moving through a high density medium. To further investigate this line of argument we look at the relative motion and position of the \object{$\sigma$ Ori} cluster and L1630-S in Sect.~\ref{sec:groups}.

Concluding, feedback in Orion, and, in particular, the Orion-BB event that took place about 6\,Myr ago, has had a fundamental role in shaping the gas distribution and gas dynamics, and consequently, the dynamics of the young stellar population in Orion.

\subsection{The connection to the Orion-Eridanus superbubble} \label{sec:superbubble}

Generally, the idea that the Orion clouds were and are being affected by the feedback of massive stars is not new \citep[e.g.,][]{Elmegreen1977, ReynoldsOgden1979, Cowie1979, Bally1987, Brown1994, Brown1995, Bally2008, Bally2010}. It has been proposed that several nested shells are superimposed along the line of sight, forming the so called \object{Orion-Eridanus superbubble}, a less than 10\,Myr old relic shell composed of several supernova remnants, spatially encircling the main Orion clouds \citep[e.g.,][]{LeeChen2009, Pon2014A, Pon2014B, Pon2016, Ochsendorf2015, Soler2018, Joubaud2019}. The entire Orion complex, including Orion\,X, is located within the superbubble, and the bubble expands with a speed of about 20 to 40\,km/s \citep[e.g.,][]{ReynoldsOgden1979,Ochsendorf2015} from a central position, located near Orion\,X. The whole superbubble (and Barnard's loop, see below) likely gets replenished of ionizing photons by the massive stars from the Orion OB1 association \citep[e.g.,][]{Brown1994, Brown1995, ODell2011, Ochsendorf2015}. 

Within the superbubble lies \object{Barnard's loop} \citep{Barnard1894,ODell1967,ODell2011}, which is suggested to be a more recent supernova remnant \citep[age $\sim 3 \times 10^5$\,yr,][]{Ochsendorf2015}. It is likely located behind L1622, since L1622 is seen in absorption against the H$\alpha$ background \citep[see also the H$\alpha$ map by][]{Finkbeiner2003}. This agrees with the fact that L1622 is a separate cometary structure, underlined by its relatively closer distance to the Sun ($\sim 338$\,pc) and very distinct radial velocities compared to its surroundings (Sect.~\ref{results_cartesian_3d}). This suggests that Barnard's loop is located roughly between 400\,pc (seen in projection above L1641-S, L1641-S/C, and L1630-N) and 340\,pc (behind L1622), in agreement with previous estimates.

Based on the expansion velocity and the extent of the Orion-Eridanus superbubble, \citet{Ochsendorf2015} give a dynamical age of about 4\,Myr for the superbubble, while they conclude that an age between $\sim$ 5 to 10\,Myr is more likely. \citet{Pon2016} estimate ages between 2 to 6\,Myr based on Kompaneets model. These previous estimates deliver partially lower dynamical ages, while they are overall agreeing with our results, given the uncertainties influencing all of these estimates. For example, \citet{Pon2016} point out that by neglecting momentum conservation or cooling via mass loading, their model could overestimate the expansion velocity, hence underestimate the age. 
As pointed out above (Sect.~\ref{results_relative_motions}), when investigating the tracebacks in more detail, a more recent time for the onset of expansion seems more likely (see movie links in Fig.~\ref{fig:YZ-Orion-OX} to \ref{fig:YX-Orion-OX}). Some traceback paths (especially the cometary clouds) seem to cross each other when starting at -6.8\,Myr. Concluding, the age as determined from cloud tracebacks seems to fit well with the dynamical age estimates for the superbubble.  This suggests that the progenitors of the superbubble are likely related to the Orion-BB event.

A similar finding is discussed in \citet{Kounkel2020} who use a different approach by investigating the 3D dynamics of stellar groups from \citet{Kounkel2019}. They also attribute the large scale dynamics of these young clusters to feedback from massive stars (while connecting it to Barnard's loop) also tracing this back to roughly 6\,Myr ago, independently confirming the main results of this work.
Another very recent result by \citet{Rezaei2020} confirms the bent structure of Orion\,A, using a Gaussian processes-based method to estimate the 3D structure of dust, via extinction. Additionally, they identify with their method a foreground dust ring, seen in projection in-front of Orion\,A, and partially Orion\,B, and they suggest that some of the closer YSOs toward Orion\,A are part of the foreground-ring. They determine the center of this ring to be roughly at a distance of 350\,pc, and argue that it could be a remnant of previous star formation episodes in this region. The location of this dust structure lies in the vicinity of Orion\,X and could be another signature of feedback connected to the Orion-BB event.

Regarding the source of feedback behind the Orion-BB event, much remains to be understood.
For example, \cite{Bally2008} estimates that 10 to 20 supernovae have exploded in the Orion complex over the last 10 to 15\,Myr, while also winds, photo-ionization, radiation pressure, and mass loss from evolved massive stars in the region have likely contributed \citep[e.g.,][]{Haid2018}.
Supernova(e) or other massive stellar feedback from the Orion\,X population are obvious culprits, while a more detailed analysis is critical and starts by identifying the complete population of Orion\,X, and possible other progenitor groups (Sect.\,\ref{sec:groups}), which urgently warrants a dedicated study. Orion\,X is a poorly understood stellar group, only recently identified in the Hipparcos data \citep{Bouy2015}, and an estimate of its initial mass function is needed to confirm its capacity to deliver massive stellar feedback, which will be addressed in a future paper.

We know from simulations \citep[e.g.,][]{Chevalier1999, Kim2015, Walch2015, Haid2016, Seifried2018, Lucas2020} that supernovae are not only disrupting surrounding gas but are also potentially able to shape pre-existing dense gas structures in molecular clouds when the shock wave expands in the vicinity (within few tens of pc). This distance criterion means that it is very unlikely for one single supernova to shape an entire large scale region such as Orion, which was probably already filled with dense molecular gas structures.
Given that the Orion complex is part of the much larger Radcliffe Wave gas structure \citep{Alves2020}, there is no implicit need in our scenario to ``collect-and-collapse'', but only to ``shape-and-collapse'' the pre-existing gas in the complex. 
One would expect that the smaller clouds move faster than more massive clouds from momentum conservation considerations alone (see Sects.~\ref{momentum} and \ref{discussion:mom}), and if one or several supernovae exploded within this pre-existing heterogeneously structured cloudy environment.
This is true to a point, L1622 is the fastest moving cloud (see Table~\ref{tab:momentum}), while the more massive Orion\,A and B clouds seem to be moving too fast if all subregions were affected by one single event. More likely, several events and the continuous forces from radiation and stellar winds from massive stars have instead been shaping the Orion complex. In Sect.~\ref{discussion:mom} we compare the results of the momentum analysis (Sect.~\ref{momentum}) with numerical simulations to get a first quantitative estimate of the signatures of feedback in Orion.

Concluding, figuring out the relative roles of the sources of feedback (supernovae, winds, ionization, radiation) in Orion is critical to  quantify, to understand the role of feedback in driving star formation, a critical missing piece in our understanding of star and molecular cloud formation, with far-reaching impact beyond the Local Milky Way.

\subsection{Other stellar groups in the context of the feedback scenario} \label{sec:groups}

There are other over-densities visible in the PV-diagram in Fig.~\ref{fig:PVD-OrionX}. Most of them are known groups and are listed in \citet{Chen2020}. The oldest subgroup in the Orion OB1 association is OB1a \citep[e.g.,][]{Blaauw1964, Warren1977, Brown1994,Briceno2001, Briceno2007a, Briceno2007b, Briceno2008}, which is part of a large structure called Orion\,D in \citet{Kounkel2019}. OB1a can be split up into further subgroups, as was demonstrated, for example, by \citet{Kos2019} or \citet{Chen2020}. The oldest subgroup is ASCC20 \citep{Kharchenko2013} and has an age of about 21\,Myr as reported in \citet{Kos2019}. Its location, however, largely to the west of our discussed clouds, does not fit into the scenario proposed here. Still, due to its age it could have been one of the first clusters in Orion producing significant feedback and potentially triggering star formation. It overlaps in the PV-space with the subgroup OBP-West \citep{Kubiak2017,Chen2020}, which is part of Blaauw's OB1b population \citep[e.g.,][]{Warren1978, Briceno2005}.
The OB1a population also contains other prominent subgroups, for example, 25\,Ori \citep{Briceno2007b}, but better named Brice\~no-1 \citep[also called ASCC16,][]{Kharchenko2013, Kos2019}, as the star \object{25\,Ori} is not part of the cluster \citep{Chen2020}. The line of sight velocities of this group largely follow the average predicted velocities assuming Galactic rotation, as can be seen in Fig.~\ref{fig:PVD-OrionX}, which could mean that OB1a, and Brice\~no-1 in particular, did not experience any significant perturbation prior to formation. It was long thought to be one of the oldest groups in Orion, however \citet{Kos2019} report that ASCC16 (Brice\~no-1, age$\sim$13\,Myr) is younger than ASCC20. A dedicated study of the ages of all Orion groups is clearly warranted, and will enlighten the star formation history as well as the role of stellar feedback in the region.

Several groups have been identified along the line of sight in the region surrounding the Orion Belt stars (OB1b). Additional to the mentioned OBP-West, there are further groups listed in \citet{Chen2020}, while the stellar members of three of these groups share approximately the same PV-space (OBP-d, OBP-b, OBP-far), and fall within the arc-like structure in the PV-diagram, above NGC\,2068/2071 (L1630-N). 
Only one of the OBP groups seems to follow average Galactic rotation, being located in the blue band in Fig.~\ref{fig:PVD-OrionX}, the OBP-Near group \citep{Chen2020}. Besides Orion\,X it is also a promising candidate for feedback in this region, having an age of roughly 10\,Myr (from an isochrone investigation). Therefore, we used the average 3D space motions of this cluster in Sect.\,\ref{momentum} to get additional rest velocity estimates. Since the OBP groups are located closer to Orion\,B, they are maybe better candidates for hosts of feedback acting on Orion\,B (see also Sect.~\ref{results_relative_motions}).
Such a scenario was already proposed by \citet{Chromey1989}, who suggest a bifurcation between the Orion\,A and B clouds based on the identification of a giant HI shell in the Orion Belt region, likely caused by massive stellar feedback.

Focusing on the head of Orion\,A, the region surrounding the \object{ONC} \citep[mean age $\sim$2.5\,Myr, e.g.,][]{Jeffries2011} is the most prominent in the PV-diagram. Along the same line of sight, the ONC is superimposed by the older \object{NGC\,1980} cluster, proposed to be in the front of the ONC in \citet{Alves2012} and \citet{Bouy2014}. It has an age of about 4-5 Myr and was recently identified as a probable separate group from the ONC \citep{Chen2020}.
A similar finding is presented in \citet{Beccari2017} and \citet{Jerabkova2019}, who identify three distinct stellar populations toward the ONC showing an age sequence, suggested to be three separate episodes of star formation, with the oldest lying in the front, likely related to NCG\,1980. The age difference between each episode is about 1\,Myr and could be the result of sequential star formation, which would fit the global star formation scenario proposed here.
In radial velocity, NGC\,1980 is virtually identical with the ONC \citep[][and Fig.\,\ref{fig:PVD-OrionX}]{DaRio2016}. This fact and the estimated distance between these two clusters (on the order of 10\,pc), has made it difficult to disentangle NGC\,1980 and the ONC populations. Better astrometric data from future \textit{Gaia} data releases will hopefully clarify the nature of these two fascinating clusters. 
In this paper we are not focusing on the stellar clusters in Orion but on the 3D dynamics of molecular clouds, and we defer the disentangling of the various proposed stellar groups for a future paper. 

Two other prominent over-densities in the PV-diagram in Fig.~\ref{fig:PVD-OrionX} are the clusters NGC\,1977 \citep[e.g.,][see Fig.~\ref{fig:pv-gas-ysos}]{Peterson2008} and $\sigma$~Orionis (see Sect.~\ref{results-PVD} and Fig.\,\ref{fig:orionb_regions}). Both are red-shifted in the PV-space in Fig.~\ref{fig:PVD-OrionX}, located above ONC/NGC\,1980 and NGC\,2023/2024 (L1630-S). The rich young cluster $\sigma$~Ori (age about 2 to 4\,Myr) is well-known for the massive O-star at its center giving it its name. The cluster still contains pre-main-sequence stars with circumstellar disks (Class\,II YSOs, see also Fig.~\ref{fig:orionb_regions}), while the molecular gas out of which the stars have formed has already been dissolved. A similar situation occurs for NGC\,1977, which contains a massive B-star (42\,Ori) at is center. 
As pointed out in Sect.~\ref{OA-Methods}, this massive star could be responsible for the deviating gas versus stellar radial velocities at the tip of the Orion\,A GMC \citep{Bouy2014,Pabst2020}.
Being located between Orion\,A and B, these two young compact clusters are important in the context of this paper, as they could be a result of the Orion-BB event that took place about 6 Myr ago. 
The two clusters lie at about the same distance as the head of Orion\,A and \mbox{L1630-S} ({\it d}$\sim$390\,pc, average distances from YSOs) while the clusters are red-shifted compared to the adjacent cloud subregions ($\Delta v_\mathrm{LSR} \sim 2$ to 3\,km/s, see Fig.~\ref{fig:PVD-OrionX} and Table\,\ref{tab:averages_rv}). This indicates, in the feedback scenario we are proposing here, that the progenitor clouds of NGC\,1977 and $\sigma$\,Ori could have been located closer to the blast.

We test this idea in more detail in Appendix~\ref{apx:ycc}, by looking at the tracebacks of these clusters in the same manner as was done for the cloud subregions in Figs.~\ref{fig:YZ-Orion-OX} to \ref{fig:YX-Orion-OX}. The result of this additional analysis is shown in Figs.~\ref{fig:orion-bb} and \ref{fig:ycc}.
The 3D space motions of the two young compact clusters indeed indicate that the gas out of which they formed was likely closer to the proposed point of origin (closer to the blast), suggesting that they were likely formed first compared to the adjacent cloud subregions. 
In the interactive 3D plot (see  \href{https://homepage.univie.ac.at/josefa.elisabeth.grossschedl/orion-bb.html}{link} in Fig.~\ref{fig:orion-bb}) one can see that about 6\,Myr ago NGC\,1977 and $\sigma$\,Ori were close in space in front of Orion A's head, suggesting that Orion A was a longer filament in the past. This fit is remarkable, implying that the role of feedback goes beyond the shaping of the gas in the complex, since it seemed also to be critical for the formation of young stellar clusters that we observe today. In particular for NGC\,1977 we see that the motions of the stellar members show a perfect continuation of the overall trend in Orion\,A's head (see Figs.\,\ref{fig:ysos_uvw_glon} and \ref{fig:ysos_uvw_relative}).
Looking more closely at $\sigma$\,Ori we find that it currently passes L1630-S, confirming the scenario proposed by \citet{Ochsendorf2014} that $\sigma$\,Ori is moving toward the Orion\,B cloud from the west causing a bow wave. 
Generally we find that the overall 3D motions of the two young clusters seem to fit perfectly within the proposed scenario, with their average motions following the radial expansion of the large-scale motions on 100-pc scales.

\subsection{Interpretation of the observed momenta} \label{discussion:mom}

\begin{figure*}[!ht]
    \centering
        \includegraphics[width=\linewidth]{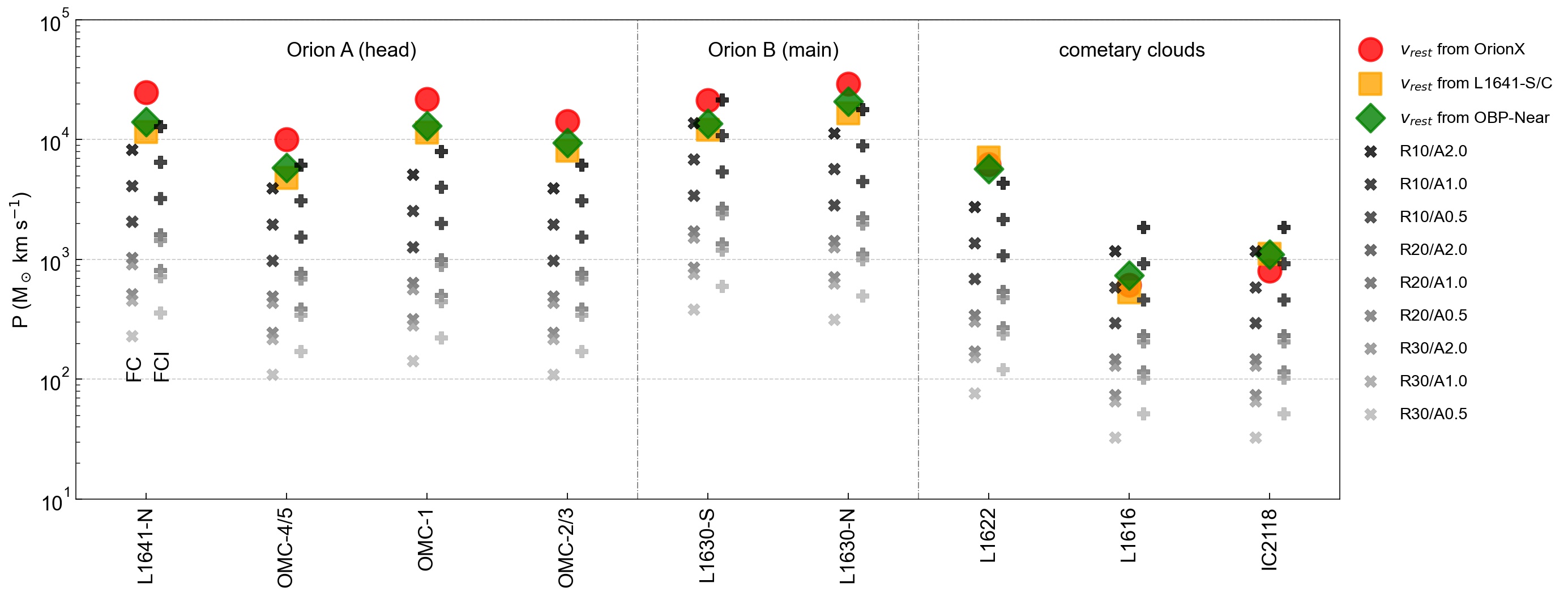}
    \caption{Momentum analysis for nine subregions. The momenta were calculated assuming three different rest velocities (from Orion\,X, red circles; L1641-S/C, orange boxes; OBP-Near, green diamonds) to calculate the relative velocities of the selected cloud parts. Additionally, theoretical estimates are shown as gray-shaded cross and plus symbols as determined from simulations by \citet{Walch2015}. The short-cuts in the legend indicate nine different cases: First, the supernova shell radius (R) was set to 10\,pc, 20\,pc, and 30\,pc at time of impact on the cloud. Second, the affected cloud areas (solid angle as seen from source of explosion) were varied as follows; case A1.0 uses the projected observed cloud surface area at the cloud's distance (Appendix~\ref{apx:masses}), and cases A2.0 and A0.5 use twice and half of the projected area, respectively. The simulation run FC (fractal+cooling) was set up for a fractal cloud with an average density of $n_0 = 100\,\mathrm{cm}^{-3}$, including cooling, and is shown representatively with cross-symbols in the legend for the nine different cases. The plus symbols are shown for the same cases, but for a model that includes initial ionization from the supernova progenitor star (FCI). See text for more explanations.}
    \label{fig:momentum}
\end{figure*}

To interpret the results from Sect.~\ref{momentum} we compare with estimates from the literature \citep{Kim2015, Walch2015, Haid2016} who numerically simulate the impact of a supernova explosion on the surrounding medium and estimate the momenta and energy input into molecular clouds.
\citet{Walch2015} test several setups where they calculate the effect of a supernova\footnote{With a total energy of $E_\mathrm{SN}=1 \times 10^{51}$\,erg and initial radial momentum of $p_\mathrm{SN}= 2.77 \times 10^{4}$\,M$_{\odot}$\,km\,s$^{-1}$.}, which explodes within a fractal molecular cloud environment with an average density of $n_0 = \SI{100}{cm^{-3}}$.
They find that after the Sedov-Taylor phase (after about 0.2\,Myr) the radial momentum output of one supernova to its surroundings from their setups with cooling (FC) or when including ionization (FCI) is about:
\begin{equation}
\begin{aligned}
  & p_\mathrm{WN15,FC} \approx 2.47 \times 10^5 \,\mathrm{M}_\odot\,\mathrm{km}\,\mathrm{s}^{-1}, \\
  & p_\mathrm{WN15,FCI} \approx 3.88 \times 10^5 \,\mathrm{M}_\odot\,\mathrm{km}\,\mathrm{s}^{-1}. \\
\end{aligned}
\end{equation}
\citet{Kim2015} determine from simulations of supernova explosions in a two-phase medium that the momentum of a supernova output can be written as:
\begin{equation}
\begin{aligned}
  & p = 2.8 \times 10^5  \,\mathrm{M}_\odot\,\mathrm{km}\,\mathrm{s}^{-1} \, n_0^{-0.17}, \\
  & p_\mathrm{KO15} \approx 1.28 \times 10^5\,\mathrm{M}_\odot\,\mathrm{km}\,\mathrm{s}^{-1} \,\,(\mathrm{with}~n_0=100).
\end{aligned}
\end{equation}
Both simulations deliver results agreeing within a factor of two when assuming a similar ambient inter stellar medium (ISM) density. 
Moreover, \citet{Haid2016} perform simulations investigating the role of turbulence at the moment of a supernova impact into a molecular cloud. They find that very high Mach numbers (highly supersonic turbulent medium) can boost the momentum transfer by 60\%.

In the following we focus on a comparison of the empirically estimated momenta with theoretical estimates. To estimate the fraction of the momentum from a supernova blast that acts on molecular clouds comparable in size to the Orion subregions, we calculate the momenta per area within a sphere by setting the supernova at different distances (R) from a molecular cloud: $\mathrm{R10}=10$\,pc, $\mathrm{R20}=20$\,pc, $\mathrm{R30}=30$\,pc\footnote{The radius range was chosen based on the relative distances of the studied subregions to each other at time = $-6.8$\,Myr, which are roughly between 20 to 80\,pc.}. This sets the shell radius within which the momentum gets transported radially outwards at given densities. The cloud surface areas are derived from dust emission or extinction maps, along with the cloud masses, as described in Appendix~\ref{apx:masses}. The surface areas are an estimate from today's projected cloud sizes, while any projection effects are ignored and also the initial surface areas are unknown. For this reason, we additionally look at scenarios where the cloud surface areas ($a_\mathrm{today}$) are half or twice as large as the projected ones, to get a measure for the influence of this parameter, namely: $\mathrm{A2.0} = 2 a_\mathrm{today}$, $\mathrm{A1.0} = a_\mathrm{today}$, $\mathrm{A0.5} = 0.5 a_\mathrm{today}$. With this we predict momenta using $p_\mathrm{WN15,FC}$ and $p_\mathrm{WN15,FCI}$ from \citet{Walch2015} with the results shown in Fig.~\ref{fig:momentum}. We exclude Orion\,A tail regions from this analysis (L1647-S to L1641-C), since these regions are likely unperturbed by the Orion-BB event, as discussed before.

The results in Fig.~\ref{fig:momentum} show that the observed momenta, as determined in Sect.~\ref{momentum} (see Table~\ref{tab:momentum}), are generally higher than the predicted ones and only match if assuming close distances or by overestimating the affected cloud surface areas. Including initial ionization from the progenitor star increases the radial momentum, but still close distances are needed for one single supernova to explain the observations. 
If using $P_\mathrm{KO15}$ as estimated by \citet{Kim2015} the predicted momenta would be even lower.
Generally, assuming higher initial cloud densities or larger distances from the expanding supernova shell would lower the energy transfer.

From the investigation of simulations we see that smaller surface areas are resulting in lower observed momenta, which roughly agrees with the trend of the observations. We would like to note that the cometary clouds L1616 and IC\,2118 show some small deviations from the overall trend, but this could be due to different reasons, including different initial distances to the blast event and an underestimation of the surface areas and the masses\footnote{This is likely, because additional cloud structures near L1616 and IC\,2118 were excluded. Especially the Witch Head Nebula shows a larger extension to the south.}. As pointed out in Appendix~\ref{apx:masses}, the cloud masses are likely underestimated by a factor of two for all regions, which would result in a larger difference of observed versus predicted momenta in Fig.~\ref{fig:momentum}. This suggests that the energy of likely more than one supernova is need to explain the observed 3D cloud kinematics.

\citet{Walch2015} also investigated binary explosions were the second supernova takes place 
with different offsets in time (varied between 0 to 100\,kyr). They find, the longer the time delay in between two explosions, the more efficient is the second one. This is because the first supernova already had sufficient time to excavate its surroundings (similar when including ionization from the progenitor star), hence the Sedov-Taylor phase of the second supernova lasts longer resulting in a larger final momentum.
This is supported by simulations from \citet{Lucas2020} who demonstrate that a supernova has only a moderate effect on dense gas, while the influence would be increased if other forms of feedback\footnote{Radiative feedback (photoionization, radiation pressure) or mechanical feedback (winds, outflows, or earlier supernovae).} acted prior to an explosion. This combination of feedback mechanisms, which originated in one or several more evolved stellar groups in Orion, is likely the best explanation for the observed scenario.

Considering the other forms of massive stellar feedback, it was shown that ionizing radiation in an expanding H\,II region could produce a total momentum injection of up to $10^5$\,M$_\odot$\,km\,s$^{-1}$ to the surrounding cold neutral medium (CNM, $n_0 \sim 100$\,cm$^{-3}$) after about 1.5 to several Myr, depending on the stellar mass and the environmental conditions \citep{Haid2018}. This is similar to the momentum output of one supernova, while different timescales and size-scales apply (the energy of a supernova can fill larger volumes on shorter timescales). The contribution of stellar winds is highly debated and is generally assumed to be less effective compared to ionization \citep[e.g.,][]{Dale2013}, especially in denser environments \citep[e.g.,][]{Haid2018}. Several authors \citep[e.g.,][]{Harper-Clark2009, RogersPittard2013, Dale2013, Walch2013, Rosen2014} have shown that the energy and winds tend to escape into rarefied regions.
Recent observational work has shown the disruption of OMC-1 by an expanding shell, suggested to be powered by the winds of \object{tet1 Ori C} 
\citep{Pabst2019}. This indicates that the mechanical impact of winds could still be critical in the context of our observations.
In general, and before a possible supernova explosion, the massive stellar feedback energy can clear its local environment \citep{RogersPittard2013, Dale2014, Dale2017, Haid2018}, which makes a subsequent supernova more effective \citep{Walch2015, Lucas2020}.

Finally, we would like to consider the influence of gravity acting between or within the GMCs. We find that gravitational attraction between Orion\,A and B would generate velocities of less than 1\,km/s after about 6\,Myr. The observed relative velocities (as determined from three different rest velocities), which point radially away from a rough common center, are on the order of about 4 to 10\,km/s. This indicates that gravitational interactions are not dominating the observed radial motions. Self-gravity within a single GMC could cause additional velocity changes that are difficult to quantify, while they are likely not dominant either. 
Further simulations are needed to test the triggered star formation scenario in the context of Orion, as presented in this paper, to quantify the various forces acting in an Orion-like star-forming complex. The complexity of such simulations go beyond the scope of this work, where we focus on the observational signatures. In a future paper (G.\,Herbst-Kiss et al.,~in prep.) we present numerical simulations using smoothed particle hydrodynamics (SPH) to investigate the role of feedback on Orion\,A-like GMCs, and finally in the whole Orion region.

Concluding, the momentum analysis suggests that the energy of a few to several supernovae are needed to produce the energy required to explain the gas kinematics, while a combination of various feedback mechanisms seems more likely. This agrees with the scenario proposed in this paper. Despite the necessary assumptions made for the momentum analysis (see also Sect.~\ref{momentum}), we find good agreement between our measured values and simulations. The presented results offer a first, but unique and promising look into future possibilities, in combining 3D positional and kinematic data of stars and gas.

\subsection{Implications for the cloud-cloud collision scenario for the Orion complex}

Several recent papers have argued for a cloud-cloud collision scenario for the formation of the ONC \citep{Fukui2018, Lim2020}, L1641-N \citep{Nakamura2012}, NGC\,2023 \citep{Yamada2020}, NGC\,2024 \citep{Enokiya2020}, and NGC\,2068/2071 \citep{Fujita2020} in the Orion complex. The cloud-cloud collision argument is made based on the analysis of the radial velocity of the CO gas, in particular a jump in the velocities at particular star formation rich regions, such as the head of the Orion\,A cloud. We note that the scenario we propose here naturally explains the CO observables in Orion\,A without the need for a second molecular cloud being involved. The mechanical feedback-driven scenario presented here is still technically a collision, likely a shock, between an existing molecular cloud and a feedback flow of atomic and ionized hydrogen. We can not rule out the cloud-cloud collision with the present data, but argue, given the dynamical status of the clouds studied here, that even for a collision of two molecular clouds, the driver behind this collision is likely the stellar feedback forces.

\section{Summary} \label{Summary}

We were able to measure the 3D space motions of molecular clouds in the Orion star-forming region for the first time, using the 3D space motion of the YSOs as a proxy for the motion of the gas. The main results of this work are as follows:

\begin{enumerate}

\item We confirm that radial velocities of YSOs and that of the hosting molecular clouds are essentially the same, as recently found in the literature. Therefore, the average proper motions of YSOs can be used to reasonably estimate the 3D motions of their parental clouds. 

\item We report the discovery of coherent radial cloud motions on 100-pc scales in the Orion complex. The best explanation for the observables is the existence of a major feedback event in Orion about 6\,Myr ago. This feedback event that we name the Orion-BB event, shaped, in part, the distribution and kinematics of the gas we observe today and might also have triggered the formation of the young compact clusters NGC\,1977 and $\sigma$\,Orionis. Our results suggest that the dynamics of the gas and young stars in Orion carry the memory of its feedback-driven star formation history. 

\item We associate the origin of the Orion-BB event with the Orion\,X population, recently identified by \citet{Bouy2015} and \citet{Chen2020}. We also argue that Orion-BB is unlikely to be the only major feedback event in the region and that feedback processes over the last 10 Myr (radiation pressure, photoionization, mass loss, stellar winds from massive stars, supernova explosions) originating from slightly more evolved stellar populations have been continuously shaping the gas distribution, gas dynamics, and the star formation rate in the Orion complex. 

\item From the momentum estimate we find that a few to several supernovae are needed to explain the observed 3D motions, assuming that there were no other forces involved. Additional feedback from massive stars, such as photoionization and winds, could also contribute to the final momentum and would lower the number of supernovae needed to explain our observations. Realistically, a combination of the different feedback processes must have contributed to the observed shapes and kinematics of the molecular clouds in Orion.

\item We argue, based on kinematics, that the majority of the young stars in Orion are a product of large-scale feedback-driven triggering, which can raise the star formation rate in a cloud by about an order of magnitude, as for the head of Orion\,A (the Integral Shape Filament). Our results imply that at the genesis of the Orion Nebula Cluster (and NGC\,2023/2024 in Orion\,B) lies a feedback, compression, and triggering process, broadly confirming the classical feedback-driven scenario proposed in \citet{Elmegreen1977}.

\item We were able to estimate for the first time the 3D dynamics of star-forming molecular clouds on the scale of an entire cloud complex. Such an analysis is a crucial missing piece to understand the clouds' formation and dissolution mechanisms, their dynamics, and their mass distribution. A similar analysis will be available for most nearby cloud complexes with existing and upcoming \textit{Gaia} data, combined with existing and upcoming radial velocity and proper motion surveys. 

\end{enumerate}

The new dynamical view of Orion presented in this paper is yet another example of how \textit{Gaia} is opening a 3D window not only on the topology \citepalias{Grossschedl2018} but also the dynamics of the dense star-forming ISM, a critical missing ingredient in our understanding of star formation. 
In the future, the superior \textit{Gaia} DR3 data coming in 2022 and the proper motions of embedded sources arising from the ESO VISIONS Public Survey\footnote{\url{http://visions.univie.ac.at}}, will make the Orion complex a benchmark region to quantify the impact of feedback by massive stars, a fundamental but poorly constrained physical process.

\begin{acknowledgements}
We thank the referee, John Bally, for the wise and insightful comments that improved immensely our manuscript.
We thank Marina Kounkel for a helpful discussion on cloud evolution scenarios in Orion.
We thank Stefanie Walch and Seamus Clarke for their very helpful input on massive stellar feedback.
Finally, we thank Bruce Elmegreen, Alyssa Goodman, \'Alvaro Hacar, and Catherine Zucker for fruitful comments and discussions. 
J.\,Gro\ss schedl acknowledges funding by the Austrian Science Fund (FWF) under project number P 26718-N27.
J.\,Gro\ss schedl acknowledges funding by the Austrian Research Promotion Agency (FFG) under project number 873708.
This work has made use of data from the European Space Agency (ESA) mission {\it Gaia} (\url{https://www.cosmos.esa.int/gaia}), processed by the {\it Gaia} Data Processing and Analysis Consortium (DPAC, \url{https://www.cosmos.esa.int/web/gaia/dpac/consortium}). Funding for the DPAC has been provided by national institutions, in particular the institutions participating in the {\it Gaia} Multilateral Agreement. 
This research has made use of Planck (\url{http://www.esa.int/Planck}), an ESA science mission with instruments and contributions directly funded by ESA Member States, NASA, and Canada.
This research has made use of Python, \url{https://www.python.org}, of \textit{Astropy}, a community-developed core Python package for Astronomy \citep{Astropy2013, Astropy2018}, NumPy \citep{Walt2011}, Matplotlib \citep{Hunter2007}, Galpy \citep{Bovy2015}, and Plotly \citep{plotly2015}. 
This research has made use of the SIMBAD database operated at CDS, Strasbourg, France \citep{Wenger2000}, of the VizieR catalog access tool, CDS, Strasbourg, France \citep{Ochsenbein2000}, and of ``Aladin sky atlas'' developed at CDS, Strasbourg Observatory, France \citep{Bonnarel2000, Boch2014}. This research has made use of TOPCAT, an interactive graphical viewer and editor for tabular data \citep{Taylor2005}.
\end{acknowledgements}

\begin{flushleft}
\bibliographystyle{aa}
\bibliography{biblio} 
\end{flushleft}


\begin{appendix}

\section{Detailed description of YSO sample selection} \label{apx:sample-selection}

In this appendix we give a detailed description of YSO sample selection. First, we define the quality criteria for \textit{Gaia} DR2 and \mbox{APOGEE-2} data, next we explain the YSO selection criteria when using WISE and 2MASS photometry, and then we present the final sample selections for the subregions in Orion\,B and the outlying clouds based on position and motion criteria. Orion\,A is not discussed here separately, because the necessary steps were already explained in the main part of this paper. 

\subsection{Gaia quality criteria} \label{apx:gaia}

We apply the following quality criteria to all our YSO samples, while for individual subregions the criteria are further fine-tuned (see Sects.~\ref{apx:regions}):
\begin{equation} \label{equ:gaia}
\begin{aligned}
  &  200\,\mathrm{pc} < d < 700\,\mathrm{pc}, \\
  & |\mu_{\alpha*}| < \SI{10}{mas/yr}, \\
  & |\mu_{\delta}| < \SI{10}{mas/yr}, \\
  & \mathrm{err}\_\mu_{\alpha*}, \, \mathrm{err}\_\mu_{\delta} < \SI{1}{mas/yr}, \\
  & \mathrm{err}\_v_{\alpha}, \, \mathrm{err}\_v_{\delta} < \SI{2}{km/s}, \\
  & \mathrm{err}_\varpi / \varpi \leq 0.1, \\
  & \mathrm{ruwe} < 1.4, \\
  & G_\mathrm{err} < \SI{0.05}{mag}, \\
  & \mathtt{visibility\_periods\_used} > 6. 
\end{aligned}
\end{equation}
These conditions select sources within a distance interval enclosing the Orion star forming complex. Moreover, we preselect sources within a proper motion range of $\pm$10\,mas/yr. Sources showing larger proper motions either do not belong to the Orion population or have peculiar motions with respect to the average motion and are therefore not needed for our analysis. The latter conditions are quality criteria to reduce contamination by inferior data. For details on the \textit{Gaia} DR2 parameters and optimal usage of the data see \citet{Arenou2018}, \citet{Lindegren2018}, \url{https://gea.esac.esa.int/archive/documentation/index.html}, and \url{https://www.cosmos.esa.int/web/gaia/dr2-known-issues}. We also include a cut on the tangential velocity errors ($\mathrm{err}\_v_{\alpha}, \, \mathrm{err}\_v_{\delta}$). Tangential velocities and their errors were determined as follows:
\begin{equation} \label{equ:tangential}
\begin{aligned}
 &  v_x  = 4.74047 \cdot \mu_x\,\mathrm{(mas/yr)} / \varpi\,\mathrm{(mas)},  \\
 &  \mathrm{err}\_v_x  = 4.74047 \cdot (\mathrm{err}\_\mu_x^2/\varpi^2 + \mu_x^2 \mathrm{err}_\varpi^2 / \varpi^4)^{0.5}.
\end{aligned}
\end{equation}
We allowed errors as high as $v_x < 2$\,km/s, which is a factor of four higher as what is allowed for \mbox{APOGEE-2} radial velocities (see below). As a consequence, the velocity uncertainties in tangential direction are overall larger compared to radial velocities.

When using \textit{Gaia} DR2 photometry (as in Fig.~\ref{fig:hrd-orionx}) we first correct for systematic errors, as proposed in \citet{Evans2018} and \citet{MaizApellaniz2018}. For clarity we list the corrections here, as also shown in \citet{Zari2019}. The corrected photometry is written with a single prime symbol for $G'$ and $G'_\mathrm{RP}$\footnote{We do not use $G'_\mathrm{BP}$ due to the resulting larger scatter in the CMD for low-mass stars when using $G'_\mathrm{BP} - G'_\mathrm{RP}$.}. \\

\noindent If $(2 \leq G \leq 6)$\,mag:\\
$G' = -0.047344 + 1.16405 G - 0.046799 G^2 + 0.0035015 G^3$\\

\noindent If $(6 < G \leq 16)$\,mag: \\
$G' = G - 0.0032(G - 6)$\\

\noindent If $G > 16$\,mag: \\
$G' = G - 0.032$\\

\noindent If $(2 \leq G \leq 3.5)$\,mag:\\
$G'_\mathrm{RP} = G_\mathrm{RP} - 13.946 + 14.239 G_\mathrm{RP} - 4.23 G_\mathrm{RP}^2 + 0.4532 G_\mathrm{RP}^3$

\subsection{APOGEE-2 quality criteria} \label{apx:apogee-cut}

For regions that were covered by the \mbox{APOGEE-2} SDSS-DR16 survey we apply the following criteria:
\begin{equation} \label{equ:apogee}
\begin{aligned}
  & \SI{10}{km/s} < \mathtt{VHELIO\_AVG} < \SI{40}{km/s}, \\
  & \mathtt{VERR} < \SI{0.5}{km/s}, \\
  & \mathtt{VERR\_MED} < \SI{0.5}{km/s}, \\
  & \mathtt{VSCATTER} \leq \SI{1}{km/s}.
\end{aligned}
\end{equation}
To get \vhel of the YSOs we use the \mbox{APOGEE-2} parameter called \texttt{VHELIO\_AVG}, which is the signal-to-noise ratio (S/N) weighted average velocity as determined from the combined spectra.
The first cut includes a rough selection in \vhel space, to pre-exclude outliers with untypical Orion radial velocities.
To get reliable stellar radial velocities we use the S/N weighted uncertainty \texttt{VERR} and the median visit radial velocity uncertainty \texttt{VERR\_MED}, while they are known to be underestimated. For some cases \texttt{VSCATTER} might represent a better estimate of the overall measurement precision. If \texttt{VSCATTER} is much larger than \texttt{VERR\_MED} then this could be an indicator that the star is in a stellar binary. Therefore, we apply an additional cut using \texttt{VSCATTER}. The quality criteria are based on the \mbox{APOGEE-2} tutorial on how to use radial velocities; see \url{https://www.sdss.org/dr16/irspec/use-radial-velocities}.

\begin{figure*}[!ht]
    \centering
    \small
    \includegraphics[width=0.95\linewidth]{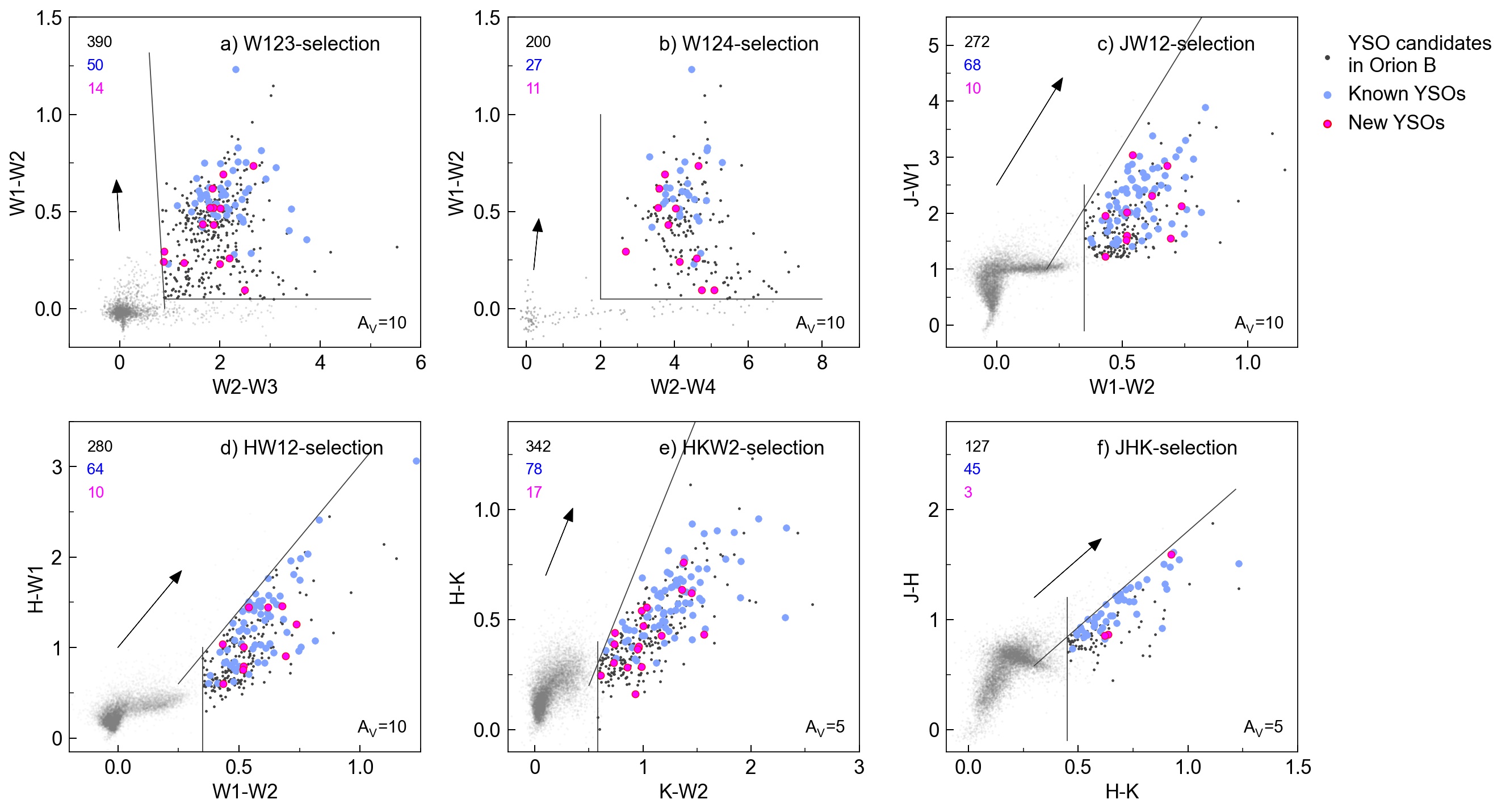}
    \caption
    {Six selected color-color diagrams showing the WISE-2MASS selection criteria to identify YSOs with infrared-excess. The gray dots in the background are all sources toward Orion\,B that pass the \textit{Gaia} quality criteria plus the additional individual WISE and/or 2MASS selection conditions, as given in Sect.~\ref{apx:yso-wise-selection} in the subpoints \textit{a} to \textit{f}. The black lines indicate the selection-borders and the black dots are selected YSO candidates toward the whole Orion\,B region as displayed in Fig.~\ref{fig:orionb_regions}. The blue dots are known YSO candidates \citep{Megeath2012,Megeath2016} and the magenta dots are the new YSO candidates that fall within the three subregions (see Fig.~\ref{fig:orionb_regions}). The numbers of selected YSO candidates are given in the respective color in the upper left corners (see also the legend on the right). The black arrows show the extinction vectors, with the individual length given in the lower right corners in A$_V$\,(mag).
    }
    \label{fig:ccd_orionb}
\end{figure*}

\begin{figure}[!ht]
    \centering
    \small
    \includegraphics[width=1\linewidth]{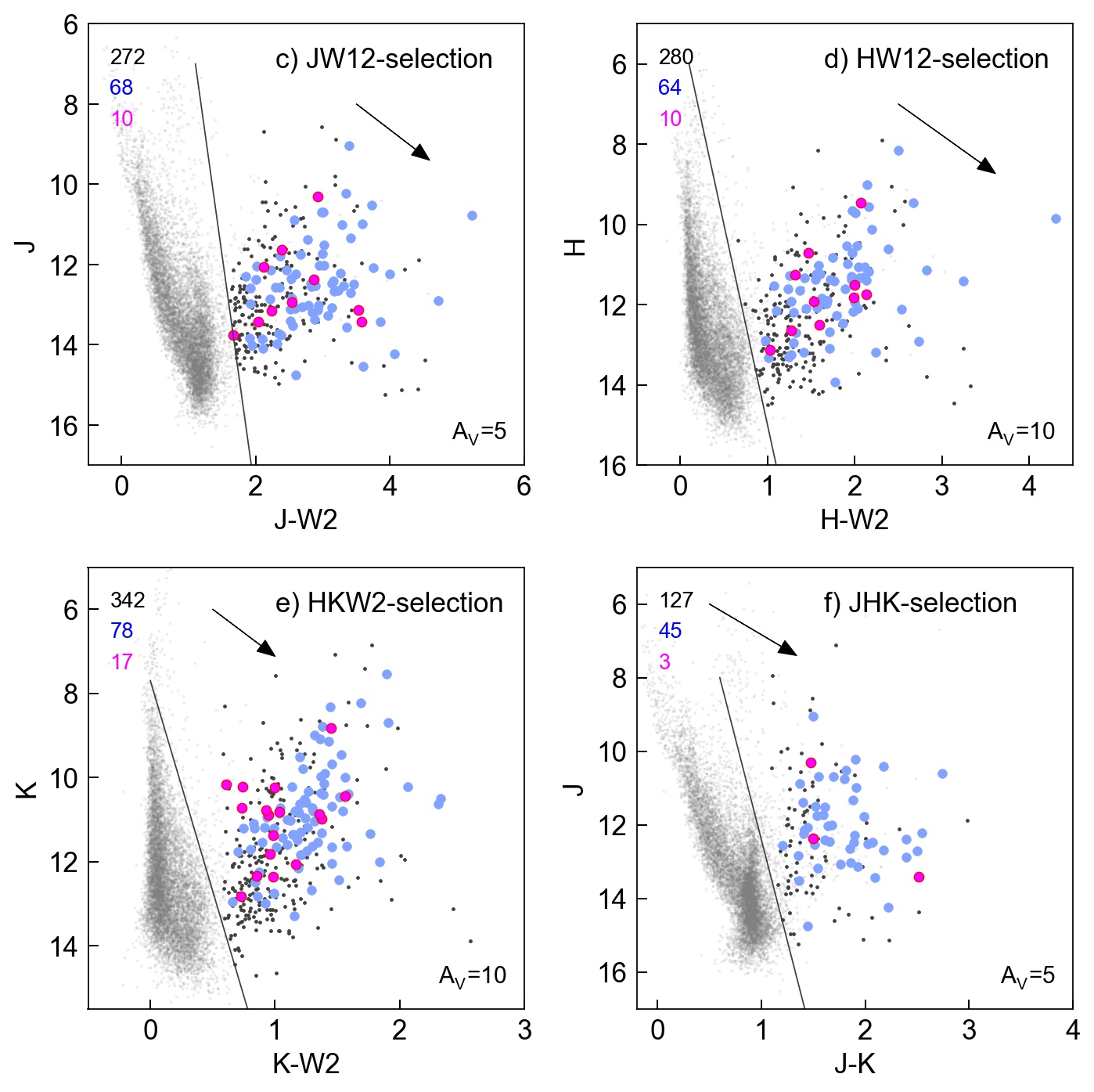}
    \caption
    {Color-magnitude diagrams, showing additional conditions for the selections \textit{c} to \textit{f}. See Fig.~\ref{fig:ccd_orionb} and text for more details. 
    }
    \label{fig:cmd_orionb}
\end{figure}

\begin{figure}[!ht]
    \centering
    \begin{minipage}[t]{1\linewidth}
        \centering
        \includegraphics[width=\linewidth]{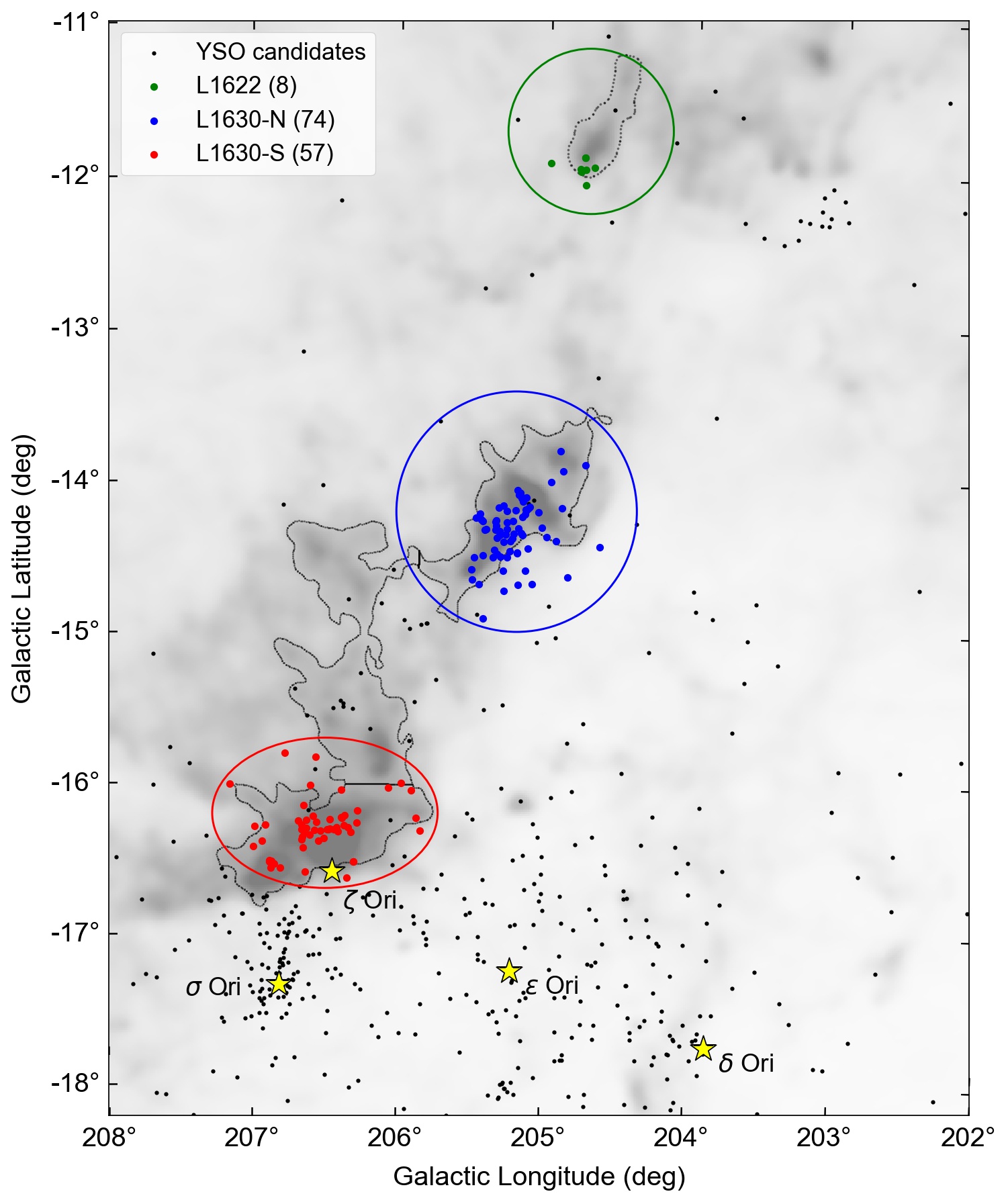}
    \end{minipage}%
    \vfill
    \begin{minipage}[t]{1\linewidth}
        \centering
        \includegraphics[width=\linewidth]{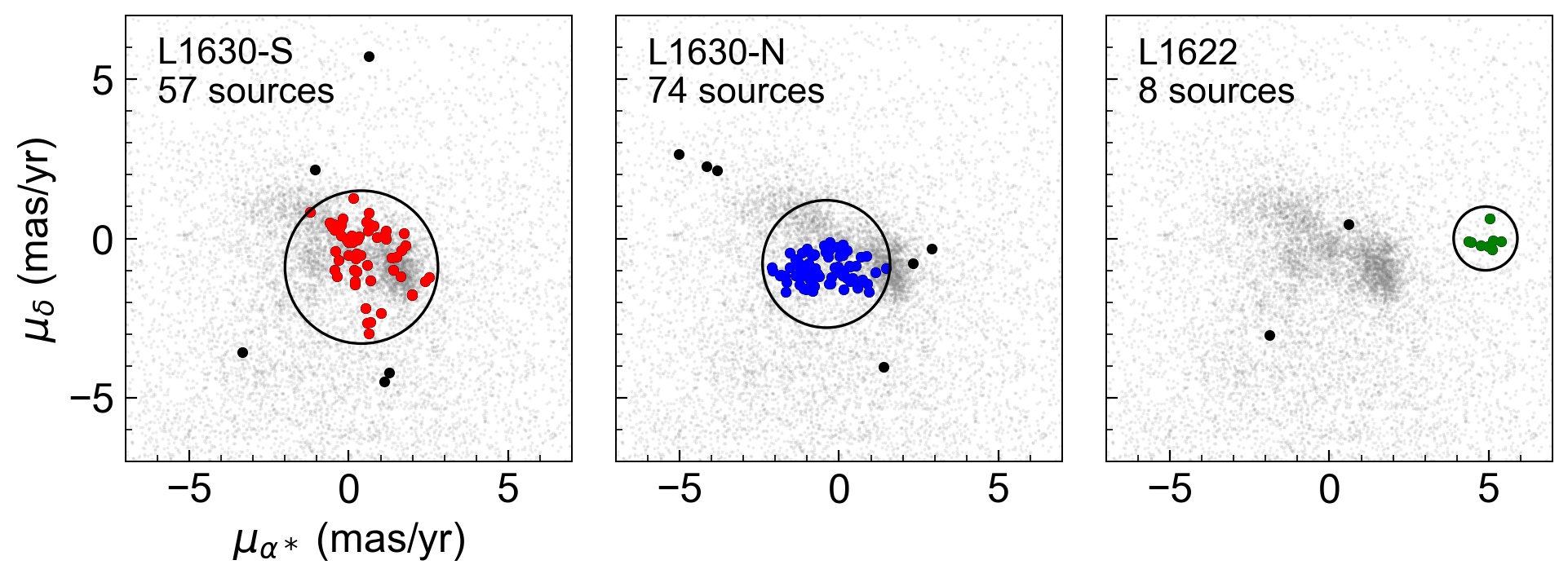}
    \end{minipage}%
    \caption{%
    \textit{Top:} Region selection for YSOs in the three Orion\,B subregions. The background is a Planck HFI image showing dust emission at 545\,GHz, while dark-gray shadings indicate regions of higher dust column-density \citep{Planck2014}. The subregions are highlighted by separate colors: L1630-S (red), L1630-N (blue), and L1622 (green). The black dots are all selected YSO candidates within the whole displayed region that pass the \textit{Gaia} quality criteria within 300\,pc$<d<$550\,pc. Only those YSOs within the circular selected regions are shown in color if they also pass the proper motion selection (bottom panels). For orientation the three belt stars and $\sigma$\,Ori are labeled and shown with yellow star symbols. The dotted lines are extinction contours (from a Herschel map) and outline the clouds of interest (see also Fig.~\ref{fig:pv-gas-ysos_orionb}).
    \textit{Bottom:} YSO proper motion selection for the three Orion\,B subregions. The colored dots indicate the proper motion selection and are the same as in the top panel. The black dots are all YSOs within the individual region selections (from top panel) but are excluded by the proper motion selection. The gray background are all sources toward Orion\,B that pass the quality criteria. See text for more details.%
    } 
    \label{fig:orionb_regions}
\end{figure}

\begin{figure}[!ht]
    \centering
    \begin{minipage}[t]{1\linewidth}
        \centering
        \includegraphics[width=\linewidth]{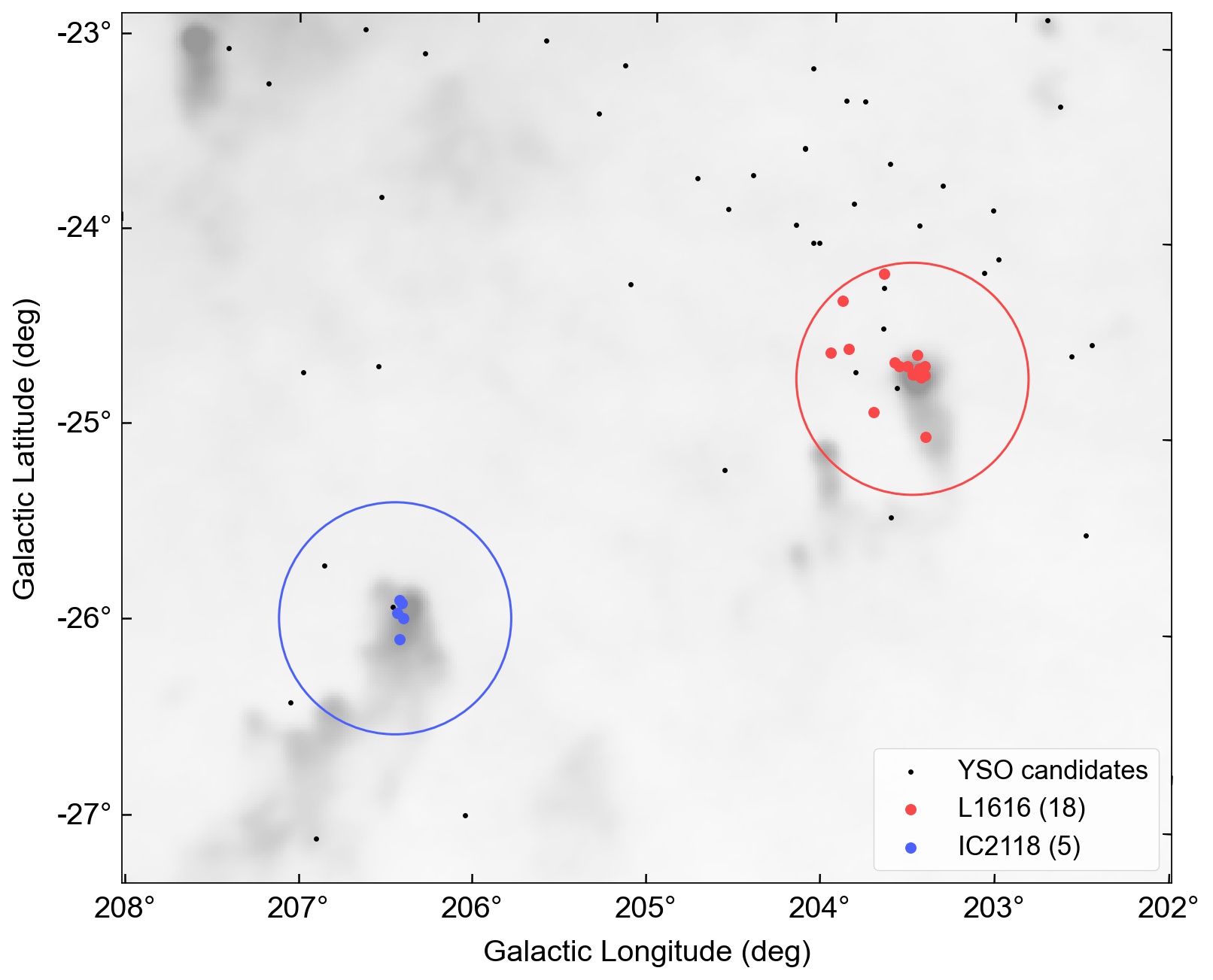}
    \end{minipage}%
    \vfill
    \begin{minipage}[t]{1\linewidth}
        \centering
        \includegraphics[width=0.95\linewidth]{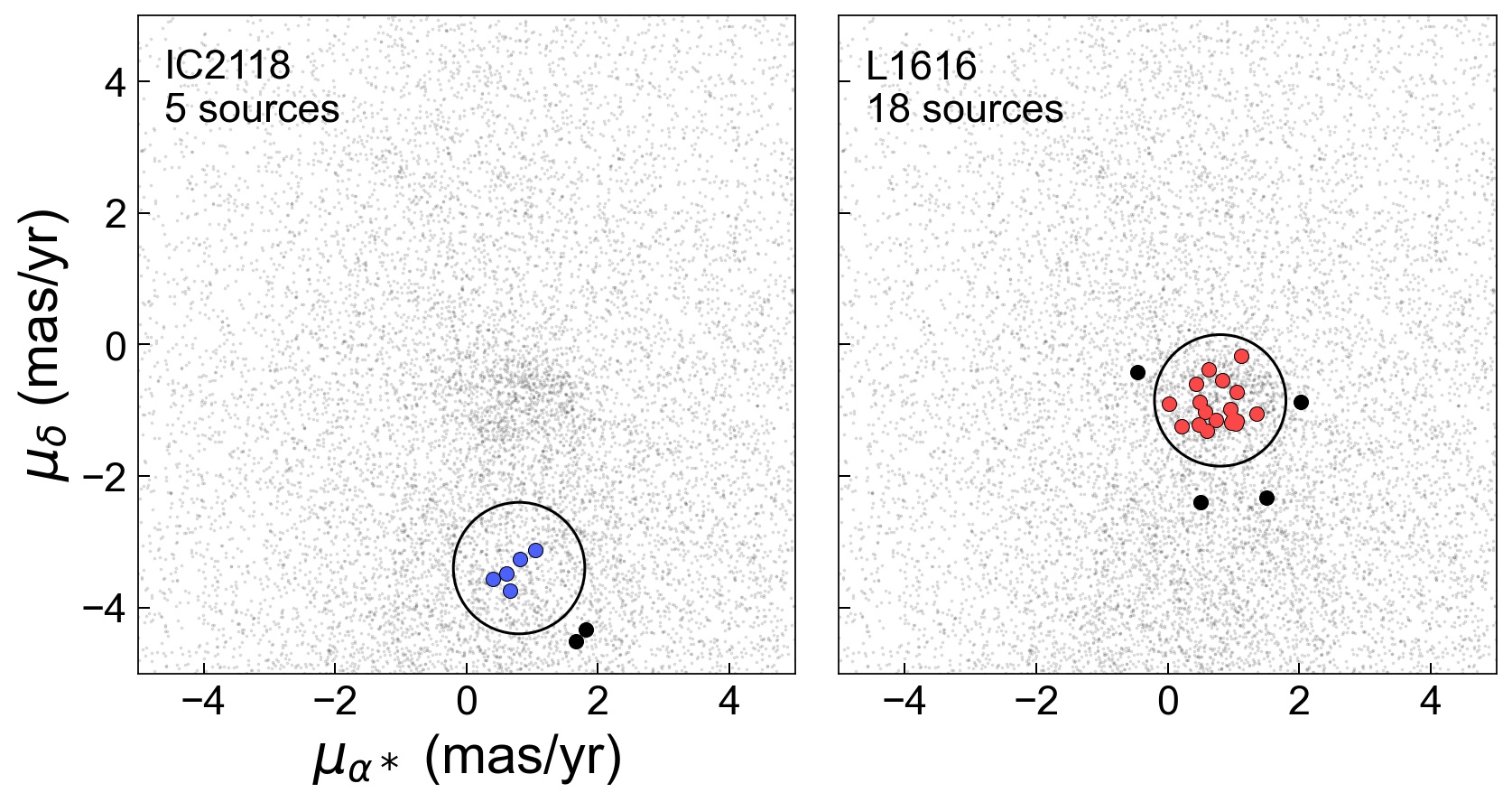}
    \end{minipage}%
    \caption{%
    \textit{Top:} Region selection for YSOs in the two outlying clouds. The subregion are highlighted separately: IC\,2118 (blue, Witch Head Nebula) and L1616 (red). 
    The black dots are all selected YSO candidates within the whole displayed region that pass the \textit{Gaia} quality criteria within 200\,pc$<d<$550\,pc.
    Only those YSOs within the circular selected regions are shown in color if they pass the proper motion selection. 
    \textit{Bottom:} YSO proper motion selection for the two outlying clouds. The colored dots indicate the proper motion selection and are the same as in the top panel. The black dots are all YSOs within the individual region selections (from top panel) but are excluded from the final samples by the proper motion cut. 
    The gray background are all sources toward the outlying clouds that pass the quality criteria. 
    Similar to Fig.~\ref{fig:orionb_regions}. See text for more details.%
    }
    \label{fig:outlying-regions}
\end{figure}

\subsection{Selecting additional YSOs with WISE and 2MASS photometry} \label{apx:yso-wise-selection}

We select additional YSO candidates, on top of the ones from the literature (see Sect.~\ref{Data}), to get more robust statistics. To this end, we apply selections in color-color and color-magnitude diagrams using photometry spanning from the near- to the mid-infrared, including data from the all-sky surveys 2MASS (near-infrared) and AllWISE (mid-infrared). These data are provided by the NASA/IPAC Infrared Science Archive \citep{Rebull2018}\footnote{\url{https://irsa.ipac.caltech.edu}}. 

Usually, an infrared based YSO selection tends to be contaminated by extra-galactic sources. Due to the requirement of using sources measured by \textit{Gaia} we are able to preselect in distance (see Equ.~\ref{equ:gaia}). This reduces fore- and background contamination substantially (e.g., AGB stars, galaxies, AGNs). Such a \textit{Gaia} preselection might not give the most complete sample of YSOs, since the optical \textit{Gaia} mission misses the youngest embedded sources, but a complete sample is not necessary for our purposes, while we strive to increase the numbers for good statistics. 

Before applying selection criteria in color and magnitude space we apply several quality criteria to the near- and mid-infrared photometry from 2MASS and WISE, sometimes used in combination. 
The WISE photometry (W1, W2, W3, W4 at $\SI{3.4}{\micro\metre}$, $\SI{4.5}{\micro\metre}$, $\SI{12}{\micro\metre}$, $\SI{22}{\micro\metre}$, respectively) is prone to be contaminated by extended emission close to star-forming regions (nebulosities, outflows), especially the two longer wavelength bands W3 and W4. To mitigate this, we include quality criteria that consider extended sources:
\begin{equation} \label{equ:ir_cut}
\begin{aligned}
  &  \mathrm{w1snr}, \, \mathrm{w2snr} > 10, \, \mathrm{w3snr}, \, \mathrm{w4snr} > 7,  \\
  &  \mathrm{w\#\_sigmpro} < 0.2,  \,\,  \mathrm{w\#rchi2} < 20,  \\
  &  \mathrm{w\#nm/w\#m} > 0.1, \\
  &  0 < \mathrm{w\#mag\_1-w\#mag\_6} < \SI{2}{mag},  \\
  &  \mathrm{w\#cc\_map\_str} \neq \mathrm{D}, \, \mathrm{H}, \, \mathrm{O}, \, \mathrm{P},  \\
  &  \mathrm{jsig}, \, \mathrm{hsig}, \, \mathrm{ksig} < \SI{0.1}{mag}.  \\
\end{aligned}
\end{equation}
The symbol ``\#'' is a placeholder for the four WISE bands (1, 2, 3, or 4) if the condition is equal for all\footnote{For more details on the AllWISE parameters see \url{http://wise2.ipac.caltech.edu/docs/release/allwise/expsup/sec2_1a.html} and \citet{Grossschedl2019A}, for the 2MASS parameters see \url{https://old.ipac.caltech.edu/2mass/releases/allsky/doc/sec2_2a.html}}.

We apply selection criteria within six different color spaces to select sources with infrared excess. The above mentioned quality criteria (Equ.~\ref{equ:ir_cut}) are applied only to those bands used in the individual selection, while all include the \textit{Gaia} criteria from Equ.~\ref{equ:gaia}. In some cases we apply a cut parallel to the extinction vector to exclude sources that are solely reddened due to foreground extinction and are consequently shifted above the main-sequence. For the used reddening law see \citet{Meingast2018} and \citet{Grossschedl2019A}. The selection criteria are as follows:

\begin{enumerate}[a)]

    \item W123-selection: WISE selection including the bands W1, W2, and W3. The value $-4.273$ represents the slope of the extinction vector in the W123 color space (see Fig.~\ref{fig:ccd_orionb}.a).
        \begin{equation}
        \begin{aligned}
            & \mathrm{W1}-\mathrm{W2} > \SI{0.05}{mag},  \\
            & \mathrm{W1}-\mathrm{W2} > - 4.273 \times (\mathrm{W2}-\mathrm{W3} - 0.7).
        \end{aligned}
        \end{equation}

    \item W124-selection: WISE selection including the bands W1, W2, and W4 (see Fig.~\ref{fig:ccd_orionb}.b).
        \begin{equation}
        \begin{aligned}
            & \mathrm{W1}-\mathrm{W2} > 0.05, \,\, \mathrm{W2}-\mathrm{W4} > 2.
        \end{aligned}
        \end{equation}
        
    \item JW12-selection: Combined WISE and 2MASS selection including the bands J, W1, and W2. The value $7.2894$ represents the slope of the extinction vector in the JW12 color space (see Fig.~\ref{fig:ccd_orionb}.c). The latter condition additionally excludes main-sequence sources in the JW2 color-magnitude diagram (Fig.~\ref{fig:cmd_orionb}.c).
        \begin{equation}
        \begin{aligned}
            & \mathrm{W1}-\mathrm{W2} > 0.35,  \\
            & \mathrm{J}-\mathrm{W1} < 7.2894 \times (\mathrm{W1}-\mathrm{W2} - 0.2) + 1, \\
            & \mathrm{J} < 12 \times (\mathrm{J}-\mathrm{W2} - 1.1) + 7.
        \end{aligned}
        \end{equation}
        
    \item HW12-selection: Combined WISE and 2MASS selection including the bands H, W1, and W2. The value $3.2298$ represents the slope of the extinction vector in the HW12 color space (see Fig.~\ref{fig:ccd_orionb}.d). The latter condition additionally excludes main-sequence sources in the HW2 color-magnitude diagram (Fig.~\ref{fig:cmd_orionb}.d).
        \begin{equation}
        \begin{aligned}
            & \mathrm{W1}-\mathrm{W2} > 0.35,  \\
            & \mathrm{H}-\mathrm{W1} < 3.2298 \times (\mathrm{W1}-\mathrm{W2} - 0.25) + 0.6, \\
            & \mathrm{H} < 10 \times (\mathrm{H}-\mathrm{W2} - 0.1) + 6.
        \end{aligned}
        \end{equation}
        
    \item HKW2-selection: Combined WISE and 2MASS selection including the bands H, K, and W2. The value $1.2188$ represents the slope of the extinction vector in the HKW2 color space (see Fig.~\ref{fig:ccd_orionb}.e). The latter condition additionally excludes main-sequence sources in the KW2 color-magnitude diagram (Fig.~\ref{fig:cmd_orionb}.e).
        \begin{equation}
        \begin{aligned}
            & \mathrm{K}-\mathrm{W2} > 0.58,  \\
            & \mathrm{H}-\mathrm{K} < 1.2188 \times (\mathrm{K}-\mathrm{W2} - 0.5) + 0.2, \\
            & \mathrm{K} < 10 \times (\mathrm{K}-\mathrm{W2}) + 7.7.
        \end{aligned}
        \end{equation}
        
    \item JHK-selection: 2MASS selection including the bands J, H, and K. The value 1.7473 represents the slope of the extinction vector in the JHK color space (see Fig.~\ref{fig:ccd_orionb}.f). The latter condition additionally excludes main-sequence sources in the JK color-magnitude diagram (Fig.~\ref{fig:cmd_orionb}.f).
        \begin{equation}
        \begin{aligned}
            & \mathrm{H}-\mathrm{K} > 0.45,   \\
            & \mathrm{J}-\mathrm{H} < 1.7473 \times (\mathrm{H}-\mathrm{K} - 0.3) + 0.58, \\
            & \mathrm{J} < 11 \times (\mathrm{J}-\mathrm{K} - 0.6) + 8.
        \end{aligned}
        \end{equation}

\end{enumerate}
These selections are representatively demonstrated for the Orion\,B region in Figs.~\ref{fig:ccd_orionb} and \ref{fig:cmd_orionb}. 
The whole investigated region in these figures extends beyond the Orion\,B clouds and includes $\sigma$\,Ori and parts of OBP, as can be seen in Fig.~\ref{fig:orionb_regions}, where the black dots are all selected YSO candidates.
The small cluster of YSOs in the top-right corner of Fig.~\ref{fig:orionb_regions}  ($l,b \sim$ 203.2,-12.2) is not obviously related to any gas, while it lies very close to the cometary clouds L1617.
The same procedure was applied for the region surrounding the outlying clouds (see Fig.~\ref{fig:outlying-regions}), while the YSO selection criteria of the sources in the outlying clouds in WISE-2MASS color-color or color-magnitude diagrams is not shown explicitly here. 

For the three Orion\,B subregions, the combined WISE-2MASS YSO selections deliver 136 YSO candidates within the three regions of interest (see Sect.~\ref{apx:regions}), which include 106 sources (78\%) that were previously identified with Spitzer \citep{Megeath2012, Megeath2016}. So we were able to add 30 additional YSO candidates for the  three Orion\,B regions. 

For the outlying clouds we select in total 19 YSOs, from which 14 are located in L1616, and five are in IC\,2118. Of the 14 sources, 11 were already known pre-main-sequence stars as reported in \citet{Alcala2004}, thus, we were able to add three additional YSOs for L1616. The five sources in IC\,2118 are already known YSOs listed in \citet{Guieu2010}. Any additional new WISE-2MASS YSOs in these subregions did not make it into the final samples because of the quality criteria or the final individual selection criteria for each region (see below, Sect.~\ref{apx:regions}).

\begin{table}[!t] 
\begin{center}
\begin{small}
\caption{Parameters for selections in position ($l$, $b$) and proper motion ($\mu_{\alpha}*$, $\mu_\delta$) space for Orion\,B and the outlying clouds, including in the last two rows the young compact clusters from Appendix~\ref{apx:ycc}.}  
\renewcommand{\arraystretch}{1.2}
\begin{tabular}{lrrcrrc}
\hline \hline
\multicolumn{1}{c}{} &
\multicolumn{3}{c}{($x$, $y$) = ($l$, $b$)} &
\multicolumn{3}{c}{($x$, $y$) = ($\mu_{\alpha*}$, $\mu_{\delta}$)} \\
\multicolumn{1}{c}{} &
\multicolumn{3}{c}{(\si{\degree})} &
\multicolumn{3}{c}{(mas/yr)} \\
\cmidrule{2-4}
\cmidrule{5-7}
\multicolumn{1}{c}{subregion} &
\multicolumn{1}{c}{$x_0$} &
\multicolumn{1}{c}{$y_0$} &
\multicolumn{1}{c}{$r$} &
\multicolumn{1}{c}{$x_0$} &
\multicolumn{1}{c}{$y_0$} &
\multicolumn{1}{c}{$r$} \\
\hline
L1630-S  & 206.50 & -16.20 & 0.75, 0.5\tablefootmark{a}  & 0.40 & -0.90 & 2.4  \\
L1630-N  & 205.20 & -14.20 & 0.80 &    -0.40 & -0.80 & 2.0  \\
L1622    & 204.73 & -11.70 & 0.55 &     4.90 &  0.00 & 1.0  \\

L1616    & 203.60 & -24.60 & 0.50 &   0.80 & -0.85 & 1.0  \\
IC\,2118   & 206.40 & -26.00 & 0.50 &   0.80 & -3.40  & 1.0  \\

NGC\,1977         & 208.50 & -19.10 & 0.22 &   1.20 & -0.70 & 1.3  \\
$\sigma$\,Ori   & 206.81 & -17.33 & 0.45 &   1.55 & -0.65  & 1.6  \\

\hline
\end{tabular}
\renewcommand{\arraystretch}{1}
\label{tab:region-selection}
\tablefoot{For circular selections applies $(x-x_0)^2 + (y-y_0)^2 < r^2$, with $x_0$, $y_0$ being the center positions of the circle and $r$ the radius. 
\tablefoottext{a}{For L1630-S we select sources within an elliptical region,  $(x-x_0)^2/a_x^2 + (y-y_0)^2/b_y^2 < 1$. The numbers given in $r$ are here the semi-major and semi-minor axis in $l$ and $b$, respectively.}
}
\end{small}
\end{center}
\end{table}

\subsection{Description of detailed sample selection for the subregions in Orion B and the Outlying Clouds} \label{apx:regions}

For Orion\,A we use the whole region surrounding the GMC as already demonstrated in \citet{Grossschedl2018} and as described in the main part of this paper. For the subregions in Orion\,B and the outlying clouds we apply selections as follows. First we select sources within circular (or elliptical) regions, which enclose the cloud parts of interest.
Such selections are applied in position ($l$, $b$) and proper motion space ($\mu_{\alpha*}$, $\mu_{\delta}$)\footnote{We use proper motions, not tangential velocities, since the sources are selected to be close in space, which makes selections in either space virtually identical.}, with selection conditions given in Table~\ref{tab:region-selection}. 
The detailed selections are further demonstrated in Figs.~\ref{fig:orionb_regions} and \ref{fig:outlying-regions}. 

We defined these selections by individually investigating each region to select YSOs that are close to the molecular clouds in projection and that show an over-density in proper motion space. For three regions, which are the cometary clouds (L1622, L1616, IC\,2118), there exists a pronounced peak in proper motion space, with a small scatter of about $\SI{0.3}{mas/yr}$ (see Figs.~\ref{fig:orionb_regions} and \ref{fig:outlying-regions}, bottom panels).
For the two regions L1630-S (NGC\,2023/2024) and L1630-N (NGC\,2068/2071) there is no such typical proper motion peak and the scatter tends to be larger, up to $\SI{1}{mas/yr}$. This is also because we allowed a larger radius in proper motion space, due to the lack of a clear peak.
For NGC\,2068/2071 (L1630-N) we find an elongated structure in proper motion space, maybe a signature of two proper motions peaks. When splitting the elongated proper motion structure in the middle, the two resulting samples roughly separate in north-south direction. This could be a signature of the two involved clusters. Nevertheless, we kept the two structures combined and used the average motions of all YSOs, since we are interested in the bulk motion of this cloud region and not the internal dynamics.
The YSOs in NGC\,2023/2024 (L1630-S) also do not show a clear peak in proper motion space. The seemingly chaotic motions could be a signature of perturbation. On the other hand, there are signs of three distinct groups within our selected L1630-S sample, while we do not attempt to separate these in this paper. As for L1630-N, we are interested in the bulk-motion of the cloud part, hence a separation seems not feasible at this stage due to low statistics when further substructuring. 

For Orion\,B regions the following additional cuts were applied. First, we applied a more stringent distance criterion after investigating the dominating distances for that regions: $\SI{300}{pc} < d < \SI{550}{pc}$.
Second, for sources observed by \mbox{APOGEE-2} (in L1630-S/N) we apply an additional cut in \vhel, to exclude obvious outliers: $\SI{21}{km/s} < \mathtt{VHELIO\_AVG} < \SI{35}{km/s}$.
For the outlying clouds we also applied additional distance cuts: for L1616, $\SI{350}{pc} < d < \SI{450}{pc}$; for IC\,2118, $\SI{200}{pc} < d < \SI{400}{pc}$.
The resulting sample sizes and average parameters are listed in Tables~\ref{tab:averages_rv_pm} and \ref{tab:averages_rv}.

The here presented selection approach is rather simple and a clustering algorithm might deliver a more robust cluster selection. However, for our purposes this basic approach is sufficient, since we are not interested in a complete cluster membership, but in the average bulk motions and positions of only the youngest stellar cloud members.

\section{The young compact clusters NGC1977 and $\sigma$Ori} \label{apx:ycc}

\begin{figure*}[!ht]
    \centering
    \begin{minipage}[t]{1\linewidth}
        \centering
        \includegraphics[width=\linewidth]{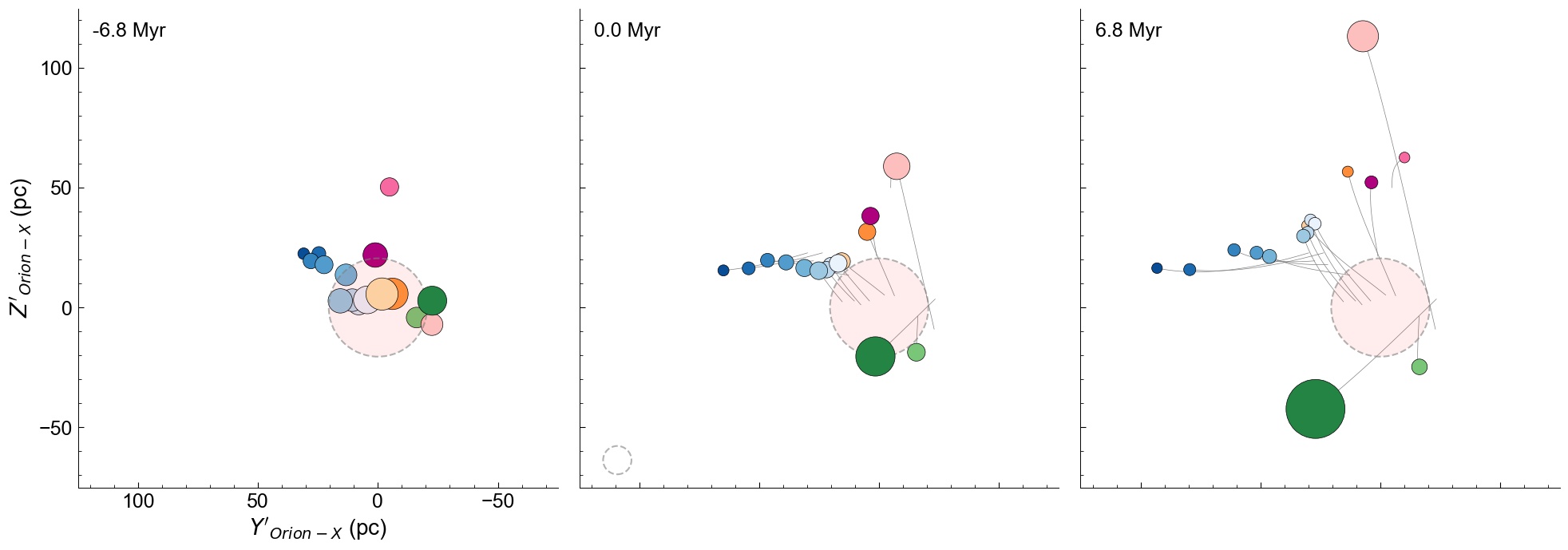}
    \end{minipage}%
    \hfill
    \begin{minipage}[t]{1\linewidth}
        \centering
        \includegraphics[width=\linewidth]{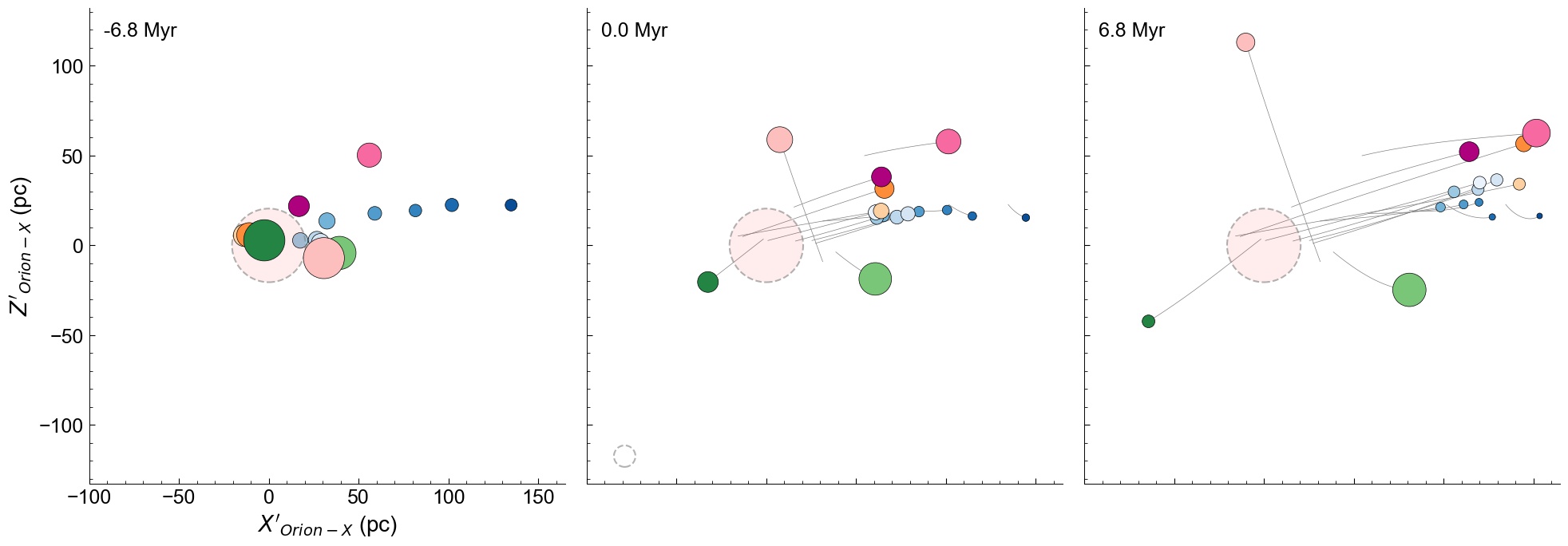}
    \end{minipage}%
    \hfill
    \begin{minipage}[t]{1\linewidth}
        \centering
        \includegraphics[width=\linewidth]{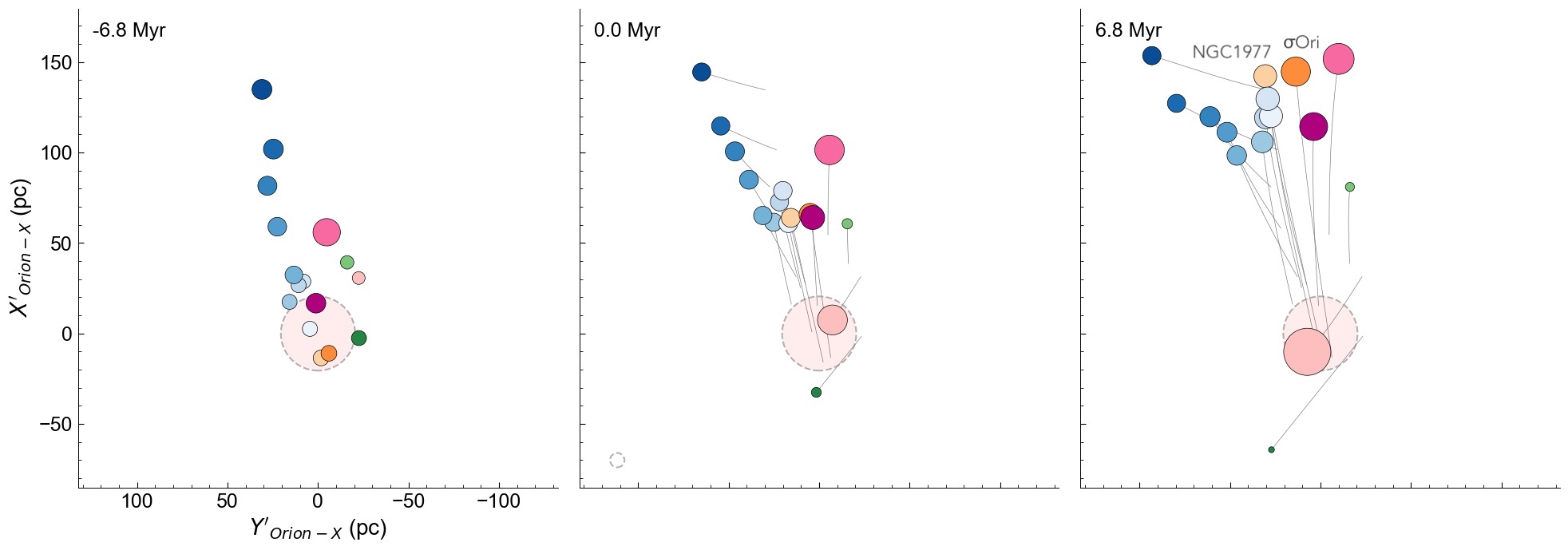}
    \end{minipage}%
    \caption{Three time snapshots showing the relative 3D space motions of the 14 cloud subregions, similar to Figs.~\ref{fig:YZ-Orion-OX} to \ref{fig:YX-Orion-OX}, now also including the two young compact clusters NGC\,1977 (light-orange) and $\sigma$\,Ori (dark-orange). The time snapshots are shown only for $-6.8$\,Myr, Today, and $+6.8$\,Myr. 
    \textit{Top:} Front view in the $Z'_\mathrm{Orion-X}$/$Y'_\mathrm{Orion-X}$ plane. A movie versions is available online at: 
    \url{https://homepage.univie.ac.at/grossschej23/figureB1-YZ-orion-timelapse-clusters.mp4}.
    \textit{Middle:} Side view in the $Z'_\mathrm{Orion-X}$/$X'_\mathrm{Orion-X}$ plane. A movie versions is available online at: 
    \url{https://homepage.univie.ac.at/grossschej23/figureB1-XZ-orion-timelapse-clusters.mp4}.
    \textit{Bottom:} Top-down view in the $X'_\mathrm{Orion-X}$/$Y'_\mathrm{Orion-X}$ plane. A movie versions is available online at: 
    \url{https://homepage.univie.ac.at/grossschej23/figureB1-YX-orion-timelapse-clusters.mp4}.
    See also the \href{https://homepage.univie.ac.at/josefa.elisabeth.grossschedl/orion-bb.html}{link} in Fig.~\ref{fig:orion-bb} for an interactive version.
    }
    \label{fig:ycc}
\end{figure*}

The well-known compact clusters NGC\,1977 and $\sigma$\,Ori can be put into the new context presented in this paper. As pointed out in Sect.~\ref{sec:groups}, they are likely connected to the Orion-BB event. In the main part of the paper we focused on the 3D space properties of molecular clouds, therefore we added these two clusters subsequently to our analysis. Both clusters still harbor a wealth of pre-main-sequence stars with circumstellar disks (Class\,II), hence are younger than about 4\,Myr (see Sect.~\ref{sec:groups}).
To understand their connection with the proposed Orion-BB event we calculated the tracebacks from their average 3D motions, as was done for the 14 subregions. The clusters were selected similarly to the methods described in Appendix~\ref{apx:regions} (see, e.g., Fig.~\ref{fig:orionb_regions}). We used only YSOs with infrared-excess (mainly Class\,II) to be consistent with the methods for the cloud subregions. To get YSOs in NGC\,1977 we used the \citet{Grossschedl2019A} catalog, and for $\sigma$\,Ori we mainly used the YSO sample from \citet{Hernandez2007} including some YSOs from the additional WISE-2MASS selection. Both clusters can be seen as over-densities in projection, while NGC\,1977 is visible in Fig.~\ref{fig:pv-gas-ysos} and $\sigma$\,Ori in Fig.~\ref{fig:orionb_regions}. Similar to the approach for Orion\,B and the Outlying Clouds (Sect.\,\ref{apx:regions}), we applied circular selections in position and proper motion space, with the conditions listed with the other subregions in Table~\ref{tab:region-selection}. After applying \textit{Gaia} quality criteria we ended up with 92 and 78 YSOs in NGC\,1977 and $\sigma$\,Ori, respectively. This includes an additional distance cut of $d<500$\,pc. Both regions were covered by \mbox{APOGEE-2}. After applying radial velocity quality criteria we applied additional radial velocity range cuts for NGC\,1977 ($\SI{22}{km/s} < \mathtt{VHELIO\_AVG} < \SI{37}{km/s}$) and $\sigma$\,Ori ($\SI{24}{km/s} < \mathtt{VHELIO\_AVG} < \SI{36}{km/s}$). This delivers 44 and 40 YSOs, respectively, and allows us to determine average cluster radial velocities. The results for the average parameters are given in Table\,\ref{tab:ycc}.

The average 6D parameters of the two clusters were determined the same way as for the subregions, while we used average stellar \vhel from APOGEE-2 as starting condition for \textit{Galpy}, since these clusters are not directly associated with any cloud structures. However, it was shown that \mbox{APOGEE-2} and \citetalias{Nishimura2015} $^{12}$CO radial velocities do not show a clear trend of being either blue- or red-shifted when compared to each other (see Figs.~\ref{fig:pv-gas-ysos}, \ref{fig:pv-gas-ysos_orionb}, \ref{fig:pv-sigma-oriona}, or \ref{fig:pv-sigma-orionb}), hence there is likely not a significant systematic shift between these observations.

The resulting 3D space motions, relative to the position of Orion\,X, are presented in Fig.~\ref{fig:ycc}, similar to Figs.~\ref{fig:YZ-Orion-OX} to \ref{fig:YX-Orion-OX}. NGC\,1977 is represented by the light-orange and $\sigma$\,Ori by the dark-orange filled circle. Both move with the adjacent cloud subregions (OMC-2/3 and L1630-S, currently at about the same distance), while they have a larger relative velocity, which is expected based on the red-shifted radial velocities compared to the adjacent cloud subregions (see Fig.~\ref{fig:PVD-OrionX}). The tracebacks reveal that the progenitor clouds, out of which the clusters formed (about 4\,Myr ago), were likely closer to the Orion-BB event. This fact might explain why they have a higher velocity compared to the Orion A and B clouds, as they would have been pushed and compressed more effectively. See the Discussion in Sect.~\ref{sec:groups}.  

\begin{table}[!t] 
\begin{center}
\begin{small}
\caption{Average properties of the two young compact clusters.}  
\renewcommand{\arraystretch}{1.2}
\begin{tabular}{lcc}
\hline \hline
\multicolumn{1}{l}{Parameter} &
\multicolumn{1}{c}{NGC\,1977} &
\multicolumn{1}{c}{$\sigma$\,Ori}  \\
\hline
$l,b$\,(deg) & $\SI{208.47}{\degree},\SI{-19.11}{\degree}$  & $\SI{206.80}{\degree},\SI{-17.33}{\degree}$  \\
$\varpi$\,(mas) & $2.567\pm0.123$ & $2.551\pm0.160$ \\
$d$\,(pc) & $390\pm19$ & $392\pm25$ \\
$\mu_{\alpha*}$\,(mas/yr) & $1.25\pm0.47$ & $1.55\pm0.51$ \\
$\mu_{\delta}$\,(mas/yr) & $-0.80\pm0.55$ & $-0.59\pm0.50$ \\
$v_{\alpha}$\,(km/s) & $2.30\pm0.86$ & $2.89\pm0.95$ \\
$v_{\delta}$\,(km/s) & $-1.47\pm1.02$ & $-1.09\pm0.90$ \\
$v_\mathrm{HEL}$\,(km/s) & $31.06\pm1.51$ & $31.09\pm1.41$ \\
\hline
\end{tabular}
\renewcommand{\arraystretch}{1}
\label{tab:ycc}
\end{small}
\end{center}
\end{table}

\section{Coordinate system definitions and LSR conversion} \label{apx:lsr}

\begin{table*}[!ht] 
\begin{center}
\small
\caption{Properties of the Sun relative to the Galactic center, as used by \texttt{Astropy\,4.0}.}
\renewcommand{\arraystretch}{1.2}
\begin{tabular}{lll}
\hline \hline
\multicolumn{1}{c}{Description} &
\multicolumn{1}{c}{Values} &
\multicolumn{1}{c}{Ref.} \\
\hline
Galactocentric Frame (ICRS) & ($\alpha_\mathrm{GC}$, $\delta_\mathrm{GC}$) = (266.4051, -28.936175)\,deg & 1 \\
Galactocentric distance of Sun & $d_{\sun}$ = 8.122\,kpc & 2 \\
Distance of Sun to Galactic mid-plane & $Z_{\sun}$ = 20.8\,pc & 3 \\
Solar velocity in Galactocentric cylindrical coordinates & ($v_{R,\sun}$, $v_{\phi,\sun}$, $v_{Z,\sun}$) = (-12.9, 245.6, 7.78)\,km/s & 4, 1, 2 \\
LSR motion in Galactocentric cylindrical coordinates & ($v_{R,\mathrm{LSR}}$, $v_{\phi,\mathrm{LSR}}$, $v_{Z,\mathrm{LSR}}$) = (1.8, 233.4, 0.53)\,km/s & 4 \\
Barycentric standard solar motion relative to LSR since 2010 & (U$_{\sun}$, V$_{\sun}$, W$_{\sun}$) = (11.1, 12.24, 7.25)\,km/s & 5 \\
Barycentric standard solar motion relative to LSR since 1986\tablefootmark{a} & (U$_{\sun}$, V$_{\sun}$, W$_{\sun}$) = (10.0, 15.4, 7.8)\,km/s & 6 \\
Barycentric standard solar motion relative to LSR since 1981\tablefootmark{b} & (U$_{\sun}$, V$_{\sun}$, W$_{\sun}$) = (9.2, 12.0, 6.9)\,km/s & 7 \\
\hline
\end{tabular}
\renewcommand{\arraystretch}{1}
\label{tab:astropy}
\tablefoot{
\tablefoottext{a}{The standard solar motion of 20\,km/s toward $l=\SI{56}{\degree}$, $b=\SI{23}{\degree}$ (RA $=18^h$, Dec $=\SI{30}{\degree}$, epoch 1900) was recommended by the IAU \citep[see, e.g.,][]{Kerr1986} and is likely often used by radio observatories to convert gas radial velocities, as derived from molecular line observations, from \vhel to \vlsr, and is given here for completeness (not used by \texttt{Astropy\,4.0}).}
\tablefoottext{b}{The standard solar motion of 16.6\,km/s is reported in \citet{Mihalas1981} and was likely used by \citet{Maddalena1986} and is given here for completeness (not used by \texttt{Astropy\,4.0}).}
}
\tablebib{
(1) \citet{Reid2004};
(2) \citet{Abuter2018};
(3) \citet{Bennett2019};
(4) \citet{Drimmel2018}; 
(5) \citet{Schoenrich2010};
(6) \citet{Kerr1986};
(7) \citet{Mihalas1981}.
}
\end{center}
\end{table*}

In Table~\ref{tab:astropy} we list the parameters that were used to determine the position and motion of the Sun within the Milky Way. Generally, we used the \texttt{Astropy\,4.0} standard as default values. These values are the basis to convert to Galactocentric Cartesian coordinates or to velocities relative to the local standard of rest (LSR).

In this section we also highlight problems that come with erroneously converted values, especially concerning gas radial velocities, which are given in \vlsr in the literature. When investigating gas kinematics as determined by emission line surveys, the gas radial velocity is given by all authors relative to the LSR. However, it is often not clear which definitions for the standard solar motion were used in individual studies (i.e., conversion from \vhel to \vlsr). This introduces an additional uncertainty and a direct comparison of independent observations can not be done at face value, which was already pointed out in \citet{Hacar2016b}. On top of that, when comparing stellar kinematics with gas kinematics, one has to convert the stellar \vhel to \vlsr or vice-versa for the gas. Using different LSR conversion methods for one of these data sets can lead to biased interpretations, and it could introduce an artificial radial velocity shift between two observations or even hide a shift that would be observable otherwise. 

In Table~\ref{tab:astropy} we list the Galactic Cartesian components of three different standard solar motions that are relevant for our work. One of the first widely used standard solar motions is given in \citet{Mihalas1981} with $\SI{16.6}{km/s}$. Since the mid 1980s the standard solar motion of 20\,km/s, reported in \citetalias{Kerr1986}, seems to be common. In the late 1990s \citet{Dehnen1998} defined a new value ($\SI{13}{km/s}$) with a significantly lower V$_\sun$ component. Today, the most widely used standard was defined by \citet{Schoenrich2010}, with a standard solar motion of $\SI{18}{km/s}$, which is used also in our work. Beside the values listed here, there are about 15 further published values, which are, for example, discussed in \citet{Francis2009}. This makes it clear how difficult it is to unambiguously interpret gas velocities. We find that if converting gas velocities \vlsr in Orion back to \vhel with different LSR definitions could lead to variations of up to 5\,km/s or even more in extreme cases. Thus, if the original conversion method is unknown this introduces an additional significant error, in our case concerning the \citet[][for Orion\,A, Orion\,B]{Nishimura2015}, \citet[][for L1622]{Kun2008}, \citet[][for L1616]{Maddalena1986}, and \citet[][for IC\,2118]{Kun2001} emission line surveys. However, we point out that this additional uncertainty does not change the results presented in this paper. Particularly, the relative radial motions of the studied clouds persist even when using different standard solar motions for the conversions.

\section{Cloud mass estimate} \label{apx:masses}

We estimate the cloud masses ($M_\mathrm{cloud}$) from dust emission and extinction maps, using the Herschel-Planck map by \citet{Lombardi2014} and the 2MASS extinction map by \citet{Lombardi2011} for regions not covered by the Herschel-Planck map. Measuring near-infrared dust extinction is overall a robust tracer of column-densities since it is virtually independent of dust temperature \citep{Goodman2009} and extinction maps were used to calibrate the Herschel-Planck maps. \citet{Lombardi2014} determined the optical depth at \SI{850}{\micro\metre} by calibrating Herschel with Planck observations and then converted $\tau_{850}$ to $A_\mathrm{K}$ via an extinction map, where they found a linear relation, which was updated by \citet{Meingast2018}: $A_\mathrm{K} \mathrm{(mag)} = \tau_{850} \cdot 3050$. The final dust column-density map was constructed at Herschel resolution ($\SI{36}{\arcsec}$ at $\SI{500}{\micro\metre}$).  Extinction $A_\mathrm{K}$ was then converted to solar masses (M$_\odot$) for each pixel as follows:
\begin{equation} \label{equ:convfactor}
M_\mathrm{gas} = A_\mathrm{K} \times ( N_\mathrm{H}/A_\mathrm{K} \cdot M_\mathrm{p} \cdot \mu_\mathrm{He} \cdot a_\mathrm{pix} )
\end{equation}
with $N_\mathrm{H}/A_\mathrm{K} = 1.24 \times 10^{22}$ cm$^{-2}$ mag$^{-1}$ (\cite{Hasenberger2016}), 
proton mass $M_\mathrm{p} = 1.67 \times 10^{-27}$\,kg, 
and a correction for 10\% He abundance $\mu_\mathrm{He} = 1.37$.
The pixel area ($a_\mathrm{pix}$) was calculated individually according to the determined distance to each subregion using either the Herschel-Planck map pixel scale of $15''$ or 2MASS extinction map pixel scale of $90''$.

In Orion\,A and Orion\,B (L1630-S/N) we estimate the cloud masses for pixels with extinctions $A_K > 0.45$\,mag using the Herschel-Planck map. We would like to note that the clouds extend beyond $A_K > 0.45$\,mag, but merge with the background, in particular toward the Galactic north, making it impossible to establish reliable cloud borders. This is not the case for the cometary clouds L1616 and IC\,2118, seen in projection against a more uniform background, while L1622 overlaps with background cloud structures and also needs to be confined within an extinction threshold. For the three cometary clouds we use the 2MASS extinction map to estimate masses and slightly lower extinction thresholds of $A_K > 0.36$\,mag for L1622 and $A_K > 0.235$\,mag for the rest, without compromising cloud confusion. Overall, and because cloud borders are hard to define in the continuum, our cloud mass measurements are lower-limits to the true cloud masses.

The extinction contours also determine the projected cloud surface areas ($a_\mathrm{today}$), as used in Sects.~\ref{momentum} and \ref{discussion:mom}, in combination with individual borders for each subregion.
We correct for background emission by subtracting $\Delta_{A_\mathrm{K,HP}} = 0.04$\,mag for the Herschel-Planck map and $\Delta_{A_\mathrm{K,2M}} = 0.05$\,mag for the 2MASS map. These values were estimated from control fields located off from the clouds at regions of almost zero extinction. 
Additionally, the cometary clouds, which were only covered by the 2MASS map, were calibrated to fit the Herschel-Planck mass estimates. To this end we selected four control clumps that roughly resemble a similar sparse clumpy structure and are within both the Herschel-Planck and 2MASS maps. With this we estimated a factor of 1.1 to calibrate 2MASS-masses to Herschel-masses.

To get the total mass of each subregion we add stellar masses using the selected YSO samples. For simplicity we use the number of YSOs as tabulated in Table~\ref{tab:averages_rv_pm} ($N_\mathit{Gaia}$) and multiply by five, since about 20\% to 30\% of YSOs are left in a \textit{Gaia}-cleaned sample, and this also accounts for overall incompleteness of a YSO sample.
\begin{equation} \label{equ:nr_yso}
\begin{aligned}
  & N_\mathrm{YSO} = 5 \cdot N_\mathit{Gaia} \\
  & M_\mathrm{YSO} = 0.5 \cdot N_\mathrm{YSO} \\
  & M_\mathrm{cloud} = M_\mathrm{gas} +  M_\mathrm{YSO}
\end{aligned}
\end{equation}
The resulting masses and surface areas ($a_\mathrm{today}$) of the 14 subregions are listed in Table~\ref{tab:momentum} in Sect.~\ref{momentum}.

The mass estimates and estimates of projected surface areas include several assumptions and uncertainties, which we would like to address here. As a first caveat, the masses were extracted within extinction contours that select reliable cloud regions (higher column-densities) and exclude low density regions that could be related, but are confused with the background. As mentioned before, this approach implies that masses should be treated as lower limits since they are likely underestimated. As a second caveat, the hard borders chosen to split the main clouds could cause a further underestimation of masses. Especially the Orion\,A cloud was split up into bins along $l$, while the cloud itself shows a continuous distribution. To treat these cloud parts as separate entities is not ideal, however, the separations were chosen to account for the velocity and distance gradient, and consequently are applied consistently for the mass and surface area estimates. When setting less conservative extinction contours we find that the masses would be larger by a factor of about two. 
Finally, the last caveat comes from the mapping techniques, in that the Herschel-Planck dust column-density map and the 2MASS extinction map are not perfect at very high column-densities. For Herschel-Planck, massive stellar heating near young embedded clusters caused the Herschel observations to saturate (near the ONC, especially at Orion BN/KL). For the 2MASS extinction map, very high-column densities could be underestimated due to insufficient sampling of background stars. This map caveats are mitigated by the fact that the affected regions (very high column-densities) occupy very small solid angles of the clouds, hence do not contribute significantly to the final masses.

\section{Auxiliary figures} \label{aux:figures}

\begin{figure*}[!ht]
    \centering
        \includegraphics[width=1\linewidth]{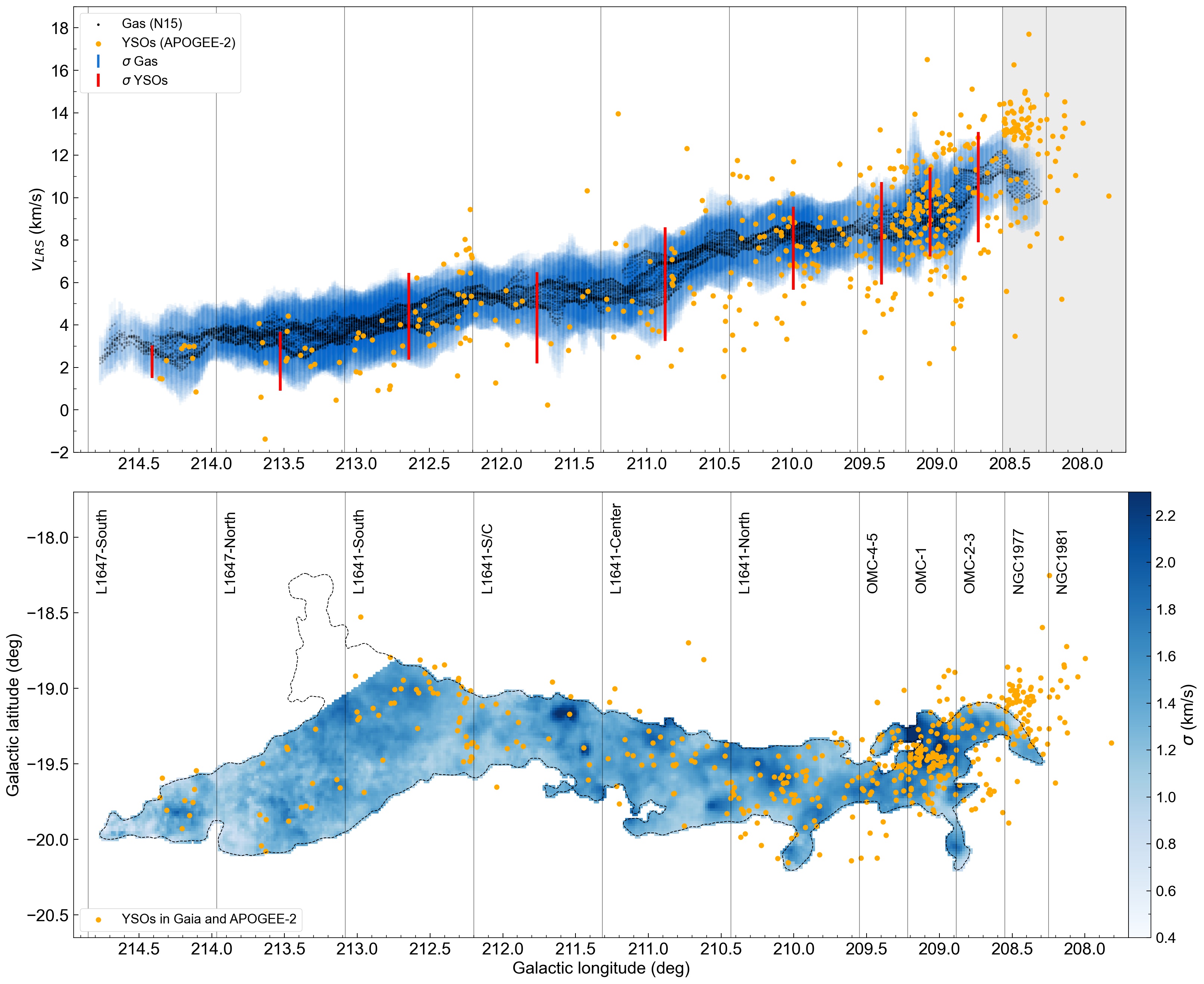}
    \caption{Velocity dispersion in Orion\,A.
    \textit{Top:} PV-diagram (\vlsr vs $l$) for CO gas and YSOs in Orion\,A (see also Fig.~\ref{fig:pv-gas-ysos}). The blue error bars represent the one-sigma velocity dispersion of the gas for each line of sight. Over-plotted are the YSOs (orange) and their velocity dispersion per region (red bars) for comparison.
    \textit{Bottom:} Sigma map, showing the velocity dispersion for each pixel in blue color scale. Over-plotted are the YSOs as orange dots. The velocity dispersion variations show that there are regions with relatively higher $\sigma$ (dark-blue). Such regions often correspond to cluster regions or regions of higher column-density.
    See Fig.~\ref{fig:pv-gas-ysos} and text (Sect.~\ref{gas-rv}) for more explanations.}
    \label{fig:pv-sigma-oriona}
\end{figure*}

\begin{figure}[!ht]
    \centering
        \includegraphics[width=1\linewidth]{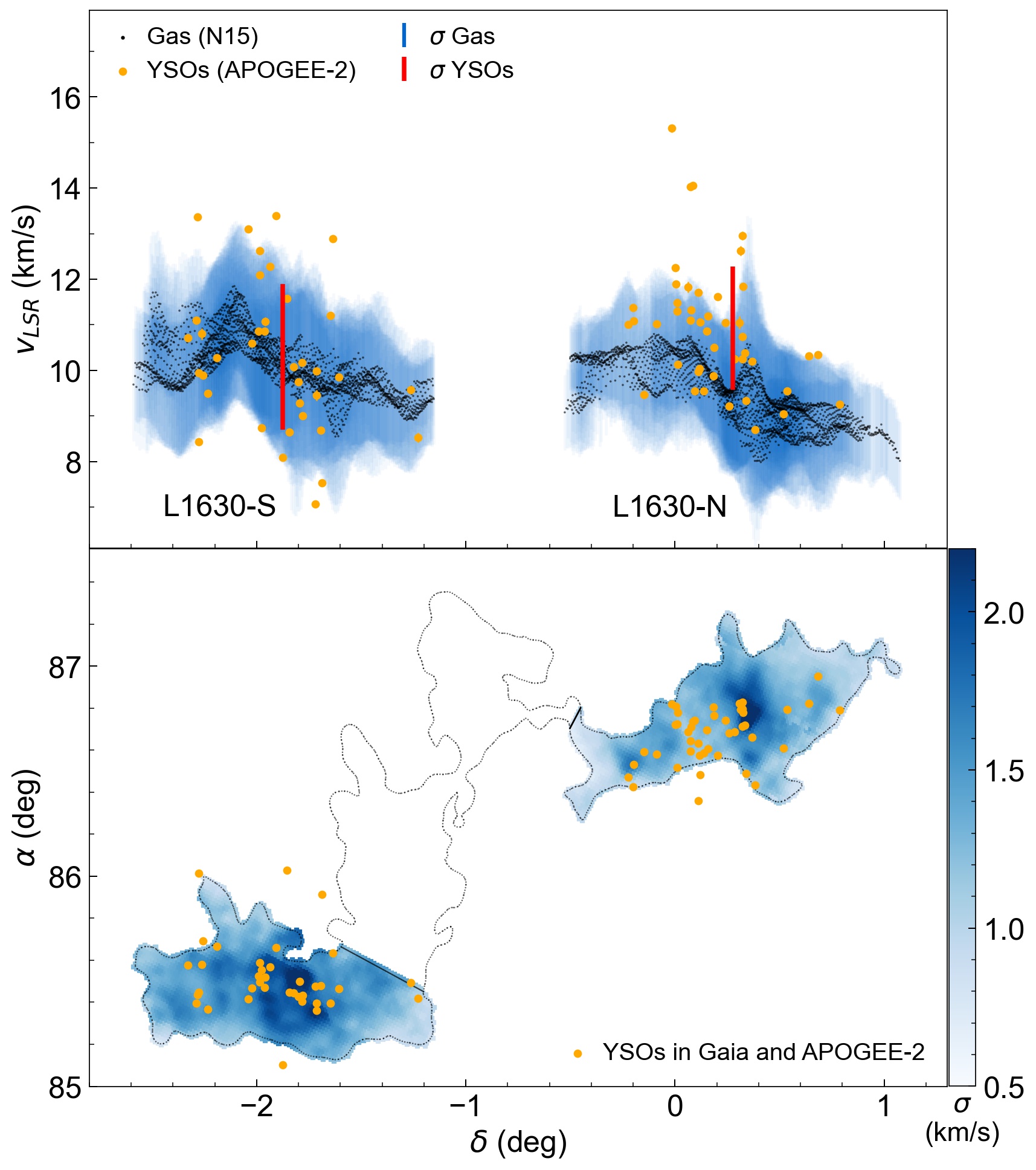}
    \caption{Velocity dispersion in Orion\,B. Similar as Fig.~\ref{fig:pv-sigma-oriona}.
    \textit{Top:} PV-diagram (\vlsr vs $\delta$) for gas and YSOs in Orion\,B (see also Fig.~\ref{fig:pv-gas-ysos_orionb}). 
    \textit{Bottom:} Sigma map, showing the velocity dispersion for each pixel in blue color scale. See Fig.\,\ref{fig:pv-gas-ysos_orionb} and text (Sect.~\ref{gas-rv}) for more explanations.}
    \label{fig:pv-sigma-orionb}
\end{figure}

In this section we provide auxiliary figures to better underline several statements included in the main part of this paper. Figures~\ref{fig:pv-sigma-oriona} and \ref{fig:pv-sigma-orionb} show PV-diagrams for gas and YSOs in Orion\,A and B, similar to Figs.~\ref{fig:pv-gas-ysos} and \ref{fig:pv-gas-ysos_orionb}, while here the velocity dispersion of the gas is highlighted. The top panels include error bars (blue), which represent the velocity dispersion ($\sigma$) for each individual pixel. These figures clearly illustrate that the velocity dispersion of the gas and the YSOs (scattered orange dots) are very similar, where only some YSOs in Orion\,A show a slightly higher dispersion compared to the gas. The calculated average velocity dispersion for gas and YSOs within each bin (subregion) are given in Table~\ref{tab:averages_rv_pm}, and are shown as bars in Figs.~\ref{fig:pv-gas-ysos} and \ref{fig:pv-gas-ysos_orionb} (top panels).

\begin{figure*}[!ht]
    \centering
    \small
    \includegraphics[width=0.99\linewidth]{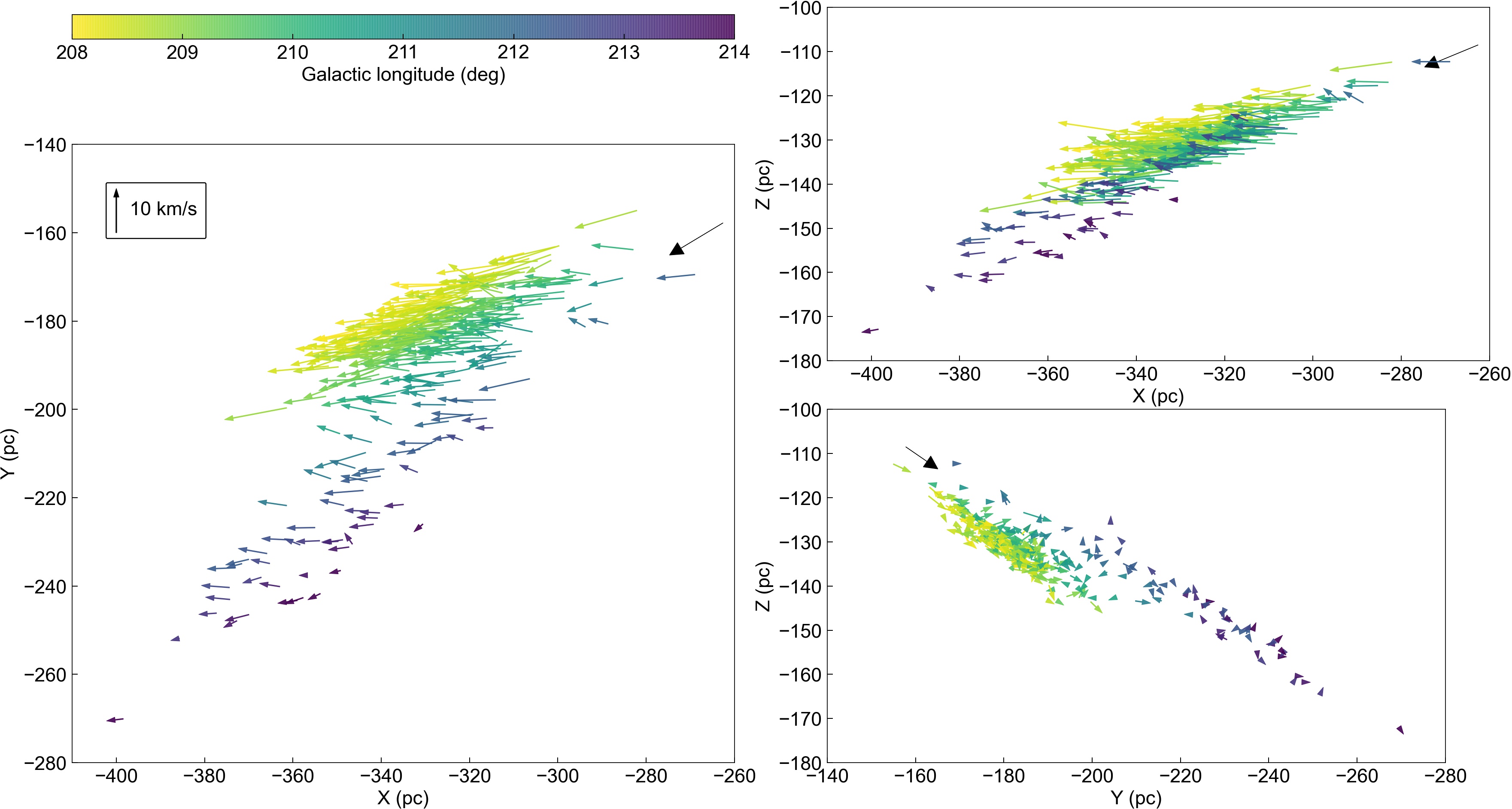}
    \caption
    [YSO UVW Orion\,A.]
    {Cartesian 3D positions and motions of YSOs in Orion\,A. The scatter plot shows YSOs with measured \textit{Gaia} DR2 astrometry and \mbox{APOGEE-2} radial velocities, projected in a Heliocentric Galactic Cartesian coordinate frame ($X$, $Y$, $Z$). The arrows are Cartesian velocity vectors ($U_\mathrm{LSR}$, $V_\mathrm{LSR}$, $W_\mathrm{LSR}$), color-coded for $l$ for orientation (head is yellow, tail is purple). A representative arrow with a length of 10\,km/s is given in the legend in the $X/Y$ panel. The black arrows in the corners of each panel indicate the line of sight from the sun, pointing toward $(l,b) = (\SI{211.0}{\degree}, \SI{-19.5}{\degree})$ and plotted from $d=$ 325\,pc to 340\,pc. 
    }
    \label{fig:ysos_uvw_glon}
\end{figure*}

\begin{figure*}[!ht]
    \centering
    \small
    \includegraphics[width=0.99\linewidth]{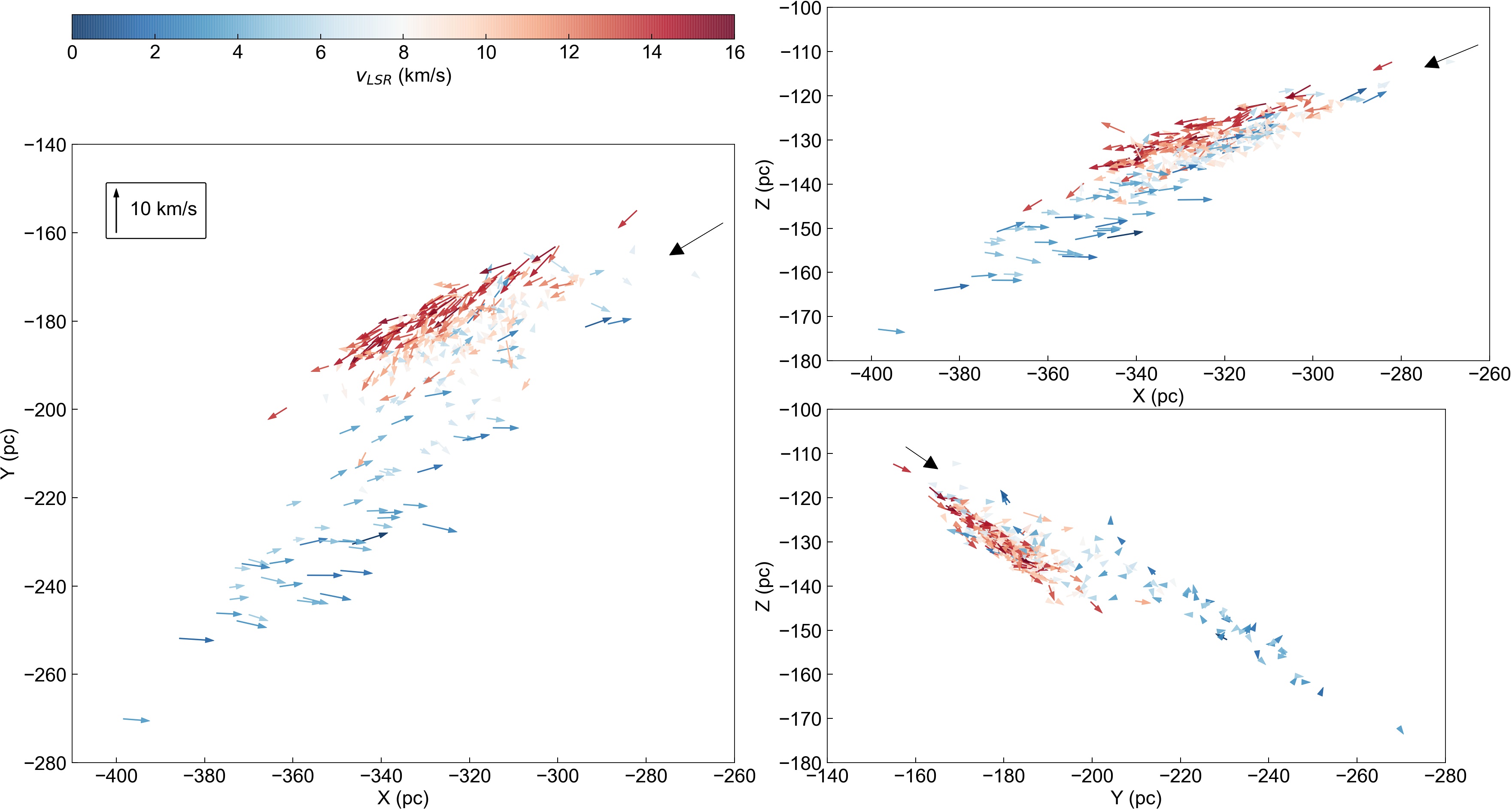}
    \caption
    [YSO UVW relative.]
    {Cartesian 3D positions and relative space motions of YSOs in Orion\,A. The Cartesian velocity vectors were calculated relative to average YSO motions of sources in Orion\,A, color-coded for \vlsr (head is red, tail is blue). See Fig.~\ref{fig:ysos_uvw_glon} and text for more details.}
    \label{fig:ysos_uvw_relative}
\end{figure*}

Figures~\ref{fig:ysos_uvw_glon} and \ref{fig:ysos_uvw_relative} show Heliocentric Galactic Cartesian representations of the YSO distribution in Orion\,A. Only YSOs which pass the \textit{Gaia} and \mbox{APOGEE-2} quality criteria are shown, while an additional parallax error cut (err$_\varpi/\varpi < 0.05$) and a flux-excess-cut \citepalias[as in][]{Grossschedl2018} were applied for a cleaner appearance. The colored arrows show the direction of the 3D velocity of each YSO, while the vector length is proportional to their
absolute velocity. Figure~\ref{fig:ysos_uvw_glon} shows the absolute $UVW_\mathrm{LSR}$ velocities in $XYZ$ directions, and Fig.~\ref{fig:ysos_uvw_relative} shows velocities relative to average $UVW_\mathrm{LSR}$ in Orion\,A. The used averages are: AVG($U_\mathrm{LSR}$) = \SI{-9.8}{km/s}, AVG($V_\mathrm{LSR}$) = \SI{0}{km/s}, AVG($W_\mathrm{LSR}$) = \SI{0}{km/s}. The relatively red-shifted YSOs near the head (including the ONC, NGC\,1977, and NGC\,1981) seem to move ``backwards'' from the Sun's point of view, supporting the proposed scenario where an external event facilitated a push on the head of Orion\,A, best visible in the top-down view ($Y/X$ plane).
In the $Z/Y$ panel it gets clear that the YSOs virtually have zero velocity in these directions. This indicates that Orion\,A has reached a minimum in its orbit around the galactic center, where the motion in $Z$-direction ($W_\mathrm{LSR}$) slowed down to a halt, hence the stars will fall back to the Galactic mid-plane in the ``near'' future. $V_\mathrm{LSR}$ near zero indicates that the stars moves similar to the LSR in $Y$-direction.

\end{appendix}
\end{document}